\documentclass[pra,twocolumn,superscriptaddress,longbibliography,nofootinbib]{revtex4-2}

\usepackage[utf8]{inputenc}
\usepackage{graphicx}

\graphicspath{{./figures/}}

\usepackage[dvipsnames]{xcolor}
\definecolor{dark-blue}{rgb}{0,0.2,0.6}

\usepackage[colorlinks,%
            pdfusetitle,%
            urlcolor=dark-blue,%
            citecolor=dark-blue,%
            linkcolor=dark-blue]{hyperref}

\usepackage{etoolbox}
\makeatletter
\pretocmd{\NAT@open}{\begingroup\color{\@citecolor}}{}{}
\apptocmd{\NAT@close}{\endgroup}{}{}
\makeatother

\usepackage{stix}
\DeclareMathAlphabet{\mathcal}{OMS}{cmsy}{m}{n}
\DeclareSymbolFont{CMAlt}{OMX}{cmex}{m}{n}
\DeclareMathSymbol{\sumop}{\mathop}{CMAlt}{"50}
\DeclareMathSymbol{\intop}{\mathop}{CMAlt}{"52}

\usepackage{amsmath}
\usepackage{upgreek}

\newcommand{\ket}[1]{\ensuremath{\left|{#1}\right\rangle}}

\newcommand{\yb}{\ensuremath{{^\text{174}\text{Yb}}}}

\newcommand{\tP}[1]{\ensuremath{{^3\text{P}_{#1}}}}
\newcommand{\tD}[1]{\ensuremath{{^3\text{D}_{#1}}}}

\newcommand{\tS}[1]{\ensuremath{{^3\text{S}_{#1}}}}

\newcommand{\sS}[1]{\ensuremath{{^1\text{S}_{#1}}}}

\newcommand{\sP}[1]{\ensuremath{{^1\text{P}_{#1}}}}

\newcommand{\mathunit}[1]{\ensuremath{\,#1}}
\newcommand{\asciimathunit}[1]{\ensuremath{\,\text{#1}}}

\newcommand{\nm}{\asciimathunit{nm}}
\newcommand{\um}{\mathunit{\upmu\text{m}}}
\newcommand{\mm}{\asciimathunit{mm}}

\newcommand{\mHz}{\asciimathunit{mHz}}
\newcommand{\Hz}{\asciimathunit{Hz}}
\newcommand{\kHz}{\asciimathunit{kHz}}
\newcommand{\MHz}{\asciimathunit{MHz}}
\newcommand{\GHz}{\asciimathunit{GHz}}
\newcommand{\THz}{\asciimathunit{THz}}

\newcommand{\mW}{\asciimathunit{mW}}

\newcommand{\Gauss}{\asciimathunit{G}}

\newcommand{\uK}{\mathunit{\upmu\text{K}}}

\newcommand{\ms}{\asciimathunit{ms}}

\newcommand{\Erec}{E_\text{rec}}

\newcommand{\subfigref}[2]{\hyperref[fig:#1]{\ref*{fig:#1}(#2)}}

\usepackage{xr}

\begin{document}


\title{State-dependent potentials for the $\sS{0}$ and $\tP{0}$ clock states of neutral ytterbium atoms}

\author{Tim~O.~H\"ohn}
\altaffiliation{These authors contributed equally to this work.}

\author{Etienne~Staub}
\altaffiliation{These authors contributed equally to this work.}

\author{Guillaume~Brochier}
\altaffiliation{Current address: Laboratoire Kastler Brossel, Coll\`ege de France, CNRS, ENS-PSL University, Sorbonne Universit\'e, 11 Place Marcelin Berthelot, 75005 Paris, France}

\author{Nelson~Darkwah Oppong}
\altaffiliation{Current address: JILA, University of Colorado and National Institute of Standards and Technology, and Department of Physics, University of Colorado, Boulder, Colorado 80309, USA}

\author{Monika~Aidelsburger}
\email{Monika.Aidelsburger@physik.uni-muenchen.de}

\affiliation{Ludwig-Maximilians-Universit{\"a}t, Schellingstra{\ss}e 4, 80799 M{\"u}nchen, Germany}
\affiliation{Munich Center for Quantum Science and Technology (MCQST), Schellingstra{\ss}e 4, 80799 M{\"u}nchen, Germany}


\date{\today}

\begin{abstract}
	
We present measurements of three distinctive state-(in)dependent wavelengths for the $\sS{0}-\tP{0}$ clock transition in \yb{} atoms. Specifically, we determine two magic wavelengths at $652.281(21)\THz$ and $542.50205(19)\THz$, where the differential light shift on the $\sS{0}-\tP{0}$ clock transition vanishes, and one tune-out wavelength at $541.8325(5)\THz$, where the polarizability of the \sS{0} ground state exhibits a zero crossing. 
The two new magic wavelengths are identified by spectroscopically interrogating cold \yb{} atoms on the clock transition in a one-dimensional optical lattice. The ground-state tune-out wavelength is determined via a parametric heating scheme. With a simple empirical model, we then extrapolate the ground and excited state polarizability over a broad range of wavelengths in the visible spectrum.
  
\end{abstract}

\maketitle


\section{Introduction}
State-dependent optical potentials play an important role in the context of quantum simulation~\cite{georgescu_quantum_2014}, computation~\cite{henriet_quantum_2020} and metrology~\cite{ludlow_optical_2015} with neutral atoms. 
The application of local differential light shifts enables high-fidelity preparation of taylored initial states for quantum simulation of out-of-equilibrium dynamics~\cite{trotzky_time_resolved_2008,weitenberg_single-spin_2011,nascimbene_experimental_2012,fukuhara_quantum_2013,dai_four-body_2017}, 
improved cooling and adiabatic state preparation schemes~\cite{robens_low-entropy_2017,yang_cooling_2020,sun_realization_2021}, 
and provides an invaluable resource for quantum-computing protocols~\cite{mandel:2003,labuhn_single-atom_2014,levine_parallel_2019,zhang_functional_2022,daley:2008,pagano_fast_2019,gonzalez-cuadra_fermionic_2023}. 
In the context of quantum simulation they further offer prospects to realize more complex model Hamiltonians, where different internal states are interpreted as independent species~\cite{riegger:2018,rubio-abadal_many-body_2019,oppong:2022}, thereby significantly reducing experimental complexity compared to atomic mixture experiments.

\begin{figure}[t!]
	\includegraphics{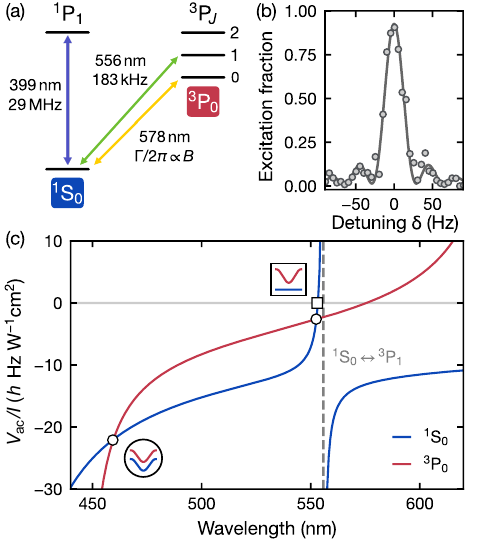}
	\caption{\textbf{Dominant transitions in $\yb$ and ac polarizabilities for the $\sS0$ and $\tP{0}$ states.}
	$\left(\text{a}\right)$ Simplified level scheme for $\yb$. The clock transition to the metastable $\tP{0}$ state is addressed 
	via magnetically-induced clock spectroscopy to admix the $\tP{1}$ state, rendering the linewidth $\Gamma$ magnetic-field ($B$-field) dependent.
	$\left(\text{b}\right)$ Fourier-limited clock spectroscopy. 
	Single-shot data points are fitted using a Rabi lineshape (solid line) to yield the linewidth and position. For this dataset we find a full-width-half-maximum of $26.7(6)$\,Hz. 
	The excitation fraction is calculated by imaging $\sS{0}$ and $\tP{0}$ independently.
	$\left(\text{c}\right)$ Intensity-normalized light shift $V_{\textrm{ac}}/I$ of the $\sS0$ and $\tP{0}$ states as a function of wavelength; here $I$ denotes the intensity. 
	Magic wavelengths measured in this work are indicated by circles.
	At these points the $\sS{0}$ and $\tP{0}$ polarizabilities are equal in magnitude and sign (schematic in round inset). 
	The square refers to the measured $\sS{0}$ tune-out wavelength, where the polarizability of the $\sS{0}$ state vanishes (schematic in square inset).}
	\label{fig:fig1}
\end{figure}

State-dependent potentials have so far been mostly realized with the bosonic Alkali atoms rubidium (Rb)~\cite{mandel:2003,gadway_superfluidity_2010,soltan-panahi_multi-component_2011,yang_spin-dependent_2017,de_hond_preparation_2022} 
and cesium (Cs)~\cite{forster_microwave_2009,belmechri_microwave_2013} due to their large fine-structure splitting $\Delta_{\textrm{FS}}$. 
The coherence time in this case is fundamentally limited by the use of near-resonant light, whose detuning must be smaller than $\Delta_{\textrm{FS}}$. 
Hence, state-dependent potentials can introduce heating due to spontaneous emission. 
Alkaline-earth(-like) atoms (AEAs), such as ytterbium (Yb) and strontium (Sr), on the other hand have two valence electrons. 
The two lowest-lying electronic states ($\sS{0}$ and $\tP{0}$) belong to the singlet and triplet manifold 
and are connected by an ultra-narrow doubly-forbidden electric dipole transition. 
This renders both states stable on typical experimental timescales~\cite{porsev_possibility_2004}
and enables state-dependent potentials with low scattering rates~\cite{daley:2008,gerbier_gauge_2010,dzuba:2010,safronova_extracting_2015,dzuba:2018}, as demonstrated in Refs.~\cite{riegger:2018,heinz:2020,oppong:2022}. This is particularly relevant for fermionic quantum simulations because all fermionic Alkali atoms have a small fine-structure splitting.

While state-dependent potentials can be harnessed for many applications, they can be detrimental for cooling and trapping of neutral atoms~\cite{mckeever:2003,hutzler_eliminating_2017,ton:2022}, in particular, for high-fidelity imaging in quantum gas microscopes~\cite{yamamoto:2016,trisnadi_design_2022} and tweezer arrays~\cite{cooper:2018,norcia:2018,saskin_narrow-line_2019,covey_2000-times_2019,okuno_high-resolution_2022}. 
Moreover, in the context of metrology with optical lattice clocks, state-dependent terms need to be finely compensated~\cite{ye:2008,ludlow_optical_2015,campbell_fermi-degenerate_2017} in order to reach high precision and accuracy~\cite{nicholson_systematic_2015,ushijima_cryogenic_2015,mcgrew_atomic_2018}.

In general, an accurate prediction of the ac polarizabilities from \textit{ab initio} calculations~\cite{mitroy_theory_2010} can be extremely challenging. 
In this work we present experimental measurements of three distinctive points of the ac polarizability for the $\sS{0}-\tP{0}$ clock transition of bosonic $^{174}$Yb atoms: one tune-out wavelength~\cite{leblanc_species-specific_2007,arora_tune-out_2011,herold_precision_2012,holmgren_measurement_2012,cheng:2013}, where the potential for the $\sS{0}$ state vanishes and two magic wavelengths~\cite{porsev_possibility_2004,mitroy_theory_2010,zhang_magic_2021}, where the potentials for $\sS{0}$ and $\tP{0}$ are identical [Fig.~\subfigref{fig1}c]. 
Our work constitutes the first measurement of a tune-out wavelength for the $\sS{0}$ state of \yb{} and complements the currently known magic wavelengths at $759.3\,$nm~\cite{barber:2008} and $397.6\,$nm~\cite{mcgrew:2020}. 
Moreover, we use our experimental results to develop a simple empirical model to estimate the $\sS{0}$ and $\tP{0}$ ac polarizabilities over a wide range of visible frequencies.

\section{AC polarizabilities and experimental setup}
\label{sec:sec2}

\begin{figure*}[t!]
	\includegraphics{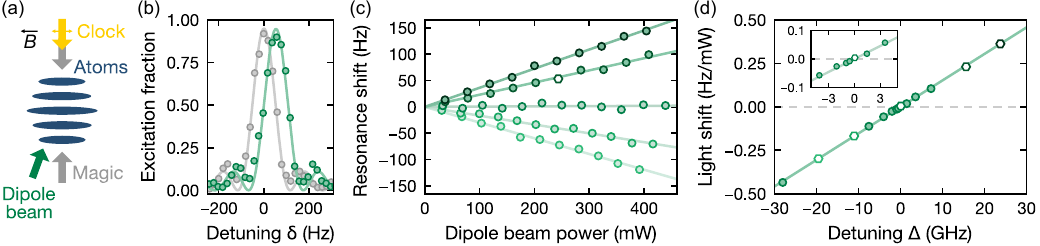}
	\caption{\textbf{Magic-wavelength measurement.}
	$\left(\text{a}\right)$ Schematic drawing of the experimental setup. 
	The 1D magic lattice (grey) and the additional dipole beam (green) intersect at an angle of $\simeq$ 1° (sketch not to scale) to avoid backreflections of the latter. 
	The spectroscopy laser (yellow) is collinear with the 1D lattice and its polarization (double arrow) is parallel to the direction of the magnetic field $B$. All beams overlap at the position of the atoms (blue).
	$\left(\text{b}\right)$ Resonance spectrum without (grey) and with dipole beam (green). 
	The resonance shift is extracted by fitting a Rabi lineshape (solid line)~\cite{SM}. The result is shown as a white hexagon in $\left(\text{c}\right)$.
	$\left(\text{c}\right)$ Resonance shift as a function of the dipole beam power. 
	Each data point consists of an average of three individual spectroscopy measurements as shown in (b). 
	The frequency dependence of the slope is evaluated using linear fits (solid lines) with vanishing offset. 
	The respective data points are shown as white hexagons in $\left(\text{d}\right)$.
	$\left(\text{d}\right)$ Fitted light shifts extracted from measurements as shown in $\left(\text{c}\right)$ 
	as a function of the frequency of the dipole beam relative to the magic wavelength $f_\mathrm{m1}$ (see main text). 
	The error bars correspond to the standard deviation (c) and $1\sigma$-fit uncertainty (d) and are smaller than the data points. Inset: Zoom-in on the region close to the magic wavelength.}
	\label{fig:fig2}
\end{figure*}

The dynamic (ac) polarizability characterizes the response of an atomic system to an applied electromagnetic field. 
An atomic state experiences a shift in energy $V_{\text{ac}}$ that is proportional to the real part of the frequency-dependent polarizability $\alpha$,
which itself depends on the frequency and strength of transitions to higher-lying energy levels.
Since the $\sS{0}$ and $\tP{0}$ states belong to different spin manifolds, they couple to different sets of higher-lying states, 
which in turn leads to different wavelength dependencies of their respective ac polarizabilities [Fig.~\subfigref{fig1}{c}].
For Yb, the ground state polarizability for visible wavelengths is dominated 
by the broad transition to the $(6s6p)\sP{1}$ state at $399\nm$ and the narrow 
intercombination transition to the $(6s6p)\tP{1}$ state at $556\nm$ (Fig.~\ref{fig:fig1}).
By contrast, the metastable excited state polarizability is mainly affected by a number of 
visible and infrared transitions to the $^3$S and $^3$D orbital angular momentum manifolds. 

To determine the precise location of the magic and tune-out wavelengths [circles and square in Fig.~\subfigref{fig1}{c}], 
we utilize both spectroscopic (Sec.~\ref{sec:sec3}) 
and parametric heating measurements (Sec.~\ref{sec:sec4}). 
In both cases the experimental sequence starts by loading $\simeq \! 10^7$ $\yb$ atoms from a 
Zeeman-slowed atomic beam into a three-dimensional (3D) magneto-optical trap (MOT) operating at the narrow \mbox{$\sS{0} - \tP{1}$} transition. 
We enhance the capture efficiency of the MOT using a set of crossed slowing beams on the \mbox{$\sS{0} - \sP{1}$} transition~\cite{plotkin-swing:2020}.
For the measurement of the tune-out (magic) wavelength, we directly load $\simeq \! 70 \times 10^3$ ($\simeq \! 300 \times 10^3$) atoms from the MOT into a $\simeq \! 390 \Erec$ ($\simeq \! 730 \Erec$) deep 1D optical lattice at the $\lambda_{\textrm{lat}}=759.3\nm$ magic wavelength~\cite{barber:2008}; here $ \Erec = h^2/(2m\lambda^2_{\textrm{lat}})$ denotes the recoil energy, $h$ Planck's constant and $m$ the mass of $^{174}$Yb.
During the measurements, the atoms are trapped at an axial temperature of $\simeq \! 12\uK$~\cite{SM} in the lattice.
Since the bosonic isotope $\yb$ exhibits zero nuclear angular momentum, neither the $\sS{0}$ nor the $\tP{0}$ 
state possess hyperfine structure, and the clock transition at $578\nm$ is induced via a magnetic field, 
which admixes a small fraction of the $\tP{1}$ state to the $\tP{0}$ state~\cite{taichenachev:2006}.
Here we apply a $B$-field of $\simeq \! 100\Gauss$ to coherently excite the $\tP{0}$ state [Fig.~\subfigref{fig1}{b}].
To obtain the excited state fraction, we separately detect the fraction of atoms in the $\sS{0}$ and $\tP{0}$ states.
To this end, atoms in the $\tP{0}$ state are repumped using the $\tP{0} - \tD{1}$ transition~\cite{hinkley:2013}.

\section{Magic wavelength measurements}
\label{sec:sec3}
At magic wavelengths the differential polarizability for a given pair of states vanishes. 
Its precise location is commonly determined by measuring the wavelength-dependent differential light shift induced by an optical potential via spectroscopy on the respective transition (see e.g. Ref.~\cite{barber:2008}). 
To this end we overlap an almost collinear additional frequency-tunable laser beam (dipole beam) with the 1D lattice [Fig.~\subfigref{fig2}{a}] and perform spectroscopy measurements with and without the dipole beam [Fig.~\subfigref{fig2}{b}]. 
The induced line shift in turn depends linearly on the applied laser power, as shown in Fig.~\subfigref{fig2}{c}, and the
corresponding slope determines the wavelength-dependent differential light shift [Fig.~\subfigref{fig2}{d}].
For small detunings $\Delta$ from the magic wavelength, the differential light shift scales approximately linearly with detuning and
the magic wavelength can be extracted with a least-squares fit of a linear function [Fig.~\subfigref{fig2}{d}].
This yields a value of
\begin{equation*}
	f_\mathrm{m1} = \left[ 542502.05 \pm 0.08_{\mathrm{stat}} \left(^{+0.01}_{-0.11}\right)_{\mathrm{sys}} \right] \GHz
\end{equation*}
for the magic wavelength close to the \mbox{$\sS{0} - \tP{1}$} transition. 
The statistical uncertainty is given by the $1\sigma$-error of the fit.

We minimize systematic errors due to slow drifts during a single measurement by randomizing the sequence of clock laser detunings $\delta$ at which the dipole beam is toggled.
Furthermore, we ensure a homogeneous resonance shift across the atomic cloud by choosing the dipole-beam 
waist $w_0= 125\um$ to be much larger than the transverse cloud size and about twice the waist of the 1D lattice $w_0 \simeq2w_{\mathrm{lat}}$.
We calibrate the dipole beam power prior to the start of each measurement 
and keep track of its spatial overlap with the lattice using a camera.
Over the course of the measurement, we observe a small position drift, 
which we estimate to result in a reduction of the intensity of the dipole beam of at most 7\% at the location of the atoms.
Assuming this drift to be linear, we obtain a conservative estimate 
for the systematic error of $-0.11 \GHz$.
A second contribution of $+0.01\GHz$ stems from the calibration uncertainty of the dipole beam intensity,
which is caused by etaloning induced by a bandpass filter mounted on the intensity stabilization photodiode~\cite{SM}.

We repeat the same measurement in the blue optical spectrum 
and obtain the following value for the magic frequency:
\begin{equation*}
	f_\mathrm{m2} = \left[ 652281 \pm 10_{\mathrm{stat}} \left(^{+11}_{-0}\right)_{\mathrm{sys}} \right] \GHz.
\end{equation*}
The corresponding data is shown in the Supplemental Material~\cite{SM}.
We note a substantially larger statistical error compared to $f_\mathrm{m1}$. 
This is due to less available dipole-beam laser power at this wavelength and a smaller slope of the differential polarizability.
To obtain the same measurement precision, we therefore require larger detunings $\Delta$ from the magic wavelength.
However, for large detunings a significant deviation from a simple linear dependence of the differential light shift is expected.
To estimate the corresponding systematic error 
we calculate the theoretically expected polarizability for the detunings sampled in the experiment 
and determine the difference between the expected zero crossing and the one extracted with a linear fit. 
We find that for a detuning range of $\pm 1 \THz$ the systematic error amounts to $11 \GHz$. 
We therefore limit the range of detunings to this regime. 
Further sources of systematic uncertainties are negligible 
compared to the dominant error sources described above~\cite{SM}.

\section{Tune-out wavelength measurement}
\label{sec:sec4}
The tune-out condition refers to a zero-crossing of the frequency-dependent polarizability for a certain state.
These points can be useful for tests of fundamental physics, since the respective frequencies can be 
determined with high precision, as they do not require exact measurements of light intensities or transition strengths~\cite{{henson:2022}}.
Several techniques have been employed in the past to locate tune-out frequencies based on Kapitza-Dirac diffraction~\cite{catani:2009,herold_precision_2012,schmidt_precision_2016,kao:2017,ratkata:2021}, 
interferometric techniques~\cite{holmgren_measurement_2012,leonard_high-precision_2015}, 
periodic modulation~\cite{henson:2015,bause:2020,heinz:2020}
or measurements of trap frequencies~\cite{henson:2022}.

\begin{figure}[t!]
	\includegraphics{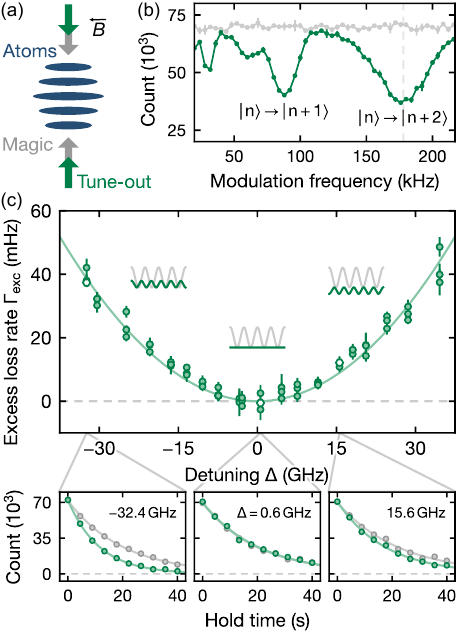}
	\caption{\textbf{Tuneout-wavelength measurement.}
		$\left(\text{a}\right)$ Sketch of the experimental setup. 
		The shallow tune-out lattice (green) is superimposed onto the deep magic lattice (grey). 
		$\left(\text{b}\right)$ Loss spectrum of $\sS{0}$ atoms as a function of the modulation frequency (green)
		after $40\times10^3$ modulation periods
		at a detuning of $327 \GHz$ from the tune-out wavelength. 
		Each data point consists of two measurements and the error bar is the corresponding standard deviation.
		Reference measurement without tune-out lattice are shown in grey. 
		Transitions to higher motional states are observed at $89 \kHz$ and $178 \kHz$, which correspond to excitations by one or two harmonic oscillator quanta, respectively. 
		The smaller resonance features correspond to third harmonics, resulting from the square-pulse amplitude modulation. 
		The lines connecting the data points are a linear interpolation.
		The vertical dashed line (grey) indicates the modulation frequency used for the following measurements.
		$\left(\text{c}\right)$ Excess loss rate $\Gamma\!_{\mathrm{exc}}$ of ground state atoms calculated by subtracting the bare loss rate $1/\tau_0$ 
		from the loss rate obtained with modulation $1/\tau$.
		The solid green line is the fitted quadratic function without offset.
		For each value of the detuning $\Delta$, three measurements are taken. The error bars for each data point correspond to the standard deviation of the fit uncertainties of the respective decay curves.
		Insets: Schematic drawings of the amplitude of the two lattices: magic (grey) and tune-out (green).
		Lower panels: The bare loss rate (grey) and the loss rate with modulation (green) are obtained from fitting the ground state atom population as a function of the modulation time with an exponential decay and zero offset.
		Displayed as solid lines are the exponential fits.
		The corresponding data points are indicated by white hexagons in the upper panel.}
	\label{fig:fig3}
\end{figure}

For our measurement we employ parametric heating induced by an additional perturbing lattice near the tune-out wavelength, similar to Ref.~\cite{heinz:2020}. 
After loading the atoms from the MOT into the $\simeq \! 390 \Erec$ deep magic lattice, we increase its depth to $\simeq 730\Erec$ within $10 \ms$.
Subsequently, we turn on the second lattice within $5\ms$, co-propagating with the main lattice as schematically shown in Fig.~\subfigref{fig3}{a}.
This lattice is several orders of magnitude weaker to ensure that the trap parameters are still dominated by the magic lattice~\cite{SM}.
By modulating the amplitude of this additional lattice we induce controlled heating, which we detect via atom-loss measurements.
Since the magic and tune-out wavelengths are incommensurate, the modulation causes both phase and amplitude modulation in the combined lattice.
Hence, the modulation introduces excitations to higher-lying vibrational bands that are separated by one or two motional quanta, in contrast to pure phase or amplitude modulation.
An example spectrum is shown in Fig.~\subfigref{fig3}{b}, where we can indeed identify two dominant resonances.
Assuming the lattice potentials are locally well described by a harmonic potential, those correspond to excitations by one, $\ket{\mathrm{n}}\!\rightarrow\!\ket{\mathrm{n+1}}$, or two, $\ket{\mathrm{n}}\!\rightarrow\!\ket{\mathrm{n+2}}$, harmonic oscillator quantum numbers, respectively.
For technical reasons, we choose to implement a square-amplitude modulation with a modulation amplitude of $100\%$, which allows us to precisely control the intensity.
This explains the additional less pronounced resonances at one third of these frequencies in Fig.~\subfigref{fig3}{b}.
To avoid systematic errors we take data with and without the perturbation lattice at each modulation frequency and randomize the order in which the data points are taken.

Selecting the $\ket{\mathrm{n}}\!\rightarrow\!\ket{\mathrm{n+2}}$ resonance, we then scan the time during which we hold and modulate the atoms in the combined lattice and observe the resulting loss in atom population from the trap [Fig.~\subfigref{fig3}{c}].
By fitting the data using an exponential decay and comparing the lifetime $\tau$ with the lifetime in the bare magic lattice without probe light $\tau_0$, we extract an excess loss rate $\Gamma\!_{\mathrm{exc}}\equiv1/\tau - 1/\tau_0$ [Fig.~\subfigref{fig3}{c}, bottom panels].
At the tune-out wavelength, the ground state experiences no trapping potential and the excess loss rate approaches zero.

Since we rely only on intensity modulation of the trap for parametric heating, we may assume a quadratic scaling behavior for the excess loss rate of the form $\Gamma\!_{\mathrm{exc}} \propto \alpha^2 \propto \Delta^2$~\cite{savard:1997,heinz:2020}.
Fitting the excess-loss-rate data using this scaling, we extract the value of the tune-out wavelength for the $\sS{0}$ state to be

\[ f_{\mathrm{to}} = \left[ 541832.49 \pm 0.23_{\mathrm{stat}} \left(^{+0.05}_{-0.24}\right)_{\mathrm{sys}} \right] \GHz,   \]

\begin{figure*}[t!]
	\includegraphics{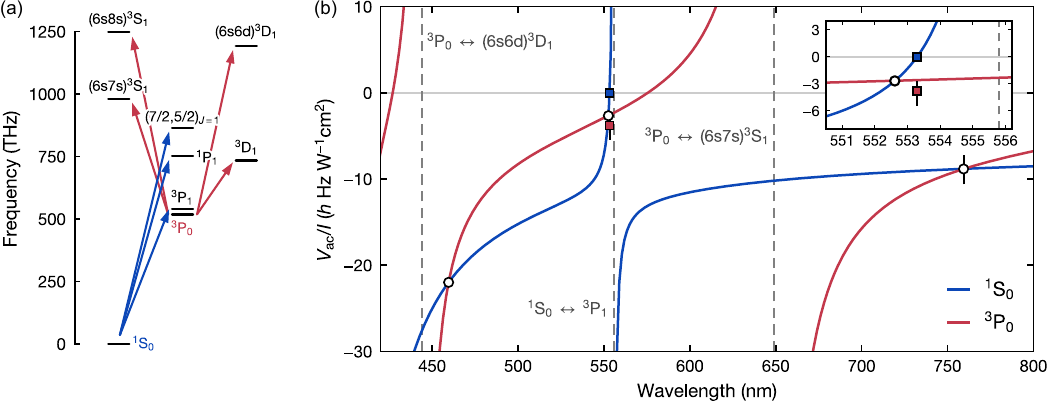}
	\caption{\textbf{Empirical model for the ac polarizability of the $\sS{0}$ and $\tP{0}$ states.}
	$\left(\text{a}\right)$ Energy levels and transitions taken into account for the empirical ac polarizability model. 
	For $\sS{0}$ we find levels beyond the core-excited state $(7/2,5/2)_{J=1}$ sufficiently captured by a global offset. 
	In the case of $\tP{0}$, we use an effective transition to absorb all states lying higher than $(6s8s)\tS{1}$.
	$\left(\text{b}\right)$ Position of measured magic and tune-out wavelengths, 
	and the magic wavelength measured in Ref.~\cite{barber:2008}.
	Note that the vertical position of the blue and green magic wavelength markers is only determined by the crossing of the polarizability curves.
	The error bar for the red magic wavelength is determined based on polarizability estimates from several experiments \cite{SM, barber:2008, riegger:2018}.
	Inset: Zoom-in on the green magic wavelength and the measured polarizabilities at the $\sS{0}$ tune-out wavelength close to the $\sS{0} - \tP{1}$ transition at $556\nm$.}
	\label{fig:fig4}
\end{figure*}

\noindent where the statistical uncertainty is given by the $1\sigma$-fit uncertainty. 

The systematic uncertainty is dominated by the effect of anharmonic deviations of the trap loss rate at large detunings.
We observe that for $\Delta\gtrsim 21 \GHz$ the measured excess loss rate is lower than expected. 
We evaluate the corresponding systematic error in two ways: 
First, we calibrate the reduction of the loss rate experimentally by increasing the
modulation amplitude at fixed detuning, 
correct the data accordingly and fit the value of the tune-out wavelength using the same quadratic function as mentioned above. 
We find that in this case the tune-out wavelength 
is shifted by $+0.05 \GHz$ with respect to $f_\mathrm{to}$.
Second, we successively limit the fit range to smaller detuning values. 
For the smallest detuning range where the deviations are negligible, i.e., $|\Delta| < 12\GHz$, 
we obtain a maximum systematic offset of $-0.24 \GHz$~\cite{SM}.
Additional systematic errors, e.g., due to pointing drifts or intensity calibrations, are found to be negligible.

In order to estimate the excited-state polarizability at the tune-out wavelength, we employ the same experimental setup as described in Sec.~\ref{sec:sec3}. 
We set the frequency of the dipole beam to be equal to the previously determined tune-out wavelength $f_\mathrm{to}$ in order to determine the light shift induced on the clock transition.
Since the tune-out frequency is almost $700\GHz$ detuned from the magic wavelength, the light shift at the maximum available power is substantial. 
We therefore limit the dipole beam power and hence the observed frequency shifts to less than $- 500 \Hz$ to mitigate broadening of the light-shifted resonance due to the non-uniformity of the Gaussian intensity profile of the dipole beam. We find the following intensity-normalized potential for the $\tP{0}$ state at the $\sS{0}$ tune-out wavelength:
\[ V_\mathrm{ac}/I = h \times \left[-3.8  \pm 0.07_{\mathrm{stat}} \left(^{+1.1}_{-1.6}\right)_{\mathrm{sys}}\right]~ \mathrm{Hz / \frac{W}{cm^2}}. \]
The statistical uncertainty is given by the $1\sigma$-error of the fit, while the systematic error stems from the uncertainty in determining the intensity of the dipole beam.  
The largest contributions arise from the estimated uncertainties in the waist of the dipole beam and its longitudinal focus position overlap with the lattice focus~\cite{SM}.

\section{Empirical model for ac polarizability}

The position of the measured tune-out and magic wavelengths entails valuable information about the ground and excited clock-state polarizability.
In particular, it allows for direct feedback to elaborate many-body perturbation (MBPT) or configuration interaction (CI) theory models to provide more accurate predictions of the ac polarizability of either state~\cite{dzuba:2010, dzuba:2018, guo:2010}.
Here, we present a simple empirical model in order to extrapolate the ac polarizability between the measured magic wavelengths at $459.6\nm$ and $759.3\nm$, covering a broad range of the visible spectrum and benchmark it with additional experimental results.

Starting from the ac polarizability in second-order perturbation theory, the normalized potential induced by a laser beam of intensity $I$ can be written as
\[ V_\mathrm{ac} / I = - \sum_{J'} \frac{\omega_{J'J}}{3 \hbar \epsilon_0 c}\ \frac{\left|\left\langle J | \! | \mathbf{d} | \! | J'\right\rangle \right|^2}{\omega_{J'J}^2- \omega^2}, \]
with $\hbar = h/2\pi$, $\epsilon_0$ the vacuum permittivity, and $c$ the speed of light.
Here we sum over all electric dipole transitions $J \leftrightarrow J'$ with corresponding angular frequency $\omega_{J'J}$ of the respective state $\ket{J}$~\cite{kien:2013, manakov:1986, SM}.
Note that due to the vanishing nuclear spin and the consequential absence of hyperfine splitting in $\yb$ the only contribution to the light shift is the scalar polarizability $\alpha^{(0)}$ and we can write $J$ and $J'$ instead of $F$ and $F'$ throughout.
We can further relate the modulus squared of the reduced matrix element $| \langle J | \! | \mathbf{d} | \! | J' \rangle |^2$ to the linewidth $\Gamma_{J'J}$ of the corresponding transition \cite{steck}.

Figure~\subfigref{fig4}{a} gives an overview of the transitions that were taken into account.
For the $\sS{0}$ state, the polarizability at the measured wavelengths is largely defined by just three transitions: the broad transition to $\sP{1}$ at $399\nm$, the narrow intercombination line to the $\tP{1}$ state at $556\nm$, and a transition to the electronic core-excited state denoted as $(7/2,5/2)_{J=1}$ at $347\nm$~\cite{blagoev:1994, takasu:2004}.
Similarly, the polarizability of the $\tP{0}$ state is dominated by the four lowest-lying transitions to the $\mathrm{S}$- and $\mathrm{D}$-manifold, in particular the transition to $(6s5d)\tD{1}$ at $1389\nm$, the \mbox{$\tP{0} - (6s7s)\tS{1}$} transition at $649\nm$, and the \mbox{$\tP{0} - (6s6d)\tD{1}$} transition at $444\nm$~\cite{beloy:2012, baumann:1985}.
However, for some of the $\tP{0}$ transitions, only the lifetime of the corresponding excited states has been measured, but the exact decay rate to the $\tP{0}$ state is not precisely known.
In that case we estimate the respective decay rate by calculating the branching ratio into all lower-lying dipole-allowed states, treating the involved states as eigenstates $\ket{J} = \ket{LSJ}$ of the angular momentum operators~\cite{SM}.
Note that, especially for the $\tP{0}$ state, there are many more transitions to higher-lying excited states that will also play a role in determining the precise polarizability at a given wavelength.
For the wavelength range we are studying, these can be treated as sufficiently far detuned such that we can describe them as a single effective transition.
Both the effective transition wavelength and linewidth are then free parameters of our $\tP{0}$ polarizability model.

For the ground state, the neglected higher-lying transitions are even further detuned such that we can absorb their effect in a small overall offset as the only free fit parameter.
This offset is fixed to $-0.8~\mathrm{Hz / \frac{W}{cm^2}}$ by requiring that the $\sS{0}$ polarizability is zero at the measured tune-out wavelength.
As this also determines the $\tP{0}$ polarizability at the experimentally known magic wavelengths, we fit an effective transition with linewidth $\Gamma_\mathrm{eff} = 2\pi \times 23\MHz$ at a wavelength of $\lambda_\mathrm{eff} = 376\nm$ to these three values.
The resulting polarizability predictions are close to the measured $\tP{0}$ polarizability at the $\sS{0}$ tune-out wavelength [Fig. \subfigref{fig4}{b}, inset].
Moreover, we benchmark our model against measured polarizability ratios at $670\nm$, $671.5\nm$, and $690.1\nm$~\cite{riegger:2018, oppong:2022}, showing excellent agreement within error bars~\cite{SM}.
In addition, we note that our $\tP{0}$ results match the CI+MBPT predictions from Ref.~\cite{dzuba:2010} well, whereas for the $\sS{0}$ state we observe a deviation for wavelengths below $\simeq \! 556\nm$, with a discrepancy of $\simeq \! 10-15\%$ in between the blue and green magic wavelength.

\section{Summary and conclusion}
In this work, we have presented measurements of three state-(in)dependent wavelengths, which open new possibilities for state-selective manipulation of the Yb $\sS{0}$ and $\tP{0}$ optical clock state pair. Using a tweezer array at the tune-out wavelength enables selective trapping of atoms in the $\tP{0}$ state while leaving $\sS{0}$ atoms unaffected. This may pave the way for novel tweezer rearrangement and local addressing techniques.
The new magic wavelengths on the other hand may prove valuable for cooling and trapping.
In particular, the blue magic wavelength at $459.6\nm$ may constitute a promising alternative~\cite{pagano_fast_2019}.
Specifically, optical tweezers at this wavelength not only benefit from the larger polarizability compared to $759.3\nm$, but also from the shorter wavelength, allowing for almost seven times deeper potentials with the same amount of laser power.
Accurate knowledge of the ac polarizabilities opens up new possibilities for novel quantum computing protocols~\cite{daley:2008,pagano_fast_2019} and applications in quantum simulation. 
This includes the engineering of artificial magnetic fields~\cite{jaksch_creation_2003,gerbier_gauge_2010}, lattice gauge theories~\cite{surace_abinitio_2023}, twisted-bilayer models~\cite{gonzalez-tudela_cold_2019,luo_spin-twisted_2021,meng_atomic_2021,du_atomic_2023} or analogue quantum chemistry simulations~\cite{arguello-luengo:2019}.
Our measurements also serve as a benchmark for existing theoretical polarizability models, enhancing the accuracy of predictions.
Experimental results for distinctive points in the ac polarizability are vital for benchmarking of transition matrix elements and to enhance the accuracy of theoretical predictions~\cite{herold_precision_2012,holmgren_measurement_2012,safronova:2015,leonard_high-precision_2015,zhang_magic_2021,ratkata:2021}. 

\begin{acknowledgments}
We thank C.~Bachorz, D.~Gr\"oters, B.~Hebbe Madhusudhana, T.~Marozsak and D.~Robledo for technical contributions to the experiment. We further acknowledge fruitful discussions with V.~A.~Dzuba, S.~Blatt, Y.~Takahashi, and G.~Pasqualetti. We thank V.A.~Dzuba, Y.~Takahashi, and F.~Scazza for careful reading of the manuscript and S.~F\"olling, G.~Pasqualetti and O.~Bettermann for providing their clock laser as a frequency reference for identifying the zero-crossing of our ULE cavity. 
We also thank the team of \textit{H\"ubner Photonics} for the broad frequency-tunable laser source, C-WAVE VIS+IR Low Power~\cite{sperling_breakthrough_2021}.
This project has received funding from the Deutsche Forschungsgemeinschaft (DFG, German Research Foundation) under Germany’s Excellence Strategy -- EXC-2111 -- 390814868, funding from the European Research Council (ERC) under the European Union’s Horizon 2020 research and innovation program (grant agreement No. 803047), from the German Federal Ministry of Education and Research (BMBF) via the funding program quantum technologies -- from basic research to market (contract number 13N15895 FermiQP) and from the Initiative \textit{Munich Quantum Valley} from the State Ministry for Science and the Arts as part of the High-Tech Agenda Plus of the Bavarian State Government.
\end{acknowledgments}

\cleardoublepage

\section*{Supplementary Information}

\renewcommand{\thefigure}{S\arabic{figure}}
\renewcommand{\theHfigure}{S\arabic{figure}}
 \setcounter{figure}{0}
\renewcommand{\theequation}{S.\arabic{equation}}
 \setcounter{equation}{0}
 \renewcommand{\thesection}{S\arabic{section}}
\setcounter{section}{0}

\section{Experimental techniques}

\subsection{Experimental sequence}

We begin our measurements by loading a magneto-optical trap (MOT) from a commercial cold atomic beam system (\textit{AOSense Yb Beam RevC}).
The MOT is operated on the narrow \mbox{$\sS{0} - \tP{1}$} transition, and we routinely load $\simeq 10\times 10^6$ \yb{} atoms into the MOT within $300\ms$.
To bring atoms from the cold atomic beam (exit velocity $\simeq 40\asciimathunit{m/s}$) within the capture velocity of our MOT, we employ a set of crossed slowing beams operated on the broad \mbox{$\sS{0} - \sP{1}$} transition following the scheme in Ref.~\cite{plotkin-swing:2020}.

After loading the MOT, the cold atomic beam and crossed slowing beams are switched off, and we compress the MOT within $100\ms$.
At the end of the compression, we keep the magnetic field gradient as well as the MOT intensity and detuning constant for another $100\ms$.
During this time, we load the one-dimensional optical lattice by linearly ramping its depth to $\simeq \! 390 \, (730) \Erec$ for the tune-out (magic) measurements.
Subsequently, the MOT light and magnetic field are turned off, and we either ramp the lattice to its final depth of $\simeq \! 730 \, \Erec$ within $10\ms$ (tune-out measurements) or keep its depth constant (magic measurements).
Depending on the initial depth, we load between $70\times 10^3$ and $300 \times 10^3$ atoms into the optical lattice, before turning off the MOT beams. 

\subsection{Initial state} \label{sec:initial-state}

\begin{figure}[t!]
	\includegraphics[width=\columnwidth]{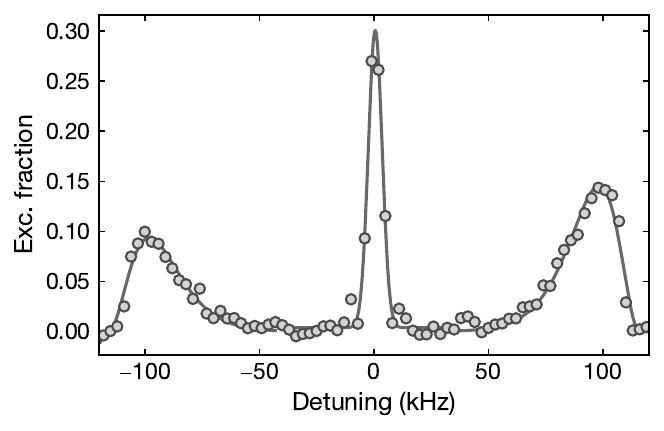}
    \caption{\textbf{Sideband spectrum showing the carrier transition and red and blue motional sidebands.} The position of the sidebands allows us to determine the longitudinal trap frequency $f_\mathrm{z}$, while the ratio of red to blue sideband area determines the corresponding longitudinal temperature $T_\mathrm{z}$. 
    By fitting \eqref{s1} to the blue sideband it is possible to extract the radial temperature $T_\mathrm{r}$.
	The solid line consists of the fitted Lorentzian lineshapes to the carrier and the sharp sideband edges as well as \eqref{s1} fitted to the red and blue shallow sideband slopes.}
	\label{fig:s1}
\end{figure}

From the MOT, atoms are loaded into a magic 1D optical lattice at $\lambda_\mathrm{lat} = 759.3\,\nm$, 
generated by a titanium:sapphire (Ti:Sapph) ring laser (\textit{Sirah Matisse CS}), 
which is digitally locked to a wavemeter (\textit{High-Finesse WS8}).
To measure the depth of our lattice and the temperature of the atoms, 
we perform sideband spectroscopy~\cite{blatt:2009}. 
The atoms in the lattice are probed with the clock laser along the tightly confined longitudindal axis of the lattice, here defined as $z$-axis.
By scanning the clock laser at high power and a large magnetic field of $400\Gauss$ over a range of detunings $\delta$ and measuring the excited state fraction we are able to resolve the carrier transition as well as the corresponding red and blue motional sidebands, as shown in Fig.~\ref{fig:s1}.
On resonance, we measure a Rabi frequency of $\Omega_0 = 2\pi\times 2.8(1)\kHz$.
It is assumed that the atomic population in different motional states follows a thermal distribution and that individual transitions between motional states may accurately be described by a Lorentzian lineshape with a width given by $\Omega_0$.
We use this lineshape to fit the sharp edge of the blue sideband and extract the longitudinal trap frequency $f_\mathrm{z}$ from the center of the Lorentzian.

The shape of the shallow sideband slope towards the carrier is given by a weighted sum of individual absorption cross sections~\cite{blatt:2009}
\begin{equation}
    \sigma\left(\delta\right) \propto \sum_{n_\mathrm{z} = 0}^{N_\mathrm{z}}e^{-E_{n_\mathrm{z}}/k_\mathrm{B}T_\mathrm{z}}\sigma_{n_\mathrm{z}}\left(\delta\right)
	\label{s1}
\end{equation}
where 
\[
    \sigma_{n_\mathrm{z}}\left(\delta\right) = \frac{\beta^2}{\widetilde{\gamma}\left(n_\mathrm{z}\right)}\left(1 - \frac{\delta}{\widetilde{\gamma}\left(n_\mathrm{z}\right)}\right)e^{-\beta\left(1 - \frac{\delta}{\widetilde{\gamma}\left(n_\mathrm{z}\right)}\right)}\Theta\left(\widetilde{\gamma}\left(n_\mathrm{z}\right) - \delta\right),
\]
with $\beta = \left(\widetilde{\gamma}\left(n_\mathrm{z}\right)/ f_{\mathrm{rec}}\right)\left(h f_z/k_\mathrm{B}T_\mathrm{r}\right)$, the Heaviside function $\Theta$, the radial temperature $T_\mathrm{r}$, the longitudinal trap frequency $f_\mathrm{z}$, and $\widetilde{\gamma}\left(n_\mathrm{z}\right) = f_\mathrm{z} - f_{\mathrm{rec}}\left(n_\mathrm{z}+1\right)$ the energy difference between longitudinal harmonic oscillator states $n_{\mathrm{z}}$, where $f_\mathrm{rec} = h/(2m\lambda^2)$ is the lattice recoil frequency, $k_\mathrm{B}$ the Boltzmann constant, and $h$ Planck's constant.
This form of the shallow sideband edge originates in the radial motion of the atoms along the weakly confined axis.
By fitting \eqref{s1} to the blue sideband we extract the radial temperature of $T_\mathrm{r} \simeq 20.0(5) \uK$ at a lattice depth of $728(2) \, \Erec$.

The longitudinal temperature $T_\mathrm{z} \simeq 11.9(2) \uK$ is obtained by numerically computing the sideband area ratio.

\subsection{State-selective absorption imaging}\label{sec:absorption-imaging}

We perform absorption imaging on the \mbox{$\sS{0} - \sP{1}$} transition along an axis tilted away from the longitudinal axis of the lattice by $\simeq 10^{\circ}$. 
Since our imaging intensity is close to the transition saturation intensity of $I_{\text{sat}} = 60$ mW/cm$^2$, we compensate the non-linear dependency of the optical density on the imaging intensity following the procedure in Ref.~\cite{reinaudi:2007}.

To detect atoms in the clock state $\tP{0}$, we employ a repumping scheme on the $\tP{0} - \tD{1}$ transition~\cite{hinkley:2013} with subsequent absorption imaging on the same transition used for the ground-state atoms.
Using two subsequent absorption images, one before repumping and one after repumping, we always determine both the atomic density distribution in $\sS{0}$ and $\tP{0}$.
The repumping efficiency for transferring population from $\tP{0}$ via the $\tP{0} - \tD{1}$ transition to the ground state is estimated by comparing the ground and excited state atom number after a resonant $\pi$-pulse when we toggle the clock beam.
Owing to the finite branching ratio from the $\tD{1}$ state to the metastable $\tP{2}$ state one would na\"ively assume a difference of $\simeq \! 3\%$ between the excited state population with the clock beam on and the ground state population with deactivated clock beam.
However, we find a repumped fraction of $84(1)\%$. 
We attribute this lower value to the rapid loss of excited state atoms due to inelastic collisions before they are repumped back to the ground state.
We account for the finite repumped fraction by correcting the excited-state population in $\tP{0}$ accordingly for all measurements reported in this work.

\subsection{Clock spectroscopy} \label{sec:clock}
To drive the clock transition, we employ a commercial laser system (\textit{Toptica TA-SHG pro}) with up to $\simeq 700\mW$ at $578\nm$.
A small fraction of the seed laser power at the fundamental wavelength of $1156\nm$ 
is used to lock the laser onto a commercial optical reference cavity 
(\textit{Menlo Systems ORC-Cylindric}) using the Pound-Drever-Hall stabilization technique~\cite{drever:1983}.
At this wavelength the finesse of the optical cavity is $>4\times 10^5$. This was determined by measuring the cavity photon lifetime 
after suddenly extinguishing the incident light using an acousto-optical modulator (AOM).
The residual linear drift of the cavity corresponds to $\simeq60\mHz/$s and is compensated with feed-forward 
on the offset lock using a fiber-coupled waveguide electro-optical modulator.
For all spectroscopy data, we fit the excitation fraction using the well-known Rabi lineshape
\begin{equation}
	\mathrm{P}\left(\delta\right)=a\frac{4}{\pi^2}\mathrm{sinc}^2\left(\frac{\pi}{2}\frac{\Omega}{\Omega_0}\right),
	\label{eq:rabilineshape}
\end{equation}
where $\Omega = \sqrt{\Omega_0 + \delta}$ is the generalized, $\Omega_0$ the resonant Rabi frequency, $a$ the amplitude and $\delta$ the detuning.

\subsection{Systematic uncertainties in the magic-wavelength measurements} \label{sec:magic}
To determine the differential light shift induced by the frequency-tunable dipole beam we perform clock spectroscopy 
with and without the additional beam and fit the spectra with Rabi lineshapes, as defined in Eq.~(\ref{eq:rabilineshape}). 
The frequency-tunable laser source for the magic-wavelength measurement in the green spectral range, 
to determine $f_\mathrm{m1}$, is a vertical-external-cavity surface-emitting laser (\mbox{VECSEL}, \textit{Vexlum VALO-SHG-SF}), 
while the one for measuring $f_\mathrm{m2}$ in the blue spectral range is a widely tunable optical parametric oscillator 
(\textit{H\"ubner Photonics C-WAVE VIS+IR Low Power}~\cite{sperling_breakthrough_2021}).

\begin{figure}[t!]
	\includegraphics[width=\columnwidth]{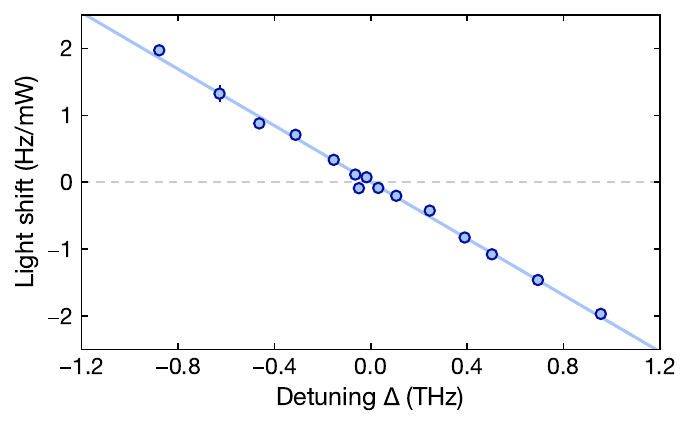}
	\caption{\textbf{Wavelength-dependent light-shift around the magic wavelength $f_\mathrm{m2}$.} Similar to Fig.~2 in the main text, we fit a linear function to the light shift and extract the magic wavelength at the zero crossing of the fit function. The error bars are determined in identical fashion to those of Fig.~2 in the main text.}
	\label{fig:s2}
\end{figure}

\paragraph{Light-shift data and systematic uncertainties in the measurement of the magic wavelength $f_\mathrm{m2}$.}
The experimental results for the wavelength-dependent light shift used to fit the value of $f_\mathrm{m2}$, 
as reported in the main text, are shown in Fig.~\ref{fig:s2} together with the least-squares fit of the linear function.
The dominant systematic uncertainty is determined by deviations from the linear approximation for the large detuning range of $\pm 1 \THz$, as discussed in the main text. 
To assess the systematic uncertainty due to relative pointing drifts between the lattice and dipole beam, 
we image the position of both beams in a plane that is approximately conjugate to the plane of the atoms.
For a quantitative analysis, we assume the measured drift to be linear in time, starting from an initially perfect overlap.
To calculate the corresponding correction to the measured light shifts, 
it is necessary to estimate the integrated dipole beam intensity reduction experienced by the atoms as the drift increases.
We therefore fit a rotated 3D Gaussian to a set of representative atom-cloud images 
to obtain the cloud size in the lattice ($\sigma_\mathrm{z} = 288(18)\um, \sigma_\mathrm{r} = 16.52(4)\um$), 
which is then integrated over the Gaussian dipole beam intensity distribution to obtain the weighted mean of the intensity experienced by the atoms.
The resulting correction to the light shifts leads to a systematic uncertainty of $0.3 \GHz$ and is much smaller than the statistical error associated with this measurement.
We note that a potential mismatch between the imaging and atomic plane would lead to an increased apparent pointing drift and thus to an overestimate of the systematic error. 

\begin{figure}[t!]
	\includegraphics[width=\columnwidth]{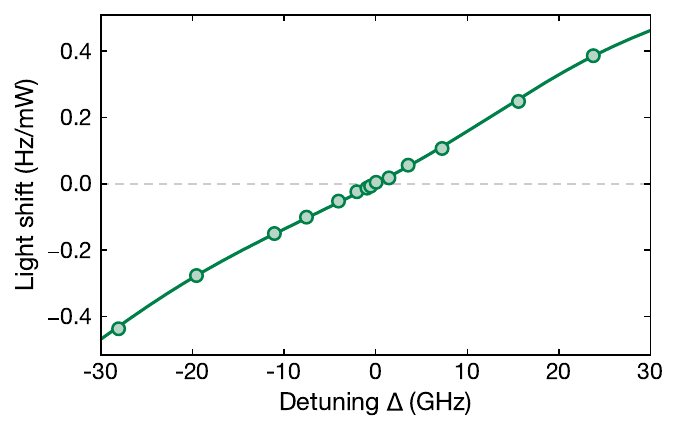}
	\caption{\textbf{Systematic uncertainty due to etaloning in the light-shift measurements for determining $f_\mathrm{m1}$.} Measured wavelength-dependent light-shifts, as in Fig.~2 in the main text, without correcting for the etaloning. The solid line shows the linear fit function taking the calibrated sinusoidal corrections due to etaloning into account.}
	\label{fig:s3}
\end{figure}

\paragraph{Systematic uncertainties in the measurement of the magic wavelength $f_\mathrm{m1}$.} 
One important systematic uncertainty for this measurement is due to relative position drifts of the lattice and the dipole beam. 
The corresponding analysis has been performed as described in the previous paragraph and gives a value of $-0.11\GHz$. 
A second contribution stems from etaloning induced by a spectral filter that is mounted onto the photodetector used to measure the intensity of the beam.
This introduces a wavelength-dependent correction for the intensity, which results in an additional sinusoidal modulation 
on top of the light-shift measurements, which are expected to scale linearly with detuning (Fig.~\ref{fig:s3}).
In order to correct for this non-linearity we simultaneously calibrate the laser beam with an integrating sphere (\textit{Thorlabs S140C}).
We find a sinusoidal dependence on the wavelength, with a relative amplitude difference of $20\%$ and a periodicity of $9.34(6)\GHz$.
We fit a sine function to this data with four free fit parameters (frequency, phase, amplitude and background)
and use this to rescale the measured light shifts at the respective laser frequencies to obtain the data shown in the main text Fig.~2(d).
In Fig.~\ref{fig:s3} we show an alternative fit function, where the sinusoidal correction is included in the fit function.
Both methods yield the same value for the magic frequency within the statistical error uncertainty of the fit.
The corresponding systematic error is computed by a linear error propagation of the fit uncertainties and amounts to $10\MHz$.

\subsection{Scattering rate at the green magic wavelength $f_\mathrm{m1}$}

To benchmark photon scattering due to the nearby \mbox{$\sS{0} - \tP{1}$} transition, 
we measure the lifetime of atoms in the ground state $\sS{0}$ in the 1D lattice with the dipole beam on and off. 
The experimental sequence is similar to the one described in the main text.
We load cold atoms from the MOT into the 1D lattice at  $\simeq \! 390 \, \Erec$.
For the measurements we then increase the depth to $\simeq \! 730 \, \Erec$ within $10\ms$ and turn on the dipole beam within $5\,$ms.
At a maximally attainable incident dipole beam power of $ \simeq \! 230 \mW$, we measure a $1/e$ decay rate of \mbox{$\Gamma = 46.3(10) \mHz$} when the dipole beam is on. 
Compared to the case where the dipole beam remains off, we measure an excess loss rate of \mbox{$\Gamma_{\!\mathrm{exc}} = 0.1(8) \mHz$} which is consistent with no lifetime difference within the error bars. 

\begin{figure*}[t!]
	\includegraphics{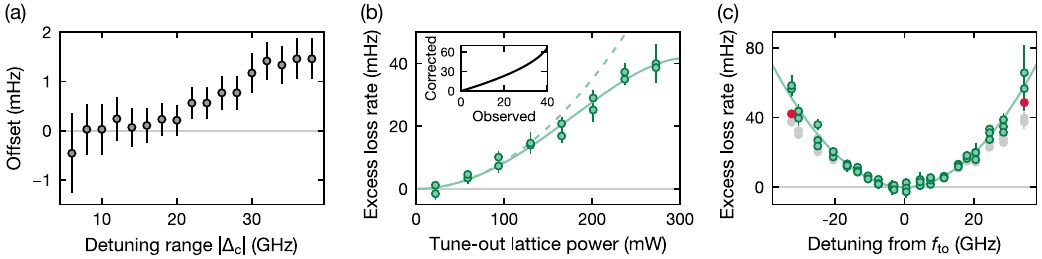}
	\caption{\textbf{Correction of the parametric heating saturation effect.} (a) Accumulation of the unphysical fit offset $c$ as a function of the detuning. As larger detunings are included in the fit, the offset increases as the deviation from a quadratic behavior around $f_{\mathrm{to}}$ increases.
	(b) Excess loss rate measurement at constant detuning but varying tune-out lattice power. For large loss rates we observe a deviation from the expected quadratic behavior (dotted line). We model this flattening behavior with a quartic function (solid line).
	Inset: Rescaling factor determined with the quartic fit to correct the observed loss rate for the anharmonicity at large detunings.
	(c) Excess loss rate $\Gamma\!_\mathrm{exc}(\Delta)$ as a function of detuning $\Delta$ from $f_\mathrm{to}$ when we employ the rescaling curve displayed in (b). Uncorrected values are shown in grey. In (b) only excess loss rates $\lesssim \! 40\mHz$ were measured, so we disregard two data points beyond this threshold (red).
	}
	\label{fig:s4}
\end{figure*}

\subsection{Parametric heating}

For the tune-out wavelength measurement we superimpose a weak frequency-tunable lattice (VECSEL) with the deep 1D lattice.
To find the parametric heating resonances we use the same sequence as described in the main text.
We load atoms into a $\simeq \!390\,\Erec$ deep lattice, 
ramp the intensity to $\simeq \! 730\,\Erec$ within 10 ms, 
turn on the shallow lattice within $5 \ms$ and hold while modulating its intensity 
using a square-wave signal generated by a function generator at a frequency $f_{\mathrm{mod}}$. 
The function generator modulates the shallow lattice by periodically turning the AOM on and off.
The modulation frequency $f_{\mathrm{mod}}$ is large enough (on the order of a few tens to few hundreds of $\kHz$) 
to prevent significant signal distortion due to the bandwidth limit of the intensity stabilization. 
This allows us to to accurately stabilize the averaged intensity of our lattice for the whole modulation frequency range.
Fourier-transforming the modulated amplitude confirms a clean rectangular modulation signal, 
where the only significant peaks appear at uneven multiples of $f_{\mathrm{mod}}$ and fall off in amplitude as $f^{-1}$.
The observed modulation resonance $f_\mathrm{res}=89\,$kHz and its first multiple $2 f_\mathrm{res}$ (cp. \ref{sec:initial-state}) 
correspond to phase and amplitude modulation in the combined lattice (see Ref.~\cite{heinz:2020} for further details).
The discrepancy between the parametric heating resonance and the lattice trap frequency $f_\mathrm{z} = 107\kHz$ independently measured using sideband spectroscopy is most likely due to the anharmonicity of the higher-lying lattice energy levels, 
which are strongly populated as the atoms are heated out of the lattice.
Note that close to the tune-out wavelength the polarizability is small and we can assume that the 1D lattice parameters 
are independent of the power of the weak perturbing lattice.
Indeed, we estimate a trap frequency of $f \simeq 40 \Hz$ for the largest detuning of $\simeq \! 35 \GHz$, 
which is more than three orders of magnitude smaller than $f_\mathrm{z}$.

\subsection{Excess loss rate and tune-out wavelength determination}

The extraction of loss rates $\Gamma$ is performed with a least-squares fit to the atom number $N(t)$ 
using a single exponential function
\begin{equation} 
N(t) = N_0 e^{-\Gamma t}. 
\end{equation}
For our lattice parameters (waist $w_{\textrm{lat}} = 66.5\um$, 
lattice spacing $\lambda_{\textrm{lat}}/2$, mean occupation number per layer $\bar{N}_\mathrm{z} \simeq 55$), 
we expect the particle densities to be too low for two- or three-body losses to play a significant role within our measured lifetimes~\cite{kitagawa:2008, fukuhara:2009}.
We test this statement by comparing the single exponential fit to super-exponential decay functions 
that take two-body or three-body losses into account.
We find that both loss rates are consistent with zero within error bars.
We therefore use a simple exponential decay for all lifetime fits in this work.
We further confirm that the lifetime in the deep 1D lattice is not modified by the presence of the static shallow tune-out lattice. 

Since we modulate the tune-out lattice at $f_\mathrm{m}=2 f_\mathrm{res}$, we expect amplitude modulation to be the main reason for heating and additional atom loss.
In this case, modulation-induced loss can be described as~\cite{savard:1997}
\begin{equation}\label{eq:1}
\Gamma\!_\mathrm{exc} = \pi^2 f_\mathrm{m}^2 S_\epsilon(2 f_\mathrm{m}) \propto f_\mathrm{m}^2 \alpha^2 I_\mathrm{to}^2,
\end{equation}
where $I_\mathrm{to}$ denotes the intensity of the tune-out lattice and we use that the one-sided modulation power spectrum, $S_\epsilon(2 f_\mathrm{m})$, is proportional to the squared modulation amplitude of the tune-out lattice, $S_\epsilon(2 f_\mathrm{m})\propto \alpha^2 I_\mathrm{to}^2$.
For constant lattice intensity this translates into $\Gamma\!_\mathrm{exc}(\Delta) \propto \alpha^2(\Delta) \propto \Delta^2$ for a small range around the tune-out frequency $f_\mathrm{to}$, where the polarizability is approximately linear in $\Delta$\cite{heinz:2020}. Within this approximation 
the tune-out frequency $f_{\textrm{to}}$ can be determined by fitting a simple quadratic function without offset.

Experimentally, however, we find that for the range of detunings probed in the experiment a simple quadratic fit does not capture the data sufficiently well. 
Adding a constant offset instead appears more accurate.
However, this offset is not supported by the vanishing measured excess loss rate close to $f_\mathrm{to}$, where we would expect the heating rate to be zero. 
In order to understand the origin of the offset, we fit the quadratic function for various detuning ranges [Fig.~\subfigref{s4}{a}] and find that the offset only becomes significant for detunings that exceed $|\Delta|\!\gtrsim\!21\GHz$. 
This seems to imply that for large detunings the loss rate is suppressed compared to the simple approximate scaling $\Gamma\!_\mathrm{exc} \propto  \Delta^2$.

To gain more insight, we measure $\Gamma\!_\mathrm{exc}$ at a fixed detuning of $\simeq \! 25\GHz$
and vary the power of the tune-out lattice [Fig.~\subfigref{s4}{b}].
Indeed, we observe a deviation from the quadratic prediction for large modulation amplitudes,
which limits the observed excess loss rates to $\simeq \! 40 \mHz$.
Empirically, we find that a quartic polynomial with an additional term $\propto I_{\textrm{to}}^4$ describes this behavior more accurately. 
In order to estimate the effect of the quartic correction on the value of the extracted tune-out frequency $f_\mathrm{to}$,
we use the quartic fit shown in Fig.~\subfigref{s4}{b} in order to rescale the measured loss rates [Fig.~\subfigref{s4}{c}]
and compare the result of quadratic fits with and without offset, yielding a difference of $53\MHz$, which we use as the positive systematic error contribution.
Furthermore, we find that within the fit uncertainty the offset indeed vanishes for the corrected data.
Note that there are two data points with excess loss rates $>40\mHz$.
Since we only calibrated loss rates of up to $40\mHz$ in Fig.~\subfigref{s4}{b} our empirical correction model 
is not applicable for these values. Therefore they are excluded from the fit.

\begin{figure}[ht!]
	\includegraphics{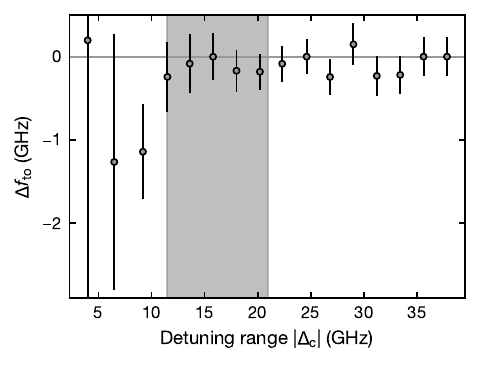}
	\caption{\textbf{Fitted tune-out frequency for constrained datasets.} We restrict the range of tune-out lattice laser detunings $|\Delta|$ to variable maximal values and fit the quadratic fitting function without offset for each dataset. For very few data points the asymmetrically spaced detunings lead to a strong bias towards lower frequencies. However, upon including data points at $|\Delta_{\textrm{c}}| > 11.5 \GHz$ the fitted minimum deviation remains within $\simeq \! 250 \MHz$ (inset). The grey shaded area displays the detuning range used to extract the systematic uncertainty from the heating saturation.}
	\label{fig:rest-dataset-shift}
\end{figure}

For the evaluation of the data shown in Fig.~3(c) in the main text, we fit a quadratic function without offset.
To evaluate the systematic uncertainty introduced by the deviations from the quadratic scaling
described above, we analyze the data for various detuning ranges $|\Delta| < |\Delta_{\textrm{c}}|$ 
around the tune-out frequency $f_{\mathrm{to}}$,
where $|\Delta_{\textrm{c}}|$ denotes the cut-off for the detuning range considered for the evaluation. 
We then perform the quadratic fit without offset for this range to obtain the minimum $\widetilde{f}_{\mathrm{to}}$ and evaluate the shift of the tune-out frequency from the value reported in the main text $\Delta f_{\mathrm{to}} = f_{\mathrm{to}} - \widetilde{f}_{\mathrm{to}}$. 
The results are shown in Fig.~\ref{fig:rest-dataset-shift}. 
The large error bars of $\Delta f_{\mathrm{to}}$ for small detuning ranges $\Delta_{\textrm{c}}$ 
are due to the limited number of data points as well as their asymmetric spacing within the detuning range.
For larger detuning ranges $|\Delta_{\textrm{c}} |\gtrsim \! 11.5 \GHz$, we find that the shift $\Delta f_{\mathrm{to}}$ 
varies around small negative shifts and does not exceed $ -250 \MHz$.

For the uncorrected data set (as used for the evaluation in the main text) large detuning ranges are unreliable for 
\mbox{$|\Delta_{\textrm{c}} | \gtrsim \! 21 \GHz$} as discussed above (Fig.~\ref{fig:s4}). 
Hence, we restrict our evaluation to detuning ranges \mbox{$11.5 \GHz \leq |\Delta_{\textrm{c}}| \leq 21 \GHz$}.
Within this range the largest deviation from zero is $-243 \MHz$ 
at $|\Delta_{\textrm{c}}| \! \simeq \! 12 \GHz$, which we identify as the dominant systematic uncertainty
of this measurement. Since we consistently observe shifts to lower frequencies, 
apart from one outlier at large detunings, we do not extract a positive systematic error contribution from this method.

\subsection{Estimating the $\tP{0}$ polarizability at the tune-out wavelength}

We examine the polarizability of the $\tP{0}$ state at the tune-out wavelength by measuring the light shift induced on the clock transition by a dipole beam operating at the tune-out wavelength. 
We make use of the fact that the $\sS{0}$ state experiences no trapping potential at the ground-state tune-out wavelength and thus any light shift induced on the clock transition at this wavelength originates from the finite ac polarizability of the $\tP{0}$ state. 
Due to the significant differential light shift at the tune-out wavelength, we expect significant broadening of the resonance for large shifts. 
To ensure that the clock resonance linewidth is only limited by the laser linewidth and not by the spatial inhomogeneity of the dipole beam we measure only at low dipole beam powers of up to $\simeq \! 35\mW$.

To compute the intensity-normalized potential $V_\mathrm{ac}/I$, we divide the measured light shift by the effective dipole beam intensity.
The latter is again estimated by integrating the dipole beam intensity distribution over the atomic cloud.
We further take the measured dipole beam waist and the small angle of $0.76(12)^\circ$ between the dipole beam and the lattice into account.

The conservatively assessed uncertainties in each of these measurements, i.e., the atom cloud size, beam waist and angle, are used to compute the linear sum of the propagated uncertainties, yielding a systematic error contribution of $0.85~ \mathrm{Hz / \frac{W}{cm^2}}$.
As the remaining sources of systematic errors, we account for a potentially imperfect focus overlap between lattice and dipole beam, the uncertainties stemming from our reference photodiode calibration, our integrating sphere and powermeter, and the finite transmission through our glass cell.

As pertains to imperfect foci overlap, we assume the mismatch to be limited to the longitudinal axis. 
This is a reasonable asssumption given that we optimize the pointing of the dipole beam (and thus the radial axis) by maximizing the induced light shift prior to our measurements. 
However, the lattice and dipole beam are focused onto the atoms through the same lens. 
Therefore, the longitudinal overlap is only determined by the relative collimation accuracy of the lattice and dipole beam prior to this focusing lens.
From our beam profiler measurements of the lattice and dipole beams, we assume a worst-case longitudinal focus mismatch of $5\mm$.
This amounts to an asymmetric error of $-0.48~ \mathrm{Hz / \frac{W}{cm^2}}$.
To calibrate the reference photodetector, we compare the measured photodetector signal with that of our integrating sphere powermeter measuring the dipole beam power incident on the atoms. 
We perform this comparison for multiple powers and fit the resulting data using a linear fit. 
The associated fit uncertainty is added to the inherent uncertainty of the integration sphere.
Finally, we measure the transmission of the dipole beam through our glass cell and assume the transmission through both windows to be the same. 
For the systematic error estimate, we conservatively assume that either the first or the second window transmits everything while the other window leads to the measured overall intensity reduction, thus either enhancing or reducing the dipole beam intensity at the atom plane.
Summing the respective absolute values of these sources of uncertainty, we obtain a systematic error contribution of $0.24~ \mathrm{Hz / \frac{W}{cm^2}}$.

\section{Derivation of the empirical ac polarizability model}

We study the ac polarizability in the semi-classical case using a classical electric light field
\begin{equation}
\mathbf{E} = \mathbf{E^{(+)}} e^{-i\omega t} + \mathrm{c.c.},
\end{equation}
where $\mathbf{E^{(+)}} = \hat{\varepsilon}E^{(+)}$ is the positive-frequency electric field amplitude vector and $\omega$ is the light's angular frequency, interacting via $V = - \mathbf{E} \cdot \mathbf{d}$ with a quantum mechanical atom with electric dipole operator $\mathbf{d}$ as discussed in depth in Refs.~\cite{kien:2013, manakov:1986}.
The second-order ac Stark shift for a state $\ket{J}$ can then be expressed as~\cite{steck}
\begin{equation} \label{eq:acstark}
\Delta E_J = -\mathrm{Re} \! \left(\alpha_J(\omega)\right) |E^{(+)}|^2
\end{equation}
with $|E^{(+)}|^2 = \frac{1}{4} |E|^2$ and
\begin{align}
\alpha_J(\omega) = & \frac{1}{3\hbar} \sum_{n'J'} (-1)^{J'+1} \left| \langle n'J'| \hat{\varepsilon} \cdot \mathbf{d} | J \rangle \right|^2 \notag \\
& \times \Big( \frac{1}{\omega_{n'J'J} + \omega + i\Gamma_{n'J'J}/2} \notag \\ 
& + \frac{1}{\omega_{n'J'J} - \omega - i\Gamma_{n'J'J}/2} \Big)
\end{align}
where $\omega_{n'J'J}$ and $\Gamma_{n'J'J}$ denote the angular frequency and linewidth of the electronic transition from $J$ to $n'J'$.
Note that we made use of the absence of hyperfine structure in $\yb$ and $J=0$ for both the $\sS{0}$ and $\tP{0}$ state to drop all but the scalar contributions to the ac polarizability.
Omitting the complex terms as we are only interested in polarizabilities far detuned from any resonance and using that only transitions $J=0 \rightarrow J'=1$ will contribute significantly, we can write~\cite{steck}
\begin{equation}
\alpha_J(\omega) = \sum_{n'J'} \frac{2 \omega_{n'J'J} \left| \langle n'J'| \hat{\varepsilon} \cdot \mathbf{d} | J \rangle \right|^2}{3 \hbar \big( \omega_{n'J'J}^2 - \omega^2 \big)}.
\end{equation}
By inserting this expression for $\alpha(\omega)$ into Eq. \eqref{eq:acstark} and using $I = \frac{\epsilon_0 c}{2} |E|^2$ we receive the formula mentioned in the main text.

The standard relation between a fine-structure transition decay rate $\Gamma_{J'J}$ and the corresponding reduced matrix element $\left| \langle J' | \! | \mathbf{d} | \! | J \rangle \right|$ is given by~\cite{steck}
\begin{equation}
\Gamma_{J'J} = \frac{\omega_{J'J}^3}{3\pi \epsilon_0 \hbar c^3} \frac{2J+1}{2J'+1} \left| \langle J | \! | \mathbf{d} | \! | J' \rangle \right|^2
\label{eq:gamma}
\end{equation}
where we can make use of $J=0$.
In the case that $\Gamma_{J'J}$ is unknown and only the lifetime $\tau$ of the excited state has been measured we estimate the decay rate to the respective state by calculating the branching ratio~\cite{riegger:2018, beloy:2012}
\begin{equation}
\beta(J',J) = \frac{\omega_{J'J}^3 \left| \langle J' | \! | \mathbf{d} | \! | J \rangle \right|^2}{\sum_{J''} \omega_{J'J''}^3 \left| \langle J' | \! | \mathbf{d} | \! | J'' \rangle \right|^2}
\end{equation}
under the $LS$ coupling approximation that we can describe the involved states as eigenstates of the angular momentum and spin operators $L$ and $S$, thus disregarding configuration mixing.
We further require $\ket{J''}$ to be a lower-lying level than $\ket{J'}$.
We can then decompose the reduced matrix element into spin and angular momentum parts~\cite{steck}
\begin{align}
\langle J' | \! | \mathbf{d} | \! | J \rangle = & \langle L' | \! | \mathbf{d} | \! | L \rangle (-1)^{J'+L+S+1} \sqrt{(2L+1)(2J'+1)} \notag \\
& \times \begin{Bmatrix}
	L & L' & 1 \\
	J' & J & S
	\end{Bmatrix} \delta_{S'S}(1-\delta_{\Pi'\Pi})
\end{align}
with $\Pi = (-1)^{l_1 + l_2}$ being the parity and $l_1$, $l_2$ the electronic orbitals of the two valence electrons such that all selection rules are respected~\cite{riegger:2018}.
Owing to the selection rule $L = L'' = L' + 1$, the generally unknown orbital angular momentum matrix elements in the numerator and denominator in the calculation of $\beta$ cancel, and $\Gamma_{J'J} = \beta(J', J) / \tau$.
In total we calculate the decay rates for two transitions from the $\sS{0}$ and five from the $\tP{0}$ state in addition to three ($\sS{0}$) and two ($\tP{0}$) experimentally known decay rates, respectively~\cite{takasu:2004, beloy:2012, blagoev:1994, baumann:1985, mcgrew:2020}.

\begin{figure}[ht!]
	\includegraphics{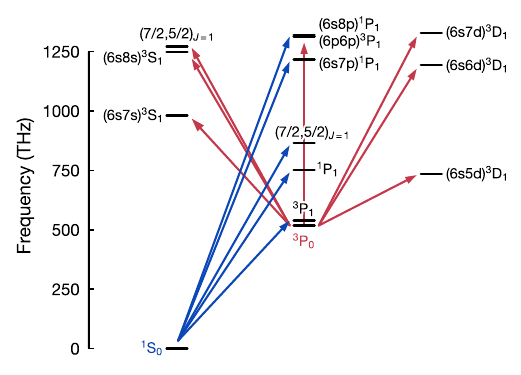}
	\caption{\textbf{Level scheme of all levels and transitions used for the empirical estimate of the $\sS{0}$ and $\tP{0}$ polarizabilities.} For clarity, the core-excited states were arbitrarily assigned to the S- and P-column, respectively. Frequency values are taken from Ref.~\cite{meggers:1978}.}
	\label{fig:s5}
\end{figure}

We treat the $\sS{0}$ polarizability as sufficiently described by the known transition wavelengths and decay rates apart from a global offset.
This offset is motivated by transitions to very high-lying energy levels that are sufficiently far blue-detuned ($>200\nm$) to leave the form of the polarizability curve unchanged, but lead to a small overall attractive shift.
Note that due to spin selection rules the strongest lowest-lying transitions from the $\sS{0}$ state go to $\sP{1}$ states, so we only take these into account aside from the $555.8\nm$ intercombination line to $(6s6p)\tP{1}$ which gives rise to the green magic and tune-out wavelength and therefore has to be included.
All contributions from higher-lying transitions to triplet states can safely be assumed to be captured in the offset as well.
We note that even without offset, the tune-out wavelength predicted with this model is accurate within $0.05\nm$.
Therefore, fitting the polarizability curve to the exact position of the tune-out wavelength only leads to a small global offset.

For $\tP{0}$ the situation is slightly more complex owing to a series of broad transitions to $\tS{1}$ and $\tD{1}$ states at wavelengths just below the blue magic wavelength~\cite{meggers:1978, baumann:1985}.
In particular the $\tP{0} \rightarrow (6s6d)\tD{1}$ transition at $444.0\nm$ is crucial, but also the $411.1\nm$ $\tP{0} \rightarrow (6s8s)\tS{1}$ transition has to be taken into consideration.
We further note that while the reduced dipole matrix element $| \langle (6s5d)\tD{1} | \! | \mathbf{d} | \! | \tP{0} \rangle |$ of the repump transition has been measured with high precision~\cite{beloy:2012}, the uncertainties of some of the measured lifetimes of higher-lying states are larger and could introduce significant systematic shifts.
Due to the proximity of the green and red magic wavelength to the corresponding transition at $649.1\nm$, for the polarizability calculation especially the lifetime of the $(6s7s)\tS{1}$ state is of importance, where the disagreement between the reported value from Ref.~\cite{lange:1970} and Ref.~\cite{baumann:1985} amounts to $\simeq \! 25\%$.
We rely on the value from Ref.~\cite{baumann:1985} as we find very good agreement of our computed decay rate with the value from Ref.~\cite{cho:2012}.
Again, we only take transitions into account where the selection rules are fulfilled, and disregard higher-order effects such as mixing.
To compensate for this simplification and our limited transition dataset we introduce the effective resonance mentioned in the main text as otherwise the intensity-normalized potential would be blue-shifted by a few $\mathrm{Hz / \frac{W}{cm^2}}$.
Taking the $\sS{0}$ polarizability curve as fixed, we can fit to its values at the three experimentally determined magic wavelengths, yielding an effective transition at $\lambda_\mathrm{eff} \simeq 379\nm$ with a linewidth of $\Gamma\!_\mathrm{eff} \simeq 16\MHz$.

\begin{figure*}[ht!]
	\includegraphics[width=0.9\linewidth]{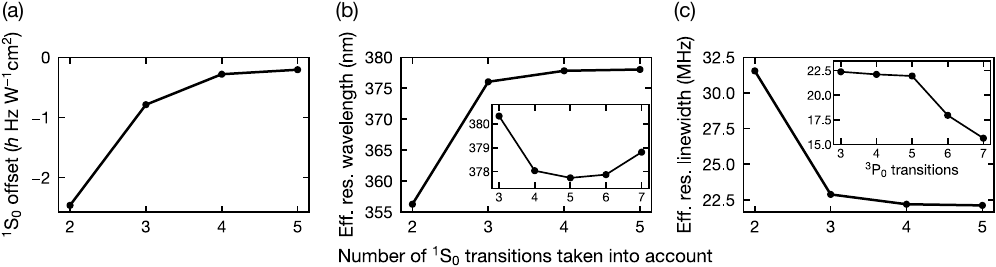}
	\caption{\textbf{Effect of reducing the number of $\sS{0}$ and $\tP{0}$ (insets) transitions on the fit parameters}, ordered by their frequency. Since the $\sS{0}$ offset is only fitted to $\sS{0}$ data, this parameter is independent of the $\tP{0}$ transition set. We observe that the $\sS{0}$ and $\tP{0}$ parameters are well decoupled if we take the transition to the core-excited state (third $\sS{0}$ transition) into account. Insets: For $\tP{0}$ the four lowest transitions are sufficient to describe the polarizability landscape above $\sim \! 420\nm$ without visible changes compared to the full dataset as higher-lying transitions are absorbed by the effective transition.}
	\label{fig:s6}
\end{figure*}

This effective resonance is thus embedded in the pair of transitions to the $(6p^2)\tP{1}$ and $(6s7d)\tD{1}$ levels at $\simeq \! 375\nm$ and we can let the actual transitions be absorbed by the effective one, increasing its linewidth accordingly.
We can further disregard the weak core-excited state transition to $4f^{13}5d6s6p(7/2, 5/2)_{J=1}$ at $397.6\nm$ described in Ref.~\cite{mcgrew:2020} such that eventually we require only four transitions to describe the $\tP{0}$ polarizability for the wavelength range $\gtrsim \! 420\nm$ [Figs.~\subfigref{s6}{b} and \subfigref{s6}{c}, insets, Table~\ref{table:1}], making this minimal model less prone to errors from, e.g., the branching ratio estimate.
Similarly, we study the effect of reducing the $\sS{0}$ transition dataset on the fitted outcome.
As expected, the strong transitions to $(6s7p)\sP{1}$ at $246.5\nm$ and $(6s8p)\sP{1}$ at $227.2\nm$ change the fitted offset slightly but do not lead to a significant deviation of the polarizability curve, which is confirmed by the small change in the fitted effective $\tP{0}$ resonance (Fig.~\ref{fig:s6}).
However, we observe a significant effect from the electronic core excitation to the $\mathrm{4f^{13}5d6s^2(7/2,5/2)_{J=1}}$ state, in line with what has been observed in complex configuration-interaction calculations~\cite{dzuba:2010}.
The final fitted parameters with the reduced transition dataset are $c_{\sS{0}} = -0.79~\mathrm{Hz / \frac{W}{cm^2}}, \lambda_\mathrm{eff} = 376.1\nm, \Gamma_\mathrm{eff} = 2\pi \! \times \! 22.9\MHz$ as described in the main text.

\begin{table}[t!]
	\centering
	\begin{tabular}{c c c c c} 
		\hline\hline
		$\ket{J}$ & $\ket{J'}$ & $\omega_{J'J} / 2\pi$ $(\mathrm{THz})$ & $\Gamma_{J'J} / 2\pi$ $(\mathrm{MHz})$ & Ref. \\ 
		\hline
		 \sS{0} & $(6s6p)\tP{1}$ & 539.386800 & 0.183 & \cite{blagoev:1994, beloy:2012} \\
		 & $(6s6p)\sP{1}$ & 751.526389 & 29.127 & \cite{takasu:2004} \\ 
		 & $(7/2,5/2)_{J=1}$ & 865.111516 & 11.052 & \cite{blagoev:1994} \\ \hline
		\tP{0} & $(6s5d)\tD{1}$ & 215.870446 & 0.308 & \cite{beloy:2012} \\
		& $(6s7s)\tS{1}$ & 461.867846 & 1.516 & \cite{baumann:1985} \\ 
		& $(6s6d)\tD{1}$ & 675.141040 & 4.081 & \cite{baumann:1985} \\
		& $(6s8s)\tS{1}$ & 729.293151 & 0.625 & \cite{baumann:1985} \\
		& Empirical & 797.204099 & 22.889 & Fit \\
		\hline\hline
	\end{tabular}
	\caption{\textbf{Overview of the transition parameters used for the empirical $\sS{0}$ and $\tP{0}$ polarizability model.} The initial state is denoted as $\ket{J}$, the final state as $\ket{J'}$, in analogy to \eqref{eq:gamma}.}
	\label{table:1}
\end{table}

Extracting the $\tP{0}/\sS{0}$ polarizability ratios from our model at $670\nm$, $671.5\nm$, and $690.1\nm$ yields the values 3.24, 3.065, and 1.939 in comparison to the measured ratios 3.3(2), 3.06(4), and 1.97(5)~\cite{riegger:2018, oppong:2022}.

Since the polarizabilities at the measured magic wavelengths are \textit{a priori} undetermined, we rely on lattice depths measured in various optical lattice clock experiments for the red magic wavelength.
In Ref.~\cite{barber:2008}, a value of $500 \, \Erec$ for a waist $ \simeq 30\um$ and power $\simeq 1\,$W is stated, corresponding to an intensity-normalized potential of $-7.0 ~ \mathrm{Hz / \frac{W}{cm^2}}$ in contrast to our model's prediction of $-8.9 ~ \mathrm{Hz / \frac{W}{cm^2}}$.
Using our lattice parameters as a second reference, we obtain a value of $-10.4 ~ \mathrm{Hz / \frac{W}{cm^2}}$, which is similar to the value we can extract from Ref.~\cite{riegger:2018}.
We therefore use the polarizability values from Ref.~\cite{barber:2008} and our experiment as bounds on the red magic wavelength polarizability in Fig.~4 of the main text.
In the case where we also include higher-lying levels and in particular the $4f^{13}5d6s6p(7/2, 5/2)_{J=1}$ core-excitation transition we can use the anti-trapping magic wavelength at $397.6\nm$ measured in Ref.~\cite{mcgrew:2020}, and again we find excellent agreement within the uncertainty of the measured value, both in frequency and polarizability.
We further note that in this wavelength range, the agreement of our $\tP{0}$ polarizability model with the \textit{ab initio} calculations in Ref.~\cite{dzuba:2010} is very high, with a maximum deviation of $\simeq 5\%$, while the corrected $\sS{0}$ polarizability calculation predicts a $\simeq 8\%$ smaller absolute value at $532\nm$.

\begin{figure*}[h!]
	\includegraphics[width=\linewidth]{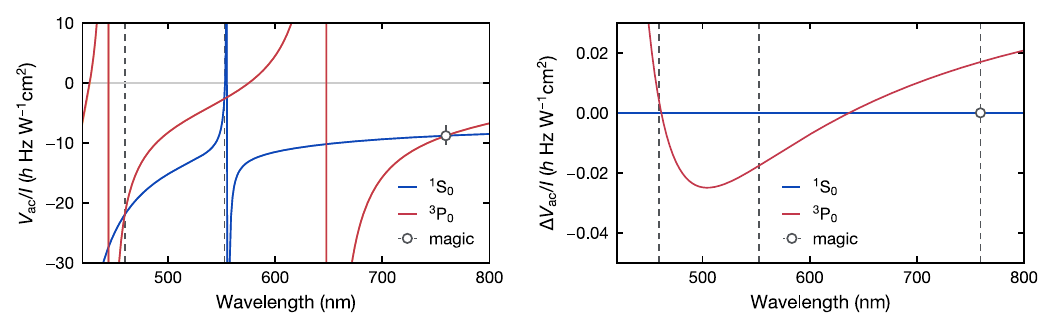}
	\caption{\textbf{Effect of including a global offset as a third free parameter for the $\tP{0}$ polarizability.} In this case, the fitted function crosses the $\sS{0}$ polarizability curve precisely at the three magic wavelengths. However, the deviation $\Delta V_\mathrm{ac}/I$ to the case without $\tP{0}$ offset is minimal and the curves lie on top of each other in the whole wavelength range above $\sim \! 420 \nm$. The deviation $\Delta V_\mathrm{ac}/I$ which is displayed on the right shows only marginal discrepancies for wavelengths above the blue magic wavelength, but starts to grow below due to a different effective resonance position, becoming sizeable for wavelengths $< 420\nm$. Since the polarizabilities at the new magic wavelengths have not been experimentally determined so far, they are shown as grey dotted vertical lines; the magic wavelength at $759.3\nm$ is displayed at the $\sS{0}$ polarizability prediction with grey dotted error bars illustrating the polarizability range obtained from Refs.~\cite{barber:2008, riegger:2018}.}
	\label{fig:s7}
\end{figure*}

As a last check, we add a global offset also to the $\tP{0}$ state fitting function, such that we fit three free parameters to three magic wavelengths.
This offset would incorporate transitions at wavelengths well below the fitted effective transition more accurately.
For wavelengths above $420\nm$ the difference between the fitted curves is only minimal, though, such that we can safely rely on the two-parameter fit [Fig.~\ref{fig:s7}].

\begin{thebibliography}{95}%
\makeatletter
\providecommand \@ifxundefined [1]{%
 \@ifx{#1\undefined}
}%
\providecommand \@ifnum [1]{%
 \ifnum #1\expandafter \@firstoftwo
 \else \expandafter \@secondoftwo
 \fi
}%
\providecommand \@ifx [1]{%
 \ifx #1\expandafter \@firstoftwo
 \else \expandafter \@secondoftwo
 \fi
}%
\providecommand \natexlab [1]{#1}%
\providecommand \enquote  [1]{``#1''}%
\providecommand \bibnamefont  [1]{#1}%
\providecommand \bibfnamefont [1]{#1}%
\providecommand \citenamefont [1]{#1}%
\providecommand \href@noop [0]{\@secondoftwo}%
\providecommand \href [0]{\begingroup \@sanitize@url \@href}%
\providecommand \@href[1]{\@@startlink{#1}\@@href}%
\providecommand \@@href[1]{\endgroup#1\@@endlink}%
\providecommand \@sanitize@url [0]{\catcode `\\12\catcode `\$12\catcode
  `\&12\catcode `\#12\catcode `\^12\catcode `\_12\catcode `\%12\relax}%
\providecommand \@@startlink[1]{}%
\providecommand \@@endlink[0]{}%
\providecommand \url  [0]{\begingroup\@sanitize@url \@url }%
\providecommand \@url [1]{\endgroup\@href {#1}{\urlprefix }}%
\providecommand \urlprefix  [0]{URL }%
\providecommand \Eprint [0]{\href }%
\providecommand \doibase [0]{https://doi.org/}%
\providecommand \selectlanguage [0]{\@gobble}%
\providecommand \bibinfo  [0]{\@secondoftwo}%
\providecommand \bibfield  [0]{\@secondoftwo}%
\providecommand \translation [1]{[#1]}%
\providecommand \BibitemOpen [0]{}%
\providecommand \bibitemStop [0]{}%
\providecommand \bibitemNoStop [0]{.\EOS\space}%
\providecommand \EOS [0]{\spacefactor3000\relax}%
\providecommand \BibitemShut  [1]{\csname bibitem#1\endcsname}%
\let\auto@bib@innerbib\@empty
\bibitem [{\citenamefont {Georgescu}\ \emph {et~al.}(2014)\citenamefont
  {Georgescu}, \citenamefont {Ashhab},\ and\ \citenamefont
  {Nori}}]{georgescu_quantum_2014}%
  \BibitemOpen
  \bibfield  {author} {\bibinfo {author} {\bibfnamefont {I.}~\bibnamefont
  {Georgescu}}, \bibinfo {author} {\bibfnamefont {S.}~\bibnamefont {Ashhab}},\
  and\ \bibinfo {author} {\bibfnamefont {F.}~\bibnamefont {Nori}},\ }\bibfield
  {title} {\bibinfo {title} {Quantum simulation},\ }\href
  {https://doi.org/10.1103/RevModPhys.86.153} {\bibfield  {journal} {\bibinfo
  {journal} {Rev. Mod. Phys.}\ }\textbf {\bibinfo {volume} {86}},\ \bibinfo
  {pages} {153} (\bibinfo {year} {2014})}\BibitemShut {NoStop}%
\bibitem [{\citenamefont {Henriet}\ \emph {et~al.}(2020)\citenamefont
  {Henriet}, \citenamefont {Beguin}, \citenamefont {Signoles}, \citenamefont
  {Lahaye}, \citenamefont {Browaeys}, \citenamefont {Reymond},\ and\
  \citenamefont {Jurczak}}]{henriet_quantum_2020}%
  \BibitemOpen
  \bibfield  {author} {\bibinfo {author} {\bibfnamefont {L.}~\bibnamefont
  {Henriet}}, \bibinfo {author} {\bibfnamefont {L.}~\bibnamefont {Beguin}},
  \bibinfo {author} {\bibfnamefont {A.}~\bibnamefont {Signoles}}, \bibinfo
  {author} {\bibfnamefont {T.}~\bibnamefont {Lahaye}}, \bibinfo {author}
  {\bibfnamefont {A.}~\bibnamefont {Browaeys}}, \bibinfo {author}
  {\bibfnamefont {G.-O.}\ \bibnamefont {Reymond}},\ and\ \bibinfo {author}
  {\bibfnamefont {C.}~\bibnamefont {Jurczak}},\ }\bibfield  {title} {\bibinfo
  {title} {Quantum computing with neutral atoms},\ }\href
  {https://doi.org/10.22331/q-2020-09-21-327} {\bibfield  {journal} {\bibinfo
  {journal} {Quantum}\ }\textbf {\bibinfo {volume} {4}},\ \bibinfo {pages}
  {327} (\bibinfo {year} {2020})}\BibitemShut {NoStop}%
\bibitem [{\citenamefont {Ludlow}\ \emph {et~al.}(2015)\citenamefont {Ludlow},
  \citenamefont {Boyd}, \citenamefont {Ye}, \citenamefont {Peik},\ and\
  \citenamefont {Schmidt}}]{ludlow_optical_2015}%
  \BibitemOpen
  \bibfield  {author} {\bibinfo {author} {\bibfnamefont {A.~D.}\ \bibnamefont
  {Ludlow}}, \bibinfo {author} {\bibfnamefont {M.~M.}\ \bibnamefont {Boyd}},
  \bibinfo {author} {\bibfnamefont {J.}~\bibnamefont {Ye}}, \bibinfo {author}
  {\bibfnamefont {E.}~\bibnamefont {Peik}},\ and\ \bibinfo {author}
  {\bibfnamefont {P.}~\bibnamefont {Schmidt}},\ }\bibfield  {title} {\bibinfo
  {title} {Optical atomic clocks},\ }\href
  {https://doi.org/10.1103/RevModPhys.87.637} {\bibfield  {journal} {\bibinfo
  {journal} {Rev. Mod. Phys.}\ }\textbf {\bibinfo {volume} {87}},\ \bibinfo
  {pages} {637} (\bibinfo {year} {2015})}\BibitemShut {NoStop}%
\bibitem [{\citenamefont {Trotzky}\ \emph {et~al.}(2008)\citenamefont
  {Trotzky}, \citenamefont {Cheinet}, \citenamefont {F\"olling}, \citenamefont
  {Feld}, \citenamefont {Schnorrberger}, \citenamefont {Rey}, \citenamefont
  {Polkovnikov}, \citenamefont {Demler}, \citenamefont {Lukin},\ and\
  \citenamefont {Bloch}}]{trotzky_time_resolved_2008}%
  \BibitemOpen
  \bibfield  {author} {\bibinfo {author} {\bibfnamefont {S.}~\bibnamefont
  {Trotzky}}, \bibinfo {author} {\bibfnamefont {P.}~\bibnamefont {Cheinet}},
  \bibinfo {author} {\bibfnamefont {S.}~\bibnamefont {F\"olling}}, \bibinfo
  {author} {\bibfnamefont {M.}~\bibnamefont {Feld}}, \bibinfo {author}
  {\bibfnamefont {U.}~\bibnamefont {Schnorrberger}}, \bibinfo {author}
  {\bibfnamefont {A.~M.}\ \bibnamefont {Rey}}, \bibinfo {author} {\bibfnamefont
  {A.}~\bibnamefont {Polkovnikov}}, \bibinfo {author} {\bibfnamefont {E.~A.}\
  \bibnamefont {Demler}}, \bibinfo {author} {\bibfnamefont {M.~D.}\
  \bibnamefont {Lukin}},\ and\ \bibinfo {author} {\bibfnamefont
  {I.}~\bibnamefont {Bloch}},\ }\bibfield  {title} {\bibinfo {title}
  {{Time}-{Resolved} {Observation} and {Control} of {Superexchange}
  {Interactions} with {Ultracold} {Atoms} in {Optical} {Lattices}},\ }\href
  {https://doi.org/10.1126/science.1150841} {\bibfield  {journal} {\bibinfo
  {journal} {Science}\ }\textbf {\bibinfo {volume} {319}},\ \bibinfo {pages}
  {295} (\bibinfo {year} {2008})}\BibitemShut {NoStop}%
\bibitem [{\citenamefont {Weitenberg}\ \emph {et~al.}(2011)\citenamefont
  {Weitenberg}, \citenamefont {Endres}, \citenamefont {Sherson}, \citenamefont
  {Cheneau}, \citenamefont {Schauß}, \citenamefont {Fukuhara}, \citenamefont
  {Bloch},\ and\ \citenamefont {Kuhr}}]{weitenberg_single-spin_2011}%
  \BibitemOpen
  \bibfield  {author} {\bibinfo {author} {\bibfnamefont {C.}~\bibnamefont
  {Weitenberg}}, \bibinfo {author} {\bibfnamefont {M.}~\bibnamefont {Endres}},
  \bibinfo {author} {\bibfnamefont {J.~F.}\ \bibnamefont {Sherson}}, \bibinfo
  {author} {\bibfnamefont {M.}~\bibnamefont {Cheneau}}, \bibinfo {author}
  {\bibfnamefont {P.}~\bibnamefont {Schauß}}, \bibinfo {author} {\bibfnamefont
  {T.}~\bibnamefont {Fukuhara}}, \bibinfo {author} {\bibfnamefont
  {I.}~\bibnamefont {Bloch}},\ and\ \bibinfo {author} {\bibfnamefont
  {S.}~\bibnamefont {Kuhr}},\ }\bibfield  {title} {\bibinfo {title}
  {Single-spin addressing in an atomic {Mott} insulator},\ }\href
  {https://doi.org/10.1038/nature09827} {\bibfield  {journal} {\bibinfo
  {journal} {Nature}\ }\textbf {\bibinfo {volume} {471}},\ \bibinfo {pages}
  {319} (\bibinfo {year} {2011})}\BibitemShut {NoStop}%
\bibitem [{\citenamefont {Nascimbène}\ \emph {et~al.}(2012)\citenamefont
  {Nascimbène}, \citenamefont {Chen}, \citenamefont {Atala}, \citenamefont
  {Aidelsburger}, \citenamefont {Trotzky}, \citenamefont {Paredes},\ and\
  \citenamefont {Bloch}}]{nascimbene_experimental_2012}%
  \BibitemOpen
  \bibfield  {author} {\bibinfo {author} {\bibfnamefont {S.}~\bibnamefont
  {Nascimbène}}, \bibinfo {author} {\bibfnamefont {Y.-A.}\ \bibnamefont
  {Chen}}, \bibinfo {author} {\bibfnamefont {M.}~\bibnamefont {Atala}},
  \bibinfo {author} {\bibfnamefont {M.}~\bibnamefont {Aidelsburger}}, \bibinfo
  {author} {\bibfnamefont {S.}~\bibnamefont {Trotzky}}, \bibinfo {author}
  {\bibfnamefont {B.}~\bibnamefont {Paredes}},\ and\ \bibinfo {author}
  {\bibfnamefont {I.}~\bibnamefont {Bloch}},\ }\bibfield  {title} {\bibinfo
  {title} {Experimental {Realization} of {Plaquette} {Resonating}
  {Valence}-{Bond} {States} with {Ultracold} {Atoms} in {Optical}
  {Superlattices}},\ }\href {https://doi.org/10.1103/PhysRevLett.108.205301}
  {\bibfield  {journal} {\bibinfo  {journal} {Phys. Rev. Lett.}\ }\textbf
  {\bibinfo {volume} {108}},\ \bibinfo {pages} {205301} (\bibinfo {year}
  {2012})}\BibitemShut {NoStop}%
\bibitem [{\citenamefont {Fukuhara}\ \emph {et~al.}(2013)\citenamefont
  {Fukuhara}, \citenamefont {Kantian}, \citenamefont {Endres}, \citenamefont
  {Cheneau}, \citenamefont {Schauß}, \citenamefont {Hild}, \citenamefont
  {Bellem}, \citenamefont {Schollwöck}, \citenamefont {Giamarchi},
  \citenamefont {Gross}, \citenamefont {Bloch},\ and\ \citenamefont
  {Kuhr}}]{fukuhara_quantum_2013}%
  \BibitemOpen
  \bibfield  {author} {\bibinfo {author} {\bibfnamefont {T.}~\bibnamefont
  {Fukuhara}}, \bibinfo {author} {\bibfnamefont {A.}~\bibnamefont {Kantian}},
  \bibinfo {author} {\bibfnamefont {M.}~\bibnamefont {Endres}}, \bibinfo
  {author} {\bibfnamefont {M.}~\bibnamefont {Cheneau}}, \bibinfo {author}
  {\bibfnamefont {P.}~\bibnamefont {Schauß}}, \bibinfo {author} {\bibfnamefont
  {S.}~\bibnamefont {Hild}}, \bibinfo {author} {\bibfnamefont {D.}~\bibnamefont
  {Bellem}}, \bibinfo {author} {\bibfnamefont {U.}~\bibnamefont {Schollwöck}},
  \bibinfo {author} {\bibfnamefont {T.}~\bibnamefont {Giamarchi}}, \bibinfo
  {author} {\bibfnamefont {C.}~\bibnamefont {Gross}}, \bibinfo {author}
  {\bibfnamefont {I.}~\bibnamefont {Bloch}},\ and\ \bibinfo {author}
  {\bibfnamefont {S.}~\bibnamefont {Kuhr}},\ }\bibfield  {title} {\bibinfo
  {title} {Quantum dynamics of a mobile spin impurity},\ }\href
  {https://doi.org/10.1038/nphys2561} {\bibfield  {journal} {\bibinfo
  {journal} {Nat. Phys.}\ }\textbf {\bibinfo {volume} {9}},\ \bibinfo {pages}
  {235} (\bibinfo {year} {2013})}\BibitemShut {NoStop}%
\bibitem [{\citenamefont {Dai}\ \emph {et~al.}(2017)\citenamefont {Dai},
  \citenamefont {Yang}, \citenamefont {Reingruber}, \citenamefont {Sun},
  \citenamefont {Xu}, \citenamefont {Chen}, \citenamefont {Yuan},\ and\
  \citenamefont {Pan}}]{dai_four-body_2017}%
  \BibitemOpen
  \bibfield  {author} {\bibinfo {author} {\bibfnamefont {H.-N.}\ \bibnamefont
  {Dai}}, \bibinfo {author} {\bibfnamefont {B.}~\bibnamefont {Yang}}, \bibinfo
  {author} {\bibfnamefont {A.}~\bibnamefont {Reingruber}}, \bibinfo {author}
  {\bibfnamefont {H.}~\bibnamefont {Sun}}, \bibinfo {author} {\bibfnamefont
  {X.-F.}\ \bibnamefont {Xu}}, \bibinfo {author} {\bibfnamefont {Y.-A.}\
  \bibnamefont {Chen}}, \bibinfo {author} {\bibfnamefont {Z.-S.}\ \bibnamefont
  {Yuan}},\ and\ \bibinfo {author} {\bibfnamefont {J.-W.}\ \bibnamefont
  {Pan}},\ }\bibfield  {title} {\bibinfo {title} {Four-body ring-exchange
  interactions and anyonic statistics within a minimal toric-code
  {Hamiltonian}},\ }\href {https://doi.org/10.1038/nphys4243} {\bibfield
  {journal} {\bibinfo  {journal} {Nat. Phys.}\ }\textbf {\bibinfo {volume}
  {13}},\ \bibinfo {pages} {1195} (\bibinfo {year} {2017})}\BibitemShut
  {NoStop}%
\bibitem [{\citenamefont {Robens}\ \emph {et~al.}(2017)\citenamefont {Robens},
  \citenamefont {Zopes}, \citenamefont {Alt}, \citenamefont {Brakhane},
  \citenamefont {Meschede},\ and\ \citenamefont
  {Alberti}}]{robens_low-entropy_2017}%
  \BibitemOpen
  \bibfield  {author} {\bibinfo {author} {\bibfnamefont {C.}~\bibnamefont
  {Robens}}, \bibinfo {author} {\bibfnamefont {J.}~\bibnamefont {Zopes}},
  \bibinfo {author} {\bibfnamefont {W.}~\bibnamefont {Alt}}, \bibinfo {author}
  {\bibfnamefont {S.}~\bibnamefont {Brakhane}}, \bibinfo {author}
  {\bibfnamefont {D.}~\bibnamefont {Meschede}},\ and\ \bibinfo {author}
  {\bibfnamefont {A.}~\bibnamefont {Alberti}},\ }\bibfield  {title} {\bibinfo
  {title} {Low-{Entropy} {States} of {Neutral} {Atoms} in
  {Polarization}-{Synthesized} {Optical} {Lattices}},\ }\href
  {https://doi.org/10.1103/PhysRevLett.118.065302} {\bibfield  {journal}
  {\bibinfo  {journal} {Phys. Rev. Lett.}\ }\textbf {\bibinfo {volume} {118}},\
  \bibinfo {pages} {065302} (\bibinfo {year} {2017})}\BibitemShut {NoStop}%
\bibitem [{\citenamefont {Yang}\ \emph {et~al.}(2020)\citenamefont {Yang},
  \citenamefont {Sun}, \citenamefont {Huang}, \citenamefont {Wang},
  \citenamefont {Deng}, \citenamefont {Dai}, \citenamefont {Yuan},\ and\
  \citenamefont {Pan}}]{yang_cooling_2020}%
  \BibitemOpen
  \bibfield  {author} {\bibinfo {author} {\bibfnamefont {B.}~\bibnamefont
  {Yang}}, \bibinfo {author} {\bibfnamefont {H.}~\bibnamefont {Sun}}, \bibinfo
  {author} {\bibfnamefont {C.-J.}\ \bibnamefont {Huang}}, \bibinfo {author}
  {\bibfnamefont {H.-Y.}\ \bibnamefont {Wang}}, \bibinfo {author}
  {\bibfnamefont {Y.}~\bibnamefont {Deng}}, \bibinfo {author} {\bibfnamefont
  {H.-N.}\ \bibnamefont {Dai}}, \bibinfo {author} {\bibfnamefont {Z.-S.}\
  \bibnamefont {Yuan}},\ and\ \bibinfo {author} {\bibfnamefont {J.-W.}\
  \bibnamefont {Pan}},\ }\bibfield  {title} {\bibinfo {title} {Cooling and
  entangling ultracold atoms in optical lattices},\ }\href
  {https://doi.org/10.1126/science.aaz6801} {\bibfield  {journal} {\bibinfo
  {journal} {Science}\ }\textbf {\bibinfo {volume} {369}},\ \bibinfo {pages}
  {550} (\bibinfo {year} {2020})}\BibitemShut {NoStop}%
\bibitem [{\citenamefont {Sun}\ \emph {et~al.}(2021)\citenamefont {Sun},
  \citenamefont {Yang}, \citenamefont {Wang}, \citenamefont {Zhou},
  \citenamefont {Su}, \citenamefont {Dai}, \citenamefont {Yuan},\ and\
  \citenamefont {Pan}}]{sun_realization_2021}%
  \BibitemOpen
  \bibfield  {author} {\bibinfo {author} {\bibfnamefont {H.}~\bibnamefont
  {Sun}}, \bibinfo {author} {\bibfnamefont {B.}~\bibnamefont {Yang}}, \bibinfo
  {author} {\bibfnamefont {H.-Y.}\ \bibnamefont {Wang}}, \bibinfo {author}
  {\bibfnamefont {Z.-Y.}\ \bibnamefont {Zhou}}, \bibinfo {author}
  {\bibfnamefont {G.-X.}\ \bibnamefont {Su}}, \bibinfo {author} {\bibfnamefont
  {H.-N.}\ \bibnamefont {Dai}}, \bibinfo {author} {\bibfnamefont {Z.-S.}\
  \bibnamefont {Yuan}},\ and\ \bibinfo {author} {\bibfnamefont {J.-W.}\
  \bibnamefont {Pan}},\ }\bibfield  {title} {\bibinfo {title} {Realization of a
  bosonic antiferromagnet},\ }\href
  {https://doi.org/10.1038/s41567-021-01277-1} {\bibfield  {journal} {\bibinfo
  {journal} {Nat. Phys.}\ }\textbf {\bibinfo {volume} {17}},\ \bibinfo {pages}
  {990} (\bibinfo {year} {2021})}\BibitemShut {NoStop}%
\bibitem [{\citenamefont {Mandel}\ \emph {et~al.}(2003)\citenamefont {Mandel},
  \citenamefont {Greiner}, \citenamefont {Widera}, \citenamefont {Rom},
  \citenamefont {H\"ansch},\ and\ \citenamefont {Bloch}}]{mandel:2003}%
  \BibitemOpen
  \bibfield  {author} {\bibinfo {author} {\bibfnamefont {O.}~\bibnamefont
  {Mandel}}, \bibinfo {author} {\bibfnamefont {M.}~\bibnamefont {Greiner}},
  \bibinfo {author} {\bibfnamefont {A.}~\bibnamefont {Widera}}, \bibinfo
  {author} {\bibfnamefont {T.}~\bibnamefont {Rom}}, \bibinfo {author}
  {\bibfnamefont {T.~W.}\ \bibnamefont {H\"ansch}},\ and\ \bibinfo {author}
  {\bibfnamefont {I.}~\bibnamefont {Bloch}},\ }\bibfield  {title} {\bibinfo
  {title} {Controlled collisions for multi-particle entanglement of optically
  trapped atoms},\ }\href {https://doi.org/10.1038/nature02008} {\bibfield
  {journal} {\bibinfo  {journal} {Nature}\ }\textbf {\bibinfo {volume} {425}},\
  \bibinfo {pages} {937} (\bibinfo {year} {2003})}\BibitemShut {NoStop}%
\bibitem [{\citenamefont {Labuhn}\ \emph {et~al.}(2014)\citenamefont {Labuhn},
  \citenamefont {Ravets}, \citenamefont {Barredo}, \citenamefont {Béguin},
  \citenamefont {Nogrette}, \citenamefont {Lahaye},\ and\ \citenamefont
  {Browaeys}}]{labuhn_single-atom_2014}%
  \BibitemOpen
  \bibfield  {author} {\bibinfo {author} {\bibfnamefont {H.}~\bibnamefont
  {Labuhn}}, \bibinfo {author} {\bibfnamefont {S.}~\bibnamefont {Ravets}},
  \bibinfo {author} {\bibfnamefont {D.}~\bibnamefont {Barredo}}, \bibinfo
  {author} {\bibfnamefont {L.}~\bibnamefont {Béguin}}, \bibinfo {author}
  {\bibfnamefont {F.}~\bibnamefont {Nogrette}}, \bibinfo {author}
  {\bibfnamefont {T.}~\bibnamefont {Lahaye}},\ and\ \bibinfo {author}
  {\bibfnamefont {A.}~\bibnamefont {Browaeys}},\ }\bibfield  {title} {\bibinfo
  {title} {Single-atom addressing in microtraps for quantum-state engineering
  using {Rydberg} atoms},\ }\href {https://doi.org/10.1103/PhysRevA.90.023415}
  {\bibfield  {journal} {\bibinfo  {journal} {Phys. Rev. A}\ }\textbf {\bibinfo
  {volume} {90}},\ \bibinfo {pages} {023415} (\bibinfo {year}
  {2014})}\BibitemShut {NoStop}%
\bibitem [{\citenamefont {Levine}\ \emph {et~al.}(2019)\citenamefont {Levine},
  \citenamefont {Keesling}, \citenamefont {Semeghini}, \citenamefont {Omran},
  \citenamefont {Wang}, \citenamefont {Ebadi}, \citenamefont {Bernien},
  \citenamefont {Greiner}, \citenamefont {Vuletić}, \citenamefont {Pichler},\
  and\ \citenamefont {Lukin}}]{levine_parallel_2019}%
  \BibitemOpen
  \bibfield  {author} {\bibinfo {author} {\bibfnamefont {H.}~\bibnamefont
  {Levine}}, \bibinfo {author} {\bibfnamefont {A.}~\bibnamefont {Keesling}},
  \bibinfo {author} {\bibfnamefont {G.}~\bibnamefont {Semeghini}}, \bibinfo
  {author} {\bibfnamefont {A.}~\bibnamefont {Omran}}, \bibinfo {author}
  {\bibfnamefont {T.~T.}\ \bibnamefont {Wang}}, \bibinfo {author}
  {\bibfnamefont {S.}~\bibnamefont {Ebadi}}, \bibinfo {author} {\bibfnamefont
  {H.}~\bibnamefont {Bernien}}, \bibinfo {author} {\bibfnamefont
  {M.}~\bibnamefont {Greiner}}, \bibinfo {author} {\bibfnamefont
  {V.}~\bibnamefont {Vuletić}}, \bibinfo {author} {\bibfnamefont
  {H.}~\bibnamefont {Pichler}},\ and\ \bibinfo {author} {\bibfnamefont {M.~D.}\
  \bibnamefont {Lukin}},\ }\bibfield  {title} {\bibinfo {title} {Parallel
  {Implementation} of {High}-{Fidelity} {Multiqubit} {Gates} with {Neutral}
  {Atoms}},\ }\href {https://doi.org/10.1103/PhysRevLett.123.170503} {\bibfield
   {journal} {\bibinfo  {journal} {Phys. Rev. Lett.}\ }\textbf {\bibinfo
  {volume} {123}},\ \bibinfo {pages} {170503} (\bibinfo {year}
  {2019})}\BibitemShut {NoStop}%
\bibitem [{\citenamefont {Zhang}\ \emph {et~al.}(2022)\citenamefont {Zhang},
  \citenamefont {He}, \citenamefont {Sun}, \citenamefont {Zheng}, \citenamefont
  {Liu}, \citenamefont {Luo}, \citenamefont {Wang}, \citenamefont {Zhu},
  \citenamefont {Qiu}, \citenamefont {Shen}, \citenamefont {Wang},
  \citenamefont {Lin}, \citenamefont {Yu}, \citenamefont {Li}, \citenamefont
  {Xiao}, \citenamefont {Li}, \citenamefont {Yang}, \citenamefont {Jiang},
  \citenamefont {Dai}, \citenamefont {Zhou}, \citenamefont {Ma}, \citenamefont
  {Yuan},\ and\ \citenamefont {Pan}}]{zhang_functional_2022}%
  \BibitemOpen
  \bibfield  {author} {\bibinfo {author} {\bibfnamefont {W.-Y.}\ \bibnamefont
  {Zhang}}, \bibinfo {author} {\bibfnamefont {M.-G.}\ \bibnamefont {He}},
  \bibinfo {author} {\bibfnamefont {H.}~\bibnamefont {Sun}}, \bibinfo {author}
  {\bibfnamefont {Y.-G.}\ \bibnamefont {Zheng}}, \bibinfo {author}
  {\bibfnamefont {Y.}~\bibnamefont {Liu}}, \bibinfo {author} {\bibfnamefont
  {A.}~\bibnamefont {Luo}}, \bibinfo {author} {\bibfnamefont {H.-Y.}\
  \bibnamefont {Wang}}, \bibinfo {author} {\bibfnamefont {Z.-H.}\ \bibnamefont
  {Zhu}}, \bibinfo {author} {\bibfnamefont {P.-Y.}\ \bibnamefont {Qiu}},
  \bibinfo {author} {\bibfnamefont {Y.-C.}\ \bibnamefont {Shen}}, \bibinfo
  {author} {\bibfnamefont {X.-K.}\ \bibnamefont {Wang}}, \bibinfo {author}
  {\bibfnamefont {W.}~\bibnamefont {Lin}}, \bibinfo {author} {\bibfnamefont
  {S.-T.}\ \bibnamefont {Yu}}, \bibinfo {author} {\bibfnamefont {B.-C.}\
  \bibnamefont {Li}}, \bibinfo {author} {\bibfnamefont {B.}~\bibnamefont
  {Xiao}}, \bibinfo {author} {\bibfnamefont {M.-D.}\ \bibnamefont {Li}},
  \bibinfo {author} {\bibfnamefont {Y.-M.}\ \bibnamefont {Yang}}, \bibinfo
  {author} {\bibfnamefont {X.}~\bibnamefont {Jiang}}, \bibinfo {author}
  {\bibfnamefont {H.-N.}\ \bibnamefont {Dai}}, \bibinfo {author} {\bibfnamefont
  {Y.}~\bibnamefont {Zhou}}, \bibinfo {author} {\bibfnamefont {X.}~\bibnamefont
  {Ma}}, \bibinfo {author} {\bibfnamefont {Z.-S.}\ \bibnamefont {Yuan}},\ and\
  \bibinfo {author} {\bibfnamefont {J.-W.}\ \bibnamefont {Pan}},\ }\bibfield
  {title} {\bibinfo {title} {Functional building blocks for scalable
  multipartite entanglement in optical lattices},\ }\href
  {http://arxiv.org/abs/2210.02936} {\bibfield  {journal} {\bibinfo  {journal}
  {arXiv:2210.02936}\ } (\bibinfo {year} {2022})}\BibitemShut {NoStop}%
\bibitem [{\citenamefont {Daley}\ \emph {et~al.}(2008)\citenamefont {Daley},
  \citenamefont {Boyd}, \citenamefont {Ye},\ and\ \citenamefont
  {Zoller}}]{daley:2008}%
  \BibitemOpen
  \bibfield  {author} {\bibinfo {author} {\bibfnamefont {A.~J.}\ \bibnamefont
  {Daley}}, \bibinfo {author} {\bibfnamefont {M.~M.}\ \bibnamefont {Boyd}},
  \bibinfo {author} {\bibfnamefont {J.}~\bibnamefont {Ye}},\ and\ \bibinfo
  {author} {\bibfnamefont {P.}~\bibnamefont {Zoller}},\ }\bibfield  {title}
  {\bibinfo {title} {Quantum {Computing} with {Alkaline}-{Earth}-{Metal}
  {Atoms}},\ }\href {https://doi.org/10.1103/PhysRevLett.101.170504} {\bibfield
   {journal} {\bibinfo  {journal} {Phys. Rev. Lett.}\ }\textbf {\bibinfo
  {volume} {101}},\ \bibinfo {pages} {170504} (\bibinfo {year}
  {2008})}\BibitemShut {NoStop}%
\bibitem [{\citenamefont {Pagano}\ \emph {et~al.}(2019)\citenamefont {Pagano},
  \citenamefont {Scazza},\ and\ \citenamefont {Foss-Feig}}]{pagano_fast_2019}%
  \BibitemOpen
  \bibfield  {author} {\bibinfo {author} {\bibfnamefont {G.}~\bibnamefont
  {Pagano}}, \bibinfo {author} {\bibfnamefont {F.}~\bibnamefont {Scazza}},\
  and\ \bibinfo {author} {\bibfnamefont {M.}~\bibnamefont {Foss-Feig}},\
  }\bibfield  {title} {\bibinfo {title} {Fast and scalable quantum information
  processing with two-electron atoms in optical tweezer arrays},\ }\href
  {https://doi.org/10.1002/qute.201800067} {\bibfield  {journal} {\bibinfo
  {journal} {Adv. Quantum Tech.}\ }\textbf {\bibinfo {volume} {2}},\ \bibinfo
  {pages} {1800067} (\bibinfo {year} {2019})}\BibitemShut {NoStop}%
\bibitem [{\citenamefont {Gonz\'alez-Cuadra}\ \emph {et~al.}(2023)\citenamefont
  {Gonz\'alez-Cuadra}, \citenamefont {Bluvstein}, \citenamefont {Kalinowski},
  \citenamefont {Kaubruegger}, \citenamefont {Maskara}, \citenamefont
  {Naldesi}, \citenamefont {Zache}, \citenamefont {Kaufman}, \citenamefont
  {Lukin}, \citenamefont {Pichler}, \citenamefont {Vermersch}, \citenamefont
  {Ye},\ and\ \citenamefont {Zoller}}]{gonzalez-cuadra_fermionic_2023}%
  \BibitemOpen
  \bibfield  {author} {\bibinfo {author} {\bibfnamefont {D.}~\bibnamefont
  {Gonz\'alez-Cuadra}}, \bibinfo {author} {\bibfnamefont {D.}~\bibnamefont
  {Bluvstein}}, \bibinfo {author} {\bibfnamefont {M.}~\bibnamefont
  {Kalinowski}}, \bibinfo {author} {\bibfnamefont {R.}~\bibnamefont
  {Kaubruegger}}, \bibinfo {author} {\bibfnamefont {N.}~\bibnamefont
  {Maskara}}, \bibinfo {author} {\bibfnamefont {P.}~\bibnamefont {Naldesi}},
  \bibinfo {author} {\bibfnamefont {T.~V.}\ \bibnamefont {Zache}}, \bibinfo
  {author} {\bibfnamefont {A.~M.}\ \bibnamefont {Kaufman}}, \bibinfo {author}
  {\bibfnamefont {M.~D.}\ \bibnamefont {Lukin}}, \bibinfo {author}
  {\bibfnamefont {H.}~\bibnamefont {Pichler}}, \bibinfo {author} {\bibfnamefont
  {B.}~\bibnamefont {Vermersch}}, \bibinfo {author} {\bibfnamefont
  {J.}~\bibnamefont {Ye}},\ and\ \bibinfo {author} {\bibfnamefont
  {P.}~\bibnamefont {Zoller}},\ }\bibfield  {title} {\bibinfo {title}
  {Fermionic quantum processing with programmable neutral atom arrays},\ }\href
  {http://arxiv.org/abs/2303.06985} {\bibfield  {journal} {\bibinfo  {journal}
  {arXiv:2303.06985}\ } (\bibinfo {year} {2023})}\BibitemShut {NoStop}%
\bibitem [{\citenamefont {Riegger}\ \emph {et~al.}(2018)\citenamefont
  {Riegger}, \citenamefont {Darkwah~Oppong}, \citenamefont {H\"ofer},
  \citenamefont {Fernandes}, \citenamefont {Bloch},\ and\ \citenamefont
  {F\"olling}}]{riegger:2018}%
  \BibitemOpen
  \bibfield  {author} {\bibinfo {author} {\bibfnamefont {L.}~\bibnamefont
  {Riegger}}, \bibinfo {author} {\bibfnamefont {N.}~\bibnamefont
  {Darkwah~Oppong}}, \bibinfo {author} {\bibfnamefont {M.}~\bibnamefont
  {H\"ofer}}, \bibinfo {author} {\bibfnamefont {D.~R.}\ \bibnamefont
  {Fernandes}}, \bibinfo {author} {\bibfnamefont {I.}~\bibnamefont {Bloch}},\
  and\ \bibinfo {author} {\bibfnamefont {S.}~\bibnamefont {F\"olling}},\
  }\bibfield  {title} {\bibinfo {title} {Localized {Magnetic} {Moments} with
  {Tunable} {Spin} {Exchange} in a {Gas} of {Ultracold} {Fermions}},\ }\href
  {https://doi.org/10.1103/PhysRevLett.120.143601} {\bibfield  {journal}
  {\bibinfo  {journal} {Phys. Rev. Lett.}\ }\textbf {\bibinfo {volume} {120}},\
  \bibinfo {pages} {143601} (\bibinfo {year} {2018})}\BibitemShut {NoStop}%
\bibitem [{\citenamefont {Rubio-Abadal}\ \emph {et~al.}(2019)\citenamefont
  {Rubio-Abadal}, \citenamefont {Choi}, \citenamefont {Zeiher}, \citenamefont
  {Hollerith}, \citenamefont {Rui}, \citenamefont {Bloch},\ and\ \citenamefont
  {Gross}}]{rubio-abadal_many-body_2019}%
  \BibitemOpen
  \bibfield  {author} {\bibinfo {author} {\bibfnamefont {A.}~\bibnamefont
  {Rubio-Abadal}}, \bibinfo {author} {\bibfnamefont {J.-y.}\ \bibnamefont
  {Choi}}, \bibinfo {author} {\bibfnamefont {J.}~\bibnamefont {Zeiher}},
  \bibinfo {author} {\bibfnamefont {S.}~\bibnamefont {Hollerith}}, \bibinfo
  {author} {\bibfnamefont {J.}~\bibnamefont {Rui}}, \bibinfo {author}
  {\bibfnamefont {I.}~\bibnamefont {Bloch}},\ and\ \bibinfo {author}
  {\bibfnamefont {C.}~\bibnamefont {Gross}},\ }\bibfield  {title} {\bibinfo
  {title} {Many-{Body} {Delocalization} in the {Presence} of a {Quantum}
  {Bath}},\ }\href {https://doi.org/10.1103/PhysRevX.9.041014} {\bibfield
  {journal} {\bibinfo  {journal} {Phys. Rev. X}\ }\textbf {\bibinfo {volume}
  {9}},\ \bibinfo {pages} {041014} (\bibinfo {year} {2019})}\BibitemShut
  {NoStop}%
\bibitem [{\citenamefont {Darkwah~Oppong}\ \emph {et~al.}(2022)\citenamefont
  {Darkwah~Oppong}, \citenamefont {Pasqualetti}, \citenamefont {Bettermann},
  \citenamefont {Zechmann}, \citenamefont {Knap}, \citenamefont {Bloch},\ and\
  \citenamefont {F\"olling}}]{oppong:2022}%
  \BibitemOpen
  \bibfield  {author} {\bibinfo {author} {\bibfnamefont {N.}~\bibnamefont
  {Darkwah~Oppong}}, \bibinfo {author} {\bibfnamefont {G.}~\bibnamefont
  {Pasqualetti}}, \bibinfo {author} {\bibfnamefont {O.}~\bibnamefont
  {Bettermann}}, \bibinfo {author} {\bibfnamefont {P.}~\bibnamefont
  {Zechmann}}, \bibinfo {author} {\bibfnamefont {M.}~\bibnamefont {Knap}},
  \bibinfo {author} {\bibfnamefont {I.}~\bibnamefont {Bloch}},\ and\ \bibinfo
  {author} {\bibfnamefont {S.}~\bibnamefont {F\"olling}},\ }\bibfield  {title}
  {\bibinfo {title} {Probing {Transport} and {Slow} {Relaxation} in the
  {Mass}-{Imbalanced} {Fermi}-{Hubbard} {Model}},\ }\href
  {https://doi.org/10.1103/PhysRevLett.120.143601} {\bibfield  {journal}
  {\bibinfo  {journal} {Phys. Rev. X}\ }\textbf {\bibinfo {volume} {12}},\
  \bibinfo {pages} {031026} (\bibinfo {year} {2022})}\BibitemShut {NoStop}%
\bibitem [{\citenamefont {Gadway}\ \emph {et~al.}(2010)\citenamefont {Gadway},
  \citenamefont {Pertot}, \citenamefont {Reimann},\ and\ \citenamefont
  {Schneble}}]{gadway_superfluidity_2010}%
  \BibitemOpen
  \bibfield  {author} {\bibinfo {author} {\bibfnamefont {B.}~\bibnamefont
  {Gadway}}, \bibinfo {author} {\bibfnamefont {D.}~\bibnamefont {Pertot}},
  \bibinfo {author} {\bibfnamefont {R.}~\bibnamefont {Reimann}},\ and\ \bibinfo
  {author} {\bibfnamefont {D.}~\bibnamefont {Schneble}},\ }\bibfield  {title}
  {\bibinfo {title} {Superfluidity of {Interacting} {Bosonic} {Mixtures} in
  {Optical} {Lattices}},\ }\href
  {https://doi.org/10.1103/PhysRevLett.105.045303} {\bibfield  {journal}
  {\bibinfo  {journal} {Phys. Rev. Lett.}\ }\textbf {\bibinfo {volume} {105}},\
  \bibinfo {pages} {045303} (\bibinfo {year} {2010})}\BibitemShut {NoStop}%
\bibitem [{\citenamefont {Soltan-Panahi}\ \emph {et~al.}(2011)\citenamefont
  {Soltan-Panahi}, \citenamefont {Struck}, \citenamefont {Hauke}, \citenamefont
  {Bick}, \citenamefont {Plenkers}, \citenamefont {Meineke}, \citenamefont
  {Becker}, \citenamefont {Windpassinger}, \citenamefont {Lewenstein},\ and\
  \citenamefont {Sengstock}}]{soltan-panahi_multi-component_2011}%
  \BibitemOpen
  \bibfield  {author} {\bibinfo {author} {\bibfnamefont {P.}~\bibnamefont
  {Soltan-Panahi}}, \bibinfo {author} {\bibfnamefont {J.}~\bibnamefont
  {Struck}}, \bibinfo {author} {\bibfnamefont {P.}~\bibnamefont {Hauke}},
  \bibinfo {author} {\bibfnamefont {A.}~\bibnamefont {Bick}}, \bibinfo {author}
  {\bibfnamefont {W.}~\bibnamefont {Plenkers}}, \bibinfo {author}
  {\bibfnamefont {G.}~\bibnamefont {Meineke}}, \bibinfo {author} {\bibfnamefont
  {C.}~\bibnamefont {Becker}}, \bibinfo {author} {\bibfnamefont
  {P.}~\bibnamefont {Windpassinger}}, \bibinfo {author} {\bibfnamefont
  {M.}~\bibnamefont {Lewenstein}},\ and\ \bibinfo {author} {\bibfnamefont
  {K.}~\bibnamefont {Sengstock}},\ }\bibfield  {title} {\bibinfo {title}
  {Multi-component quantum gases in spin-dependent hexagonal lattices},\ }\href
  {https://doi.org/10.1038/nphys1916} {\bibfield  {journal} {\bibinfo
  {journal} {Nat. Phys.}\ }\textbf {\bibinfo {volume} {7}},\ \bibinfo {pages}
  {434} (\bibinfo {year} {2011})}\BibitemShut {NoStop}%
\bibitem [{\citenamefont {Yang}\ \emph {et~al.}(2017)\citenamefont {Yang},
  \citenamefont {Dai}, \citenamefont {Sun}, \citenamefont {Reingruber},
  \citenamefont {Yuan},\ and\ \citenamefont {Pan}}]{yang_spin-dependent_2017}%
  \BibitemOpen
  \bibfield  {author} {\bibinfo {author} {\bibfnamefont {B.}~\bibnamefont
  {Yang}}, \bibinfo {author} {\bibfnamefont {H.-N.}\ \bibnamefont {Dai}},
  \bibinfo {author} {\bibfnamefont {H.}~\bibnamefont {Sun}}, \bibinfo {author}
  {\bibfnamefont {A.}~\bibnamefont {Reingruber}}, \bibinfo {author}
  {\bibfnamefont {Z.-S.}\ \bibnamefont {Yuan}},\ and\ \bibinfo {author}
  {\bibfnamefont {J.-W.}\ \bibnamefont {Pan}},\ }\bibfield  {title} {\bibinfo
  {title} {Spin-dependent optical superlattice},\ }\href
  {https://doi.org/10.1103/PhysRevA.96.011602} {\bibfield  {journal} {\bibinfo
  {journal} {Phys. Rev. A}\ }\textbf {\bibinfo {volume} {96}},\ \bibinfo
  {pages} {011602} (\bibinfo {year} {2017})}\BibitemShut {NoStop}%
\bibitem [{\citenamefont {de~Hond}\ \emph {et~al.}(2022)\citenamefont
  {de~Hond}, \citenamefont {Xiang}, \citenamefont {Chung}, \citenamefont
  {Cruz-Colón}, \citenamefont {Chen}, \citenamefont {Burton}, \citenamefont
  {Kennedy},\ and\ \citenamefont {Ketterle}}]{de_hond_preparation_2022}%
  \BibitemOpen
  \bibfield  {author} {\bibinfo {author} {\bibfnamefont {J.}~\bibnamefont
  {de~Hond}}, \bibinfo {author} {\bibfnamefont {J.}~\bibnamefont {Xiang}},
  \bibinfo {author} {\bibfnamefont {W.~C.}\ \bibnamefont {Chung}}, \bibinfo
  {author} {\bibfnamefont {E.}~\bibnamefont {Cruz-Colón}}, \bibinfo {author}
  {\bibfnamefont {W.}~\bibnamefont {Chen}}, \bibinfo {author} {\bibfnamefont
  {W.~C.}\ \bibnamefont {Burton}}, \bibinfo {author} {\bibfnamefont {C.~J.}\
  \bibnamefont {Kennedy}},\ and\ \bibinfo {author} {\bibfnamefont
  {W.}~\bibnamefont {Ketterle}},\ }\bibfield  {title} {\bibinfo {title}
  {Preparation of the {Spin}-{Mott} {State}: {A} {Spinful} {Mott} {Insulator}
  of {Repulsively} {Bound} {Pairs}},\ }\href
  {https://doi.org/10.1103/PhysRevLett.128.093401} {\bibfield  {journal}
  {\bibinfo  {journal} {Phys. Rev. Lett.}\ }\textbf {\bibinfo {volume} {128}},\
  \bibinfo {pages} {093401} (\bibinfo {year} {2022})}\BibitemShut {NoStop}%
\bibitem [{\citenamefont {Förster}\ \emph {et~al.}(2009)\citenamefont
  {Förster}, \citenamefont {Karski}, \citenamefont {Choi}, \citenamefont
  {Steffen}, \citenamefont {Alt}, \citenamefont {Meschede}, \citenamefont
  {Widera}, \citenamefont {Montano}, \citenamefont {Lee}, \citenamefont
  {Rakreungdet},\ and\ \citenamefont {Jessen}}]{forster_microwave_2009}%
  \BibitemOpen
  \bibfield  {author} {\bibinfo {author} {\bibfnamefont {L.}~\bibnamefont
  {Förster}}, \bibinfo {author} {\bibfnamefont {M.}~\bibnamefont {Karski}},
  \bibinfo {author} {\bibfnamefont {J.-M.}\ \bibnamefont {Choi}}, \bibinfo
  {author} {\bibfnamefont {A.}~\bibnamefont {Steffen}}, \bibinfo {author}
  {\bibfnamefont {W.}~\bibnamefont {Alt}}, \bibinfo {author} {\bibfnamefont
  {D.}~\bibnamefont {Meschede}}, \bibinfo {author} {\bibfnamefont
  {A.}~\bibnamefont {Widera}}, \bibinfo {author} {\bibfnamefont
  {E.}~\bibnamefont {Montano}}, \bibinfo {author} {\bibfnamefont {J.~H.}\
  \bibnamefont {Lee}}, \bibinfo {author} {\bibfnamefont {W.}~\bibnamefont
  {Rakreungdet}},\ and\ \bibinfo {author} {\bibfnamefont {P.~S.}\ \bibnamefont
  {Jessen}},\ }\bibfield  {title} {\bibinfo {title} {Microwave {Control} of
  {Atomic} {Motion} in {Optical} {Lattices}},\ }\href
  {https://doi.org/10.1103/PhysRevLett.103.233001} {\bibfield  {journal}
  {\bibinfo  {journal} {Phys. Rev. Lett.}\ }\textbf {\bibinfo {volume} {103}},\
  \bibinfo {pages} {233001} (\bibinfo {year} {2009})}\BibitemShut {NoStop}%
\bibitem [{\citenamefont {Belmechri}\ \emph {et~al.}(2013)\citenamefont
  {Belmechri}, \citenamefont {Förster}, \citenamefont {Alt}, \citenamefont
  {Widera}, \citenamefont {Meschede},\ and\ \citenamefont
  {Alberti}}]{belmechri_microwave_2013}%
  \BibitemOpen
  \bibfield  {author} {\bibinfo {author} {\bibfnamefont {N.}~\bibnamefont
  {Belmechri}}, \bibinfo {author} {\bibfnamefont {L.}~\bibnamefont {Förster}},
  \bibinfo {author} {\bibfnamefont {W.}~\bibnamefont {Alt}}, \bibinfo {author}
  {\bibfnamefont {A.}~\bibnamefont {Widera}}, \bibinfo {author} {\bibfnamefont
  {D.}~\bibnamefont {Meschede}},\ and\ \bibinfo {author} {\bibfnamefont
  {A.}~\bibnamefont {Alberti}},\ }\bibfield  {title} {\bibinfo {title}
  {Microwave control of atomic motional states in a spin-dependent optical
  lattice},\ }\href {https://doi.org/10.1088/0953-4075/46/10/104006} {\bibfield
   {journal} {\bibinfo  {journal} {J. Phys. B: At. Mol. Opt. Phys.}\ }\textbf
  {\bibinfo {volume} {46}},\ \bibinfo {pages} {104006} (\bibinfo {year}
  {2013})}\BibitemShut {NoStop}%
\bibitem [{\citenamefont {Porsev}\ \emph {et~al.}(2004)\citenamefont {Porsev},
  \citenamefont {Derevianko},\ and\ \citenamefont
  {Fortson}}]{porsev_possibility_2004}%
  \BibitemOpen
  \bibfield  {author} {\bibinfo {author} {\bibfnamefont {S.~G.}\ \bibnamefont
  {Porsev}}, \bibinfo {author} {\bibfnamefont {A.}~\bibnamefont {Derevianko}},\
  and\ \bibinfo {author} {\bibfnamefont {E.~N.}\ \bibnamefont {Fortson}},\
  }\bibfield  {title} {\bibinfo {title} {Possibility of an optical clock using
  the $6{}^{1}${S}$_{0}\ensuremath{\rightarrow}6{}^{3}${P}$_{0}^{o}$ transition
  in ${}^{171,173}\mathrm{Yb}$ atoms held in an optical lattice},\ }\href
  {https://doi.org/10.1103/PhysRevA.69.021403} {\bibfield  {journal} {\bibinfo
  {journal} {Phys. Rev. A}\ }\textbf {\bibinfo {volume} {69}},\ \bibinfo
  {pages} {021403} (\bibinfo {year} {2004})}\BibitemShut {NoStop}%
\bibitem [{\citenamefont {Gerbier}\ and\ \citenamefont
  {Dalibard}(2010)}]{gerbier_gauge_2010}%
  \BibitemOpen
  \bibfield  {author} {\bibinfo {author} {\bibfnamefont {F.}~\bibnamefont
  {Gerbier}}\ and\ \bibinfo {author} {\bibfnamefont {J.}~\bibnamefont
  {Dalibard}},\ }\bibfield  {title} {\bibinfo {title} {Gauge fields for
  ultracold atoms in optical superlattices},\ }\href
  {https://doi.org/10.1088/1367-2630/12/3/033007} {\bibfield  {journal}
  {\bibinfo  {journal} {New J. Phys.}\ }\textbf {\bibinfo {volume} {12}},\
  \bibinfo {pages} {033007} (\bibinfo {year} {2010})}\BibitemShut {NoStop}%
\bibitem [{\citenamefont {Dzuba}\ and\ \citenamefont
  {Derevianko}(2010)}]{dzuba:2010}%
  \BibitemOpen
  \bibfield  {author} {\bibinfo {author} {\bibfnamefont {V.~A.}\ \bibnamefont
  {Dzuba}}\ and\ \bibinfo {author} {\bibfnamefont {A.}~\bibnamefont
  {Derevianko}},\ }\bibfield  {title} {\bibinfo {title} {Dynamic
  polarizabilities and related properties of clock states of the ytterbium
  atom},\ }\href {https://doi.org/10.1088/0953-4075/43/7/074011} {\bibfield
  {journal} {\bibinfo  {journal} {J. Phys. B: At. Mol. Opt. Phys.}\ }\textbf
  {\bibinfo {volume} {43}},\ \bibinfo {pages} {074011} (\bibinfo {year}
  {2010})}\BibitemShut {NoStop}%
\bibitem [{\citenamefont {Safronova}\ \emph
  {et~al.}(2015{\natexlab{a}})\citenamefont {Safronova}, \citenamefont
  {Zuhrianda}, \citenamefont {Safronova},\ and\ \citenamefont
  {Clark}}]{safronova_extracting_2015}%
  \BibitemOpen
  \bibfield  {author} {\bibinfo {author} {\bibfnamefont {M.~S.}\ \bibnamefont
  {Safronova}}, \bibinfo {author} {\bibfnamefont {Z.}~\bibnamefont
  {Zuhrianda}}, \bibinfo {author} {\bibfnamefont {U.~I.}\ \bibnamefont
  {Safronova}},\ and\ \bibinfo {author} {\bibfnamefont {C.~W.}\ \bibnamefont
  {Clark}},\ }\bibfield  {title} {\bibinfo {title} {Extracting transition rates
  from zero-polarizability spectroscopy},\ }\href
  {https://doi.org/10.1103/PhysRevA.92.040501} {\bibfield  {journal} {\bibinfo
  {journal} {Phys. Rev. A}\ }\textbf {\bibinfo {volume} {92}},\ \bibinfo
  {pages} {040501} (\bibinfo {year} {2015}{\natexlab{a}})}\BibitemShut
  {NoStop}%
\bibitem [{\citenamefont {Dzuba}\ \emph {et~al.}(2018)\citenamefont {Dzuba},
  \citenamefont {Flambaum},\ and\ \citenamefont {Schiller}}]{dzuba:2018}%
  \BibitemOpen
  \bibfield  {author} {\bibinfo {author} {\bibfnamefont {V.~A.}\ \bibnamefont
  {Dzuba}}, \bibinfo {author} {\bibfnamefont {V.~V.}\ \bibnamefont
  {Flambaum}},\ and\ \bibinfo {author} {\bibfnamefont {S.}~\bibnamefont
  {Schiller}},\ }\bibfield  {title} {\bibinfo {title} {Testing physics beyond
  the standard model through additional clock transitions in neutral
  ytterbium},\ }\href {https://doi.org/10.1103/PhysRevA.98.022501} {\bibfield
  {journal} {\bibinfo  {journal} {Phys. Rev. A}\ }\textbf {\bibinfo {volume}
  {98}},\ \bibinfo {pages} {022501} (\bibinfo {year} {2018})}\BibitemShut
  {NoStop}%
\bibitem [{\citenamefont {Heinz}\ \emph {et~al.}(2020)\citenamefont {Heinz},
  \citenamefont {Park}, \citenamefont {{\v S}anti{\'c}}, \citenamefont
  {Trautmann}, \citenamefont {Porsev}, \citenamefont {Safronova}, \citenamefont
  {Bloch},\ and\ \citenamefont {Blatt}}]{heinz:2020}%
  \BibitemOpen
  \bibfield  {author} {\bibinfo {author} {\bibfnamefont {A.}~\bibnamefont
  {Heinz}}, \bibinfo {author} {\bibfnamefont {A.~J.}\ \bibnamefont {Park}},
  \bibinfo {author} {\bibfnamefont {N.}~\bibnamefont {{\v S}anti{\'c}}},
  \bibinfo {author} {\bibfnamefont {J.}~\bibnamefont {Trautmann}}, \bibinfo
  {author} {\bibfnamefont {S.~G.}\ \bibnamefont {Porsev}}, \bibinfo {author}
  {\bibfnamefont {M.~S.}\ \bibnamefont {Safronova}}, \bibinfo {author}
  {\bibfnamefont {I.}~\bibnamefont {Bloch}},\ and\ \bibinfo {author}
  {\bibfnamefont {S.}~\bibnamefont {Blatt}},\ }\bibfield  {title} {\bibinfo
  {title} {State-dependent optical lattices for the strontium optical qubit},\
  }\href {https://doi.org/10.1103/PhysRevLett.124.203201} {\bibfield  {journal}
  {\bibinfo  {journal} {Phys. Rev. Lett.}\ }\textbf {\bibinfo {volume} {124}},\
  \bibinfo {pages} {203201} (\bibinfo {year} {2020})}\BibitemShut {NoStop}%
\bibitem [{\citenamefont {McKeever}\ \emph {et~al.}(2003)\citenamefont
  {McKeever}, \citenamefont {Buck}, \citenamefont {Boozer}, \citenamefont
  {Kuzmich}, \citenamefont {N\"agerl}, \citenamefont {Stamper-Kurn},\ and\
  \citenamefont {Kimble}}]{mckeever:2003}%
  \BibitemOpen
  \bibfield  {author} {\bibinfo {author} {\bibfnamefont {J.}~\bibnamefont
  {McKeever}}, \bibinfo {author} {\bibfnamefont {J.~R.}\ \bibnamefont {Buck}},
  \bibinfo {author} {\bibfnamefont {A.~D.}\ \bibnamefont {Boozer}}, \bibinfo
  {author} {\bibfnamefont {A.}~\bibnamefont {Kuzmich}}, \bibinfo {author}
  {\bibfnamefont {H.-C.}\ \bibnamefont {N\"agerl}}, \bibinfo {author}
  {\bibfnamefont {D.~M.}\ \bibnamefont {Stamper-Kurn}},\ and\ \bibinfo {author}
  {\bibfnamefont {H.~J.}\ \bibnamefont {Kimble}},\ }\bibfield  {title}
  {\bibinfo {title} {State-insensitive cooling and trapping of single atoms in
  an optical cavity},\ }\href {https://doi.org/10.1103/PhysRevLett.90.133602}
  {\bibfield  {journal} {\bibinfo  {journal} {Phys. Rev. Lett.}\ }\textbf
  {\bibinfo {volume} {90}},\ \bibinfo {pages} {133602} (\bibinfo {year}
  {2003})}\BibitemShut {NoStop}%
\bibitem [{\citenamefont {Hutzler}\ \emph {et~al.}(2017)\citenamefont
  {Hutzler}, \citenamefont {Liu}, \citenamefont {Yu},\ and\ \citenamefont
  {Ni}}]{hutzler_eliminating_2017}%
  \BibitemOpen
  \bibfield  {author} {\bibinfo {author} {\bibfnamefont {N.~R.}\ \bibnamefont
  {Hutzler}}, \bibinfo {author} {\bibfnamefont {L.~R.}\ \bibnamefont {Liu}},
  \bibinfo {author} {\bibfnamefont {Y.}~\bibnamefont {Yu}},\ and\ \bibinfo
  {author} {\bibfnamefont {K.-K.}\ \bibnamefont {Ni}},\ }\bibfield  {title}
  {\bibinfo {title} {Eliminating light shifts for single atom trapping},\
  }\href {https://doi.org/10.1088/1367-2630/aa5a3b} {\bibfield  {journal}
  {\bibinfo  {journal} {New J. Phys.}\ }\textbf {\bibinfo {volume} {19}},\
  \bibinfo {pages} {023007} (\bibinfo {year} {2017})}\BibitemShut {NoStop}%
\bibitem [{\citenamefont {Ton}\ \emph {et~al.}(2022)\citenamefont {Ton},
  \citenamefont {Kestler}, \citenamefont {Filin}, \citenamefont {Cheung},
  \citenamefont {Schneeweiss}, \citenamefont {Hoinkes}, \citenamefont {Volz},
  \citenamefont {Safronova}, \citenamefont {Rauschenbeutel},\ and\
  \citenamefont {Barreiro}}]{ton:2022}%
  \BibitemOpen
  \bibfield  {author} {\bibinfo {author} {\bibfnamefont {K.}~\bibnamefont
  {Ton}}, \bibinfo {author} {\bibfnamefont {G.}~\bibnamefont {Kestler}},
  \bibinfo {author} {\bibfnamefont {D.}~\bibnamefont {Filin}}, \bibinfo
  {author} {\bibfnamefont {C.}~\bibnamefont {Cheung}}, \bibinfo {author}
  {\bibfnamefont {P.}~\bibnamefont {Schneeweiss}}, \bibinfo {author}
  {\bibfnamefont {T.}~\bibnamefont {Hoinkes}}, \bibinfo {author} {\bibfnamefont
  {J.}~\bibnamefont {Volz}}, \bibinfo {author} {\bibfnamefont {M.~S.}\
  \bibnamefont {Safronova}}, \bibinfo {author} {\bibfnamefont {A.}~\bibnamefont
  {Rauschenbeutel}},\ and\ \bibinfo {author} {\bibfnamefont {J.~T.}\
  \bibnamefont {Barreiro}},\ }\bibfield  {title} {\bibinfo {title}
  {State-{Insensitive} {Trapping} of {Alkaline}-{Earth} {Atoms} in a
  {Nanofiber}-{Based} {Optical} {Dipole} {Trap}},\ }\href
  {https://arxiv.org/abs/2211.04004} {\bibfield  {journal} {\bibinfo  {journal}
  {arXiv:2211.04004}\ } (\bibinfo {year} {2022})}\BibitemShut {NoStop}%
\bibitem [{\citenamefont {Yamamoto}\ \emph {et~al.}(2016)\citenamefont
  {Yamamoto}, \citenamefont {Kobayashi}, \citenamefont {Kuno}, \citenamefont
  {Kato},\ and\ \citenamefont {Takahashi}}]{yamamoto:2016}%
  \BibitemOpen
  \bibfield  {author} {\bibinfo {author} {\bibfnamefont {R.}~\bibnamefont
  {Yamamoto}}, \bibinfo {author} {\bibfnamefont {J.}~\bibnamefont {Kobayashi}},
  \bibinfo {author} {\bibfnamefont {T.}~\bibnamefont {Kuno}}, \bibinfo {author}
  {\bibfnamefont {K.}~\bibnamefont {Kato}},\ and\ \bibinfo {author}
  {\bibfnamefont {Y.}~\bibnamefont {Takahashi}},\ }\bibfield  {title} {\bibinfo
  {title} {An ytterbium quantum gas microscope with narrow-line laser
  cooling},\ }\href {https://doi.org/10.1088/1367-2630/18/2/023016} {\bibfield
  {journal} {\bibinfo  {journal} {New J. Phys.}\ }\textbf {\bibinfo {volume}
  {18}},\ \bibinfo {pages} {023016} (\bibinfo {year} {2016})}\BibitemShut
  {NoStop}%
\bibitem [{\citenamefont {Trisnadi}\ \emph {et~al.}(2022)\citenamefont
  {Trisnadi}, \citenamefont {Zhang}, \citenamefont {Weiss},\ and\ \citenamefont
  {Chin}}]{trisnadi_design_2022}%
  \BibitemOpen
  \bibfield  {author} {\bibinfo {author} {\bibfnamefont {J.}~\bibnamefont
  {Trisnadi}}, \bibinfo {author} {\bibfnamefont {M.}~\bibnamefont {Zhang}},
  \bibinfo {author} {\bibfnamefont {L.}~\bibnamefont {Weiss}},\ and\ \bibinfo
  {author} {\bibfnamefont {C.}~\bibnamefont {Chin}},\ }\bibfield  {title}
  {\bibinfo {title} {Design and construction of a quantum matter synthesizer},\
  }\href {https://doi.org/10.1063/5.0100088} {\bibfield  {journal} {\bibinfo
  {journal} {Rev. Sci. Instrum.}\ }\textbf {\bibinfo {volume} {93}},\ \bibinfo
  {pages} {083203} (\bibinfo {year} {2022})}\BibitemShut {NoStop}%
\bibitem [{\citenamefont {Cooper}\ \emph {et~al.}(2018)\citenamefont {Cooper},
  \citenamefont {Covey}, \citenamefont {Madjarov}, \citenamefont {Porsev},
  \citenamefont {Safronova},\ and\ \citenamefont {Endres}}]{cooper:2018}%
  \BibitemOpen
  \bibfield  {author} {\bibinfo {author} {\bibfnamefont {A.}~\bibnamefont
  {Cooper}}, \bibinfo {author} {\bibfnamefont {J.~P.}\ \bibnamefont {Covey}},
  \bibinfo {author} {\bibfnamefont {I.~S.}\ \bibnamefont {Madjarov}}, \bibinfo
  {author} {\bibfnamefont {S.~G.}\ \bibnamefont {Porsev}}, \bibinfo {author}
  {\bibfnamefont {M.~S.}\ \bibnamefont {Safronova}},\ and\ \bibinfo {author}
  {\bibfnamefont {M.}~\bibnamefont {Endres}},\ }\bibfield  {title} {\bibinfo
  {title} {Alkaline-{Earth} {Atoms} in {Optical} {Tweezers}},\ }\href
  {https://doi.org/10.1103/PhysRevX.8.041055} {\bibfield  {journal} {\bibinfo
  {journal} {Phys. Rev. X}\ }\textbf {\bibinfo {volume} {8}},\ \bibinfo {pages}
  {041055} (\bibinfo {year} {2018})}\BibitemShut {NoStop}%
\bibitem [{\citenamefont {Norcia}\ \emph {et~al.}(2018)\citenamefont {Norcia},
  \citenamefont {Young},\ and\ \citenamefont {Kaufman}}]{norcia:2018}%
  \BibitemOpen
  \bibfield  {author} {\bibinfo {author} {\bibfnamefont {M.~A.}\ \bibnamefont
  {Norcia}}, \bibinfo {author} {\bibfnamefont {A.~W.}\ \bibnamefont {Young}},\
  and\ \bibinfo {author} {\bibfnamefont {A.~M.}\ \bibnamefont {Kaufman}},\
  }\bibfield  {title} {\bibinfo {title} {Microscopic {Control} and {Detection}
  of {Ultracold} {Strontium} in {Optical}-{Tweezer} {Arrays}},\ }\href
  {https://doi.org/10.1103/PhysRevX.8.041054} {\bibfield  {journal} {\bibinfo
  {journal} {Phys. Rev. X}\ }\textbf {\bibinfo {volume} {8}},\ \bibinfo {pages}
  {041054} (\bibinfo {year} {2018})}\BibitemShut {NoStop}%
\bibitem [{\citenamefont {Saskin}\ \emph {et~al.}(2019)\citenamefont {Saskin},
  \citenamefont {Wilson}, \citenamefont {Grinkemeyer},\ and\ \citenamefont
  {Thompson}}]{saskin_narrow-line_2019}%
  \BibitemOpen
  \bibfield  {author} {\bibinfo {author} {\bibfnamefont {S.}~\bibnamefont
  {Saskin}}, \bibinfo {author} {\bibfnamefont {J.}~\bibnamefont {Wilson}},
  \bibinfo {author} {\bibfnamefont {B.}~\bibnamefont {Grinkemeyer}},\ and\
  \bibinfo {author} {\bibfnamefont {J.}~\bibnamefont {Thompson}},\ }\bibfield
  {title} {\bibinfo {title} {Narrow-{Line} {Cooling} and {Imaging} of
  {Ytterbium} {Atoms} in an {Optical} {Tweezer} {Array}},\ }\href
  {https://doi.org/10.1103/PhysRevLett.122.143002} {\bibfield  {journal}
  {\bibinfo  {journal} {Phys. Rev. Lett.}\ }\textbf {\bibinfo {volume} {122}},\
  \bibinfo {pages} {143002} (\bibinfo {year} {2019})}\BibitemShut {NoStop}%
\bibitem [{\citenamefont {Covey}\ \emph {et~al.}(2019)\citenamefont {Covey},
  \citenamefont {Madjarov}, \citenamefont {Cooper},\ and\ \citenamefont
  {Endres}}]{covey_2000-times_2019}%
  \BibitemOpen
  \bibfield  {author} {\bibinfo {author} {\bibfnamefont {J.~P.}\ \bibnamefont
  {Covey}}, \bibinfo {author} {\bibfnamefont {I.~S.}\ \bibnamefont {Madjarov}},
  \bibinfo {author} {\bibfnamefont {A.}~\bibnamefont {Cooper}},\ and\ \bibinfo
  {author} {\bibfnamefont {M.}~\bibnamefont {Endres}},\ }\bibfield  {title}
  {\bibinfo {title} {2000-{Times} {Repeated} {Imaging} of {Strontium} {Atoms}
  in {Clock}-{Magic} {Tweezer} {Arrays}},\ }\href
  {https://doi.org/10.1103/PhysRevLett.122.173201} {\bibfield  {journal}
  {\bibinfo  {journal} {Phys. Rev. Lett.}\ }\textbf {\bibinfo {volume} {122}},\
  \bibinfo {pages} {173201} (\bibinfo {year} {2019})}\BibitemShut {NoStop}%
\bibitem [{\citenamefont {Okuno}\ \emph {et~al.}(2022)\citenamefont {Okuno},
  \citenamefont {Nakamura}, \citenamefont {Kusano}, \citenamefont {Takasu},
  \citenamefont {Takei}, \citenamefont {Konishi},\ and\ \citenamefont
  {Takahashi}}]{okuno_high-resolution_2022}%
  \BibitemOpen
  \bibfield  {author} {\bibinfo {author} {\bibfnamefont {D.}~\bibnamefont
  {Okuno}}, \bibinfo {author} {\bibfnamefont {Y.}~\bibnamefont {Nakamura}},
  \bibinfo {author} {\bibfnamefont {T.}~\bibnamefont {Kusano}}, \bibinfo
  {author} {\bibfnamefont {Y.}~\bibnamefont {Takasu}}, \bibinfo {author}
  {\bibfnamefont {N.}~\bibnamefont {Takei}}, \bibinfo {author} {\bibfnamefont
  {H.}~\bibnamefont {Konishi}},\ and\ \bibinfo {author} {\bibfnamefont
  {Y.}~\bibnamefont {Takahashi}},\ }\bibfield  {title} {\bibinfo {title}
  {High-resolution {Spectroscopy} and {Single}-photon {Rydberg} {Excitation} of
  {Reconfigurable} {Ytterbium} {Atom} {Tweezer} {Arrays} {Utilizing} a
  {Metastable} {State}},\ }\href {https://doi.org/10.7566/JPSJ.91.084301}
  {\bibfield  {journal} {\bibinfo  {journal} {J. Phys. Soc. Jpn.}\ }\textbf
  {\bibinfo {volume} {91}},\ \bibinfo {pages} {084301} (\bibinfo {year}
  {2022})}\BibitemShut {NoStop}%
\bibitem [{\citenamefont {Ye}\ \emph {et~al.}(2008)\citenamefont {Ye},
  \citenamefont {Kimble},\ and\ \citenamefont {Katori}}]{ye:2008}%
  \BibitemOpen
  \bibfield  {author} {\bibinfo {author} {\bibfnamefont {J.}~\bibnamefont
  {Ye}}, \bibinfo {author} {\bibfnamefont {H.~J.}\ \bibnamefont {Kimble}},\
  and\ \bibinfo {author} {\bibfnamefont {H.}~\bibnamefont {Katori}},\
  }\bibfield  {title} {\bibinfo {title} {Quantum {State} {Engineering} and
  {Precision} {Metrology} {Using} {State}-{Insensitive} {Light} {Traps}},\
  }\href {https://doi.org/10.1126/science.1148259} {\bibfield  {journal}
  {\bibinfo  {journal} {Science}\ }\textbf {\bibinfo {volume} {320}},\ \bibinfo
  {pages} {1734} (\bibinfo {year} {2008})}\BibitemShut {NoStop}%
\bibitem [{\citenamefont {Campbell}\ \emph {et~al.}(2017)\citenamefont
  {Campbell}, \citenamefont {Hutson}, \citenamefont {Marti}, \citenamefont
  {Goban}, \citenamefont {Darkwah~Oppong}, \citenamefont {McNally},
  \citenamefont {Sonderhouse}, \citenamefont {Robinson}, \citenamefont {Zhang},
  \citenamefont {Bloom},\ and\ \citenamefont
  {Ye}}]{campbell_fermi-degenerate_2017}%
  \BibitemOpen
  \bibfield  {author} {\bibinfo {author} {\bibfnamefont {S.~L.}\ \bibnamefont
  {Campbell}}, \bibinfo {author} {\bibfnamefont {R.~B.}\ \bibnamefont
  {Hutson}}, \bibinfo {author} {\bibfnamefont {G.~E.}\ \bibnamefont {Marti}},
  \bibinfo {author} {\bibfnamefont {A.}~\bibnamefont {Goban}}, \bibinfo
  {author} {\bibfnamefont {N.}~\bibnamefont {Darkwah~Oppong}}, \bibinfo
  {author} {\bibfnamefont {R.~L.}\ \bibnamefont {McNally}}, \bibinfo {author}
  {\bibfnamefont {L.}~\bibnamefont {Sonderhouse}}, \bibinfo {author}
  {\bibfnamefont {J.~M.}\ \bibnamefont {Robinson}}, \bibinfo {author}
  {\bibfnamefont {W.}~\bibnamefont {Zhang}}, \bibinfo {author} {\bibfnamefont
  {B.~J.}\ \bibnamefont {Bloom}},\ and\ \bibinfo {author} {\bibfnamefont
  {J.}~\bibnamefont {Ye}},\ }\bibfield  {title} {\bibinfo {title} {A
  {Fermi}-degenerate three-dimensional optical lattice clock},\ }\href
  {https://doi.org/10.1126/science.aam5538} {\bibfield  {journal} {\bibinfo
  {journal} {Science}\ }\textbf {\bibinfo {volume} {358}},\ \bibinfo {pages}
  {90} (\bibinfo {year} {2017})}\BibitemShut {NoStop}%
\bibitem [{\citenamefont {Nicholson}\ \emph {et~al.}(2015)\citenamefont
  {Nicholson}, \citenamefont {Campbell}, \citenamefont {Hutson}, \citenamefont
  {Marti}, \citenamefont {Bloom}, \citenamefont {McNally}, \citenamefont
  {Zhang}, \citenamefont {Barrett}, \citenamefont {Safronova}, \citenamefont
  {Strouse}, \citenamefont {Tew},\ and\ \citenamefont
  {Ye}}]{nicholson_systematic_2015}%
  \BibitemOpen
  \bibfield  {author} {\bibinfo {author} {\bibfnamefont {T.~L.}\ \bibnamefont
  {Nicholson}}, \bibinfo {author} {\bibfnamefont {S.~L.}\ \bibnamefont
  {Campbell}}, \bibinfo {author} {\bibfnamefont {R.~B.}\ \bibnamefont
  {Hutson}}, \bibinfo {author} {\bibfnamefont {G.~E.}\ \bibnamefont {Marti}},
  \bibinfo {author} {\bibfnamefont {B.~J.}\ \bibnamefont {Bloom}}, \bibinfo
  {author} {\bibfnamefont {R.~L.}\ \bibnamefont {McNally}}, \bibinfo {author}
  {\bibfnamefont {W.}~\bibnamefont {Zhang}}, \bibinfo {author} {\bibfnamefont
  {M.~D.}\ \bibnamefont {Barrett}}, \bibinfo {author} {\bibfnamefont {M.~S.}\
  \bibnamefont {Safronova}}, \bibinfo {author} {\bibfnamefont {G.~F.}\
  \bibnamefont {Strouse}}, \bibinfo {author} {\bibfnamefont {W.~L.}\
  \bibnamefont {Tew}},\ and\ \bibinfo {author} {\bibfnamefont {J.}~\bibnamefont
  {Ye}},\ }\bibfield  {title} {\bibinfo {title} {Systematic evaluation of an
  atomic clock at $2\times 10^{-18}$ total uncertainty},\ }\href
  {https://doi.org/10.1038/ncomms7896} {\bibfield  {journal} {\bibinfo
  {journal} {Nat. Commun.}\ }\textbf {\bibinfo {volume} {6}},\ \bibinfo {pages}
  {6896} (\bibinfo {year} {2015})}\BibitemShut {NoStop}%
\bibitem [{\citenamefont {Ushijima}\ \emph {et~al.}(2015)\citenamefont
  {Ushijima}, \citenamefont {Takamoto}, \citenamefont {Das}, \citenamefont
  {Ohkubo},\ and\ \citenamefont {Katori}}]{ushijima_cryogenic_2015}%
  \BibitemOpen
  \bibfield  {author} {\bibinfo {author} {\bibfnamefont {I.}~\bibnamefont
  {Ushijima}}, \bibinfo {author} {\bibfnamefont {M.}~\bibnamefont {Takamoto}},
  \bibinfo {author} {\bibfnamefont {M.}~\bibnamefont {Das}}, \bibinfo {author}
  {\bibfnamefont {T.}~\bibnamefont {Ohkubo}},\ and\ \bibinfo {author}
  {\bibfnamefont {H.}~\bibnamefont {Katori}},\ }\bibfield  {title} {\bibinfo
  {title} {Cryogenic optical lattice clocks},\ }\href
  {https://doi.org/10.1038/nphoton.2015.5} {\bibfield  {journal} {\bibinfo
  {journal} {Nat. Photon.}\ }\textbf {\bibinfo {volume} {9}},\ \bibinfo {pages}
  {185} (\bibinfo {year} {2015})}\BibitemShut {NoStop}%
\bibitem [{\citenamefont {McGrew}\ \emph {et~al.}(2018)\citenamefont {McGrew},
  \citenamefont {Zhang}, \citenamefont {Fasano}, \citenamefont {Sch\"affer},
  \citenamefont {Beloy}, \citenamefont {Nicolodi}, \citenamefont {Brown},
  \citenamefont {Hinkley}, \citenamefont {Milani}, \citenamefont {Schioppo},
  \citenamefont {Yoon},\ and\ \citenamefont {Ludlow}}]{mcgrew_atomic_2018}%
  \BibitemOpen
  \bibfield  {author} {\bibinfo {author} {\bibfnamefont {W.~F.}\ \bibnamefont
  {McGrew}}, \bibinfo {author} {\bibfnamefont {X.}~\bibnamefont {Zhang}},
  \bibinfo {author} {\bibfnamefont {R.~J.}\ \bibnamefont {Fasano}}, \bibinfo
  {author} {\bibfnamefont {S.~A.}\ \bibnamefont {Sch\"affer}}, \bibinfo
  {author} {\bibfnamefont {K.}~\bibnamefont {Beloy}}, \bibinfo {author}
  {\bibfnamefont {D.}~\bibnamefont {Nicolodi}}, \bibinfo {author}
  {\bibfnamefont {R.~C.}\ \bibnamefont {Brown}}, \bibinfo {author}
  {\bibfnamefont {N.}~\bibnamefont {Hinkley}}, \bibinfo {author} {\bibfnamefont
  {G.}~\bibnamefont {Milani}}, \bibinfo {author} {\bibfnamefont
  {M.}~\bibnamefont {Schioppo}}, \bibinfo {author} {\bibfnamefont {T.~H.}\
  \bibnamefont {Yoon}},\ and\ \bibinfo {author} {\bibfnamefont {A.~D.}\
  \bibnamefont {Ludlow}},\ }\bibfield  {title} {\bibinfo {title} {Atomic clock
  performance enabling geodesy below the centimetre level},\ }\href
  {https://doi.org/10.1038/s41586-018-0738-2} {\bibfield  {journal} {\bibinfo
  {journal} {Nature}\ }\textbf {\bibinfo {volume} {564}},\ \bibinfo {pages}
  {87} (\bibinfo {year} {2018})}\BibitemShut {NoStop}%
\bibitem [{\citenamefont {Mitroy}\ \emph {et~al.}(2010)\citenamefont {Mitroy},
  \citenamefont {Safronova},\ and\ \citenamefont {Clark}}]{mitroy_theory_2010}%
  \BibitemOpen
  \bibfield  {author} {\bibinfo {author} {\bibfnamefont {J.}~\bibnamefont
  {Mitroy}}, \bibinfo {author} {\bibfnamefont {M.~S.}\ \bibnamefont
  {Safronova}},\ and\ \bibinfo {author} {\bibfnamefont {C.~W.}\ \bibnamefont
  {Clark}},\ }\bibfield  {title} {\bibinfo {title} {Theory and applications of
  atomic and ionic polarizabilities},\ }\href
  {https://doi.org/10.1088/0953-4075/43/20/202001} {\bibfield  {journal}
  {\bibinfo  {journal} {J. Phys. B: At., Mol. Opt. Phys.}\ }\textbf {\bibinfo
  {volume} {43}},\ \bibinfo {pages} {202001} (\bibinfo {year}
  {2010})}\BibitemShut {NoStop}%
\bibitem [{\citenamefont {LeBlanc}\ and\ \citenamefont
  {Thywissen}(2007)}]{leblanc_species-specific_2007}%
  \BibitemOpen
  \bibfield  {author} {\bibinfo {author} {\bibfnamefont {L.~J.}\ \bibnamefont
  {LeBlanc}}\ and\ \bibinfo {author} {\bibfnamefont {J.~H.}\ \bibnamefont
  {Thywissen}},\ }\bibfield  {title} {\bibinfo {title} {Species-specific
  optical lattices},\ }\href {https://doi.org/10.1103/PhysRevA.75.053612}
  {\bibfield  {journal} {\bibinfo  {journal} {Phys. Rev. A}\ }\textbf {\bibinfo
  {volume} {75}},\ \bibinfo {pages} {053612} (\bibinfo {year}
  {2007})}\BibitemShut {NoStop}%
\bibitem [{\citenamefont {Arora}\ \emph {et~al.}(2011)\citenamefont {Arora},
  \citenamefont {Safronova},\ and\ \citenamefont
  {Clark}}]{arora_tune-out_2011}%
  \BibitemOpen
  \bibfield  {author} {\bibinfo {author} {\bibfnamefont {B.}~\bibnamefont
  {Arora}}, \bibinfo {author} {\bibfnamefont {M.~S.}\ \bibnamefont
  {Safronova}},\ and\ \bibinfo {author} {\bibfnamefont {C.~W.}\ \bibnamefont
  {Clark}},\ }\bibfield  {title} {\bibinfo {title} {Tune-out wavelengths of
  alkali-metal atoms and their applications},\ }\href
  {https://doi.org/10.1103/PhysRevA.84.043401} {\bibfield  {journal} {\bibinfo
  {journal} {Phys. Rev. A}\ }\textbf {\bibinfo {volume} {84}},\ \bibinfo
  {pages} {043401} (\bibinfo {year} {2011})}\BibitemShut {NoStop}%
\bibitem [{\citenamefont {Herold}\ \emph {et~al.}(2012)\citenamefont {Herold},
  \citenamefont {Vaidya}, \citenamefont {Li}, \citenamefont {Rolston},
  \citenamefont {Porto},\ and\ \citenamefont
  {Safronova}}]{herold_precision_2012}%
  \BibitemOpen
  \bibfield  {author} {\bibinfo {author} {\bibfnamefont {C.~D.}\ \bibnamefont
  {Herold}}, \bibinfo {author} {\bibfnamefont {V.~D.}\ \bibnamefont {Vaidya}},
  \bibinfo {author} {\bibfnamefont {X.}~\bibnamefont {Li}}, \bibinfo {author}
  {\bibfnamefont {S.~L.}\ \bibnamefont {Rolston}}, \bibinfo {author}
  {\bibfnamefont {J.~V.}\ \bibnamefont {Porto}},\ and\ \bibinfo {author}
  {\bibfnamefont {M.~S.}\ \bibnamefont {Safronova}},\ }\bibfield  {title}
  {\bibinfo {title} {Precision {Measurement} of {Transition} {Matrix}
  {Elements} via {Light} {Shift} {Cancellation}},\ }\href
  {https://doi.org/10.1103/PhysRevLett.109.243003} {\bibfield  {journal}
  {\bibinfo  {journal} {Phys. Rev. Lett.}\ }\textbf {\bibinfo {volume} {109}},\
  \bibinfo {pages} {243003} (\bibinfo {year} {2012})}\BibitemShut {NoStop}%
\bibitem [{\citenamefont {Holmgren}\ \emph {et~al.}(2012)\citenamefont
  {Holmgren}, \citenamefont {Trubko}, \citenamefont {Hromada},\ and\
  \citenamefont {Cronin}}]{holmgren_measurement_2012}%
  \BibitemOpen
  \bibfield  {author} {\bibinfo {author} {\bibfnamefont {W.~F.}\ \bibnamefont
  {Holmgren}}, \bibinfo {author} {\bibfnamefont {R.}~\bibnamefont {Trubko}},
  \bibinfo {author} {\bibfnamefont {I.}~\bibnamefont {Hromada}},\ and\ \bibinfo
  {author} {\bibfnamefont {A.~D.}\ \bibnamefont {Cronin}},\ }\bibfield  {title}
  {\bibinfo {title} {Measurement of a {Wavelength} of {Light} for {Which} the
  {Energy} {Shift} for an {Atom} {Vanishes}},\ }\href
  {https://doi.org/10.1103/PhysRevLett.109.243004} {\bibfield  {journal}
  {\bibinfo  {journal} {Phys. Rev. Lett.}\ }\textbf {\bibinfo {volume} {109}},\
  \bibinfo {pages} {243004} (\bibinfo {year} {2012})}\BibitemShut {NoStop}%
\bibitem [{\citenamefont {Cheng}\ \emph {et~al.}(2013)\citenamefont {Cheng},
  \citenamefont {Jiang},\ and\ \citenamefont {Mitroy}}]{cheng:2013}%
  \BibitemOpen
  \bibfield  {author} {\bibinfo {author} {\bibfnamefont {Y.}~\bibnamefont
  {Cheng}}, \bibinfo {author} {\bibfnamefont {J.}~\bibnamefont {Jiang}},\ and\
  \bibinfo {author} {\bibfnamefont {J.}~\bibnamefont {Mitroy}},\ }\bibfield
  {title} {\bibinfo {title} {Tune-out wavelengths for the alkaline-earth-metal
  atoms},\ }\href {https://doi.org/10.1103/PhysRevA.88.022511} {\bibfield
  {journal} {\bibinfo  {journal} {Phys. Rev. A}\ }\textbf {\bibinfo {volume}
  {88}},\ \bibinfo {pages} {022511} (\bibinfo {year} {2013})}\BibitemShut
  {NoStop}%
\bibitem [{\citenamefont {Zhang}\ \emph {et~al.}(2021)\citenamefont {Zhang},
  \citenamefont {Tang}, \citenamefont {Zhang},\ and\ \citenamefont
  {Shi}}]{zhang_magic_2021}%
  \BibitemOpen
  \bibfield  {author} {\bibinfo {author} {\bibfnamefont {Y.-H.}\ \bibnamefont
  {Zhang}}, \bibinfo {author} {\bibfnamefont {L.-Y.}\ \bibnamefont {Tang}},
  \bibinfo {author} {\bibfnamefont {J.-Y.}\ \bibnamefont {Zhang}},\ and\
  \bibinfo {author} {\bibfnamefont {T.-Y.}\ \bibnamefont {Shi}},\ }\bibfield
  {title} {\bibinfo {title} {Magic wavelengths for the helium
  $2^{3}\mathrm{S}_1 \rightarrow 2^{1}\mathrm{P}_1$ forbidden transition},\
  }\href {https://doi.org/10.1103/PhysRevA.103.032810} {\bibfield  {journal}
  {\bibinfo  {journal} {Phys. Rev. A}\ }\textbf {\bibinfo {volume} {103}},\
  \bibinfo {pages} {032810} (\bibinfo {year} {2021})}\BibitemShut {NoStop}%
\bibitem [{\citenamefont {Barber}\ \emph {et~al.}(2008)\citenamefont {Barber},
  \citenamefont {Stalnaker}, \citenamefont {Lemke}, \citenamefont {Poli},
  \citenamefont {Oates}, \citenamefont {Fortier}, \citenamefont {Diddams},
  \citenamefont {Hollberg}, \citenamefont {Hoyt}, \citenamefont
  {Taichenachev},\ and\ \citenamefont {Yudin}}]{barber:2008}%
  \BibitemOpen
  \bibfield  {author} {\bibinfo {author} {\bibfnamefont {Z.~W.}\ \bibnamefont
  {Barber}}, \bibinfo {author} {\bibfnamefont {J.~E.}\ \bibnamefont
  {Stalnaker}}, \bibinfo {author} {\bibfnamefont {N.~D.}\ \bibnamefont
  {Lemke}}, \bibinfo {author} {\bibfnamefont {N.}~\bibnamefont {Poli}},
  \bibinfo {author} {\bibfnamefont {C.~W.}\ \bibnamefont {Oates}}, \bibinfo
  {author} {\bibfnamefont {T.~M.}\ \bibnamefont {Fortier}}, \bibinfo {author}
  {\bibfnamefont {S.~A.}\ \bibnamefont {Diddams}}, \bibinfo {author}
  {\bibfnamefont {L.}~\bibnamefont {Hollberg}}, \bibinfo {author}
  {\bibfnamefont {C.~W.}\ \bibnamefont {Hoyt}}, \bibinfo {author}
  {\bibfnamefont {A.~V.}\ \bibnamefont {Taichenachev}},\ and\ \bibinfo {author}
  {\bibfnamefont {V.~I.}\ \bibnamefont {Yudin}},\ }\bibfield  {title} {\bibinfo
  {title} {Optical {Lattice} {Induced} {Light} {Shifts} in an {Yb} {Atomic}
  {Clock}},\ }\href {https://doi.org/10.1103/PhysRevLett.100.103002} {\bibfield
   {journal} {\bibinfo  {journal} {Phys. Rev. Lett.}\ }\textbf {\bibinfo
  {volume} {100}},\ \bibinfo {pages} {103002} (\bibinfo {year}
  {2008})}\BibitemShut {NoStop}%
\bibitem [{\citenamefont {McGrew}(2020)}]{mcgrew:2020}%
  \BibitemOpen
  \bibfield  {author} {\bibinfo {author} {\bibfnamefont {W.~F.}\ \bibnamefont
  {McGrew}},\ }\href@noop {} {\bibinfo {title} {An {Ytterbium} {Optical}
  {Lattice} {Clock} with {Eighteen} {Digits} of {Uncertainty}, {Instability},
  and {Reproducibility}}} (\bibinfo {year} {2020})\BibitemShut {NoStop}%
\bibitem [{SM()}]{SM}%
  \BibitemOpen
  \href@noop {} {\bibinfo {title} {See {Supplemental Material} which includes
  {Refs.~\cite{blatt:2009,reinaudi:2007,drever:1983,kitagawa:2008,fukuhara:2009,meggers:1978,lange:1970,cho:2012}},
  for additional information about the experimental setup, the initial state
  characterization, read-out techniques, additional data and error analysis for
  the magic and tune-out measurements, the derivation of the quadratic fitting
  function for {$\Gamma_\text{exc}$}, experimental results on the scattering
  rate at $f_\mathrm{m1}$, the error analysis for the $^3${P}$_{0}$
  polarizability at the tune-out wavelength, and the derivation of the
  empirical polarizability model.}}\BibitemShut {Stop}%
\bibitem [{\citenamefont {Plotkin-Swing}\ \emph {et~al.}(2020)\citenamefont
  {Plotkin-Swing}, \citenamefont {Wirth}, \citenamefont {Gochnauer},
  \citenamefont {Rahman}, \citenamefont {McAlpine},\ and\ \citenamefont
  {Gupta}}]{plotkin-swing:2020}%
  \BibitemOpen
  \bibfield  {author} {\bibinfo {author} {\bibfnamefont {B.}~\bibnamefont
  {Plotkin-Swing}}, \bibinfo {author} {\bibfnamefont {A.}~\bibnamefont
  {Wirth}}, \bibinfo {author} {\bibfnamefont {D.}~\bibnamefont {Gochnauer}},
  \bibinfo {author} {\bibfnamefont {T.}~\bibnamefont {Rahman}}, \bibinfo
  {author} {\bibfnamefont {K.~E.}\ \bibnamefont {McAlpine}},\ and\ \bibinfo
  {author} {\bibfnamefont {S.}~\bibnamefont {Gupta}},\ }\bibfield  {title}
  {\bibinfo {title} {Crossed-beam slowing to enhance narrow-line ytterbium
  magneto-optic traps},\ }\href {https://doi.org/10.1063/5.0011361} {\bibfield
  {journal} {\bibinfo  {journal} {Rev. Sci. Instrum.}\ }\textbf {\bibinfo
  {volume} {91}},\ \bibinfo {pages} {093201} (\bibinfo {year}
  {2020})}\BibitemShut {NoStop}%
\bibitem [{\citenamefont {Taichenachev}\ \emph {et~al.}(2006)\citenamefont
  {Taichenachev}, \citenamefont {Yudin}, \citenamefont {Oates}, \citenamefont
  {Hoyt}, \citenamefont {Barber},\ and\ \citenamefont
  {Hollberg}}]{taichenachev:2006}%
  \BibitemOpen
  \bibfield  {author} {\bibinfo {author} {\bibfnamefont {A.~V.}\ \bibnamefont
  {Taichenachev}}, \bibinfo {author} {\bibfnamefont {V.~I.}\ \bibnamefont
  {Yudin}}, \bibinfo {author} {\bibfnamefont {C.~W.}\ \bibnamefont {Oates}},
  \bibinfo {author} {\bibfnamefont {C.~W.}\ \bibnamefont {Hoyt}}, \bibinfo
  {author} {\bibfnamefont {Z.~W.}\ \bibnamefont {Barber}},\ and\ \bibinfo
  {author} {\bibfnamefont {L.}~\bibnamefont {Hollberg}},\ }\bibfield  {title}
  {\bibinfo {title} {Magnetic {Field}-{Induced} {Spectroscopy} of {Forbidden}
  {Optical} {Transitions} with {Application} to {Lattice}-{Based} {Optical}
  {Atomic} {Clocks}},\ }\href {https://doi.org/10.1103/PhysRevLett.96.083001}
  {\bibfield  {journal} {\bibinfo  {journal} {Phys. Rev. Lett.}\ }\textbf
  {\bibinfo {volume} {96}},\ \bibinfo {pages} {083001} (\bibinfo {year}
  {2006})}\BibitemShut {NoStop}%
\bibitem [{\citenamefont {Hinkley}\ \emph {et~al.}(2013)\citenamefont
  {Hinkley}, \citenamefont {Sherman}, \citenamefont {Phillips}, \citenamefont
  {Schioppo}, \citenamefont {Lemke}, \citenamefont {Beloy}, \citenamefont
  {Pizzocaro}, \citenamefont {Oates},\ and\ \citenamefont
  {Ludlow}}]{hinkley:2013}%
  \BibitemOpen
  \bibfield  {author} {\bibinfo {author} {\bibfnamefont {N.}~\bibnamefont
  {Hinkley}}, \bibinfo {author} {\bibfnamefont {J.~A.}\ \bibnamefont
  {Sherman}}, \bibinfo {author} {\bibfnamefont {N.~B.}\ \bibnamefont
  {Phillips}}, \bibinfo {author} {\bibfnamefont {M.}~\bibnamefont {Schioppo}},
  \bibinfo {author} {\bibfnamefont {N.~D.}\ \bibnamefont {Lemke}}, \bibinfo
  {author} {\bibfnamefont {K.}~\bibnamefont {Beloy}}, \bibinfo {author}
  {\bibfnamefont {M.}~\bibnamefont {Pizzocaro}}, \bibinfo {author}
  {\bibfnamefont {C.~W.}\ \bibnamefont {Oates}},\ and\ \bibinfo {author}
  {\bibfnamefont {A.~D.}\ \bibnamefont {Ludlow}},\ }\bibfield  {title}
  {\bibinfo {title} {An {Atomic} {Clock} with $\text{10}^{-\text{18}}$
  {Instability}},\ }\href {https://doi.org/10.1126/science.1240420} {\bibfield
  {journal} {\bibinfo  {journal} {Science}\ }\textbf {\bibinfo {volume}
  {341}},\ \bibinfo {pages} {1215} (\bibinfo {year} {2013})}\BibitemShut
  {NoStop}%
\bibitem [{\citenamefont {Henson}\ \emph {et~al.}(2022)\citenamefont {Henson},
  \citenamefont {Ross}, \citenamefont {Thomas}, \citenamefont {Kuhn},
  \citenamefont {Shin}, \citenamefont {Hodgman}, \citenamefont {Zhang},
  \citenamefont {Tang}, \citenamefont {Drake}, \citenamefont {Bondy},
  \citenamefont {Truscott},\ and\ \citenamefont {Baldwin}}]{henson:2022}%
  \BibitemOpen
  \bibfield  {author} {\bibinfo {author} {\bibfnamefont {B.~M.}\ \bibnamefont
  {Henson}}, \bibinfo {author} {\bibfnamefont {J.~A.}\ \bibnamefont {Ross}},
  \bibinfo {author} {\bibfnamefont {K.~F.}\ \bibnamefont {Thomas}}, \bibinfo
  {author} {\bibfnamefont {C.~N.}\ \bibnamefont {Kuhn}}, \bibinfo {author}
  {\bibfnamefont {D.~K.}\ \bibnamefont {Shin}}, \bibinfo {author}
  {\bibfnamefont {S.~S.}\ \bibnamefont {Hodgman}}, \bibinfo {author}
  {\bibfnamefont {Y.-H.}\ \bibnamefont {Zhang}}, \bibinfo {author}
  {\bibfnamefont {L.-Y.}\ \bibnamefont {Tang}}, \bibinfo {author}
  {\bibfnamefont {G.~W.~F.}\ \bibnamefont {Drake}}, \bibinfo {author}
  {\bibfnamefont {A.~T.}\ \bibnamefont {Bondy}}, \bibinfo {author}
  {\bibfnamefont {A.~G.}\ \bibnamefont {Truscott}},\ and\ \bibinfo {author}
  {\bibfnamefont {K.~G.~H.}\ \bibnamefont {Baldwin}},\ }\bibfield  {title}
  {\bibinfo {title} {Measurement of a helium tune-out frequency: an independent
  test of quantum electrodynamics},\ }\href
  {https://doi.org/10.1126/science.abk2502} {\bibfield  {journal} {\bibinfo
  {journal} {Science}\ }\textbf {\bibinfo {volume} {376}},\ \bibinfo {pages}
  {199} (\bibinfo {year} {2022})}\BibitemShut {NoStop}%
\bibitem [{\citenamefont {Catani}\ \emph {et~al.}(2009)\citenamefont {Catani},
  \citenamefont {Barontini}, \citenamefont {Lamporesi}, \citenamefont
  {Rabatti}, \citenamefont {Thalhammer}, \citenamefont {Minardi}, \citenamefont
  {Stringari},\ and\ \citenamefont {Inguscio}}]{catani:2009}%
  \BibitemOpen
  \bibfield  {author} {\bibinfo {author} {\bibfnamefont {J.}~\bibnamefont
  {Catani}}, \bibinfo {author} {\bibfnamefont {G.}~\bibnamefont {Barontini}},
  \bibinfo {author} {\bibfnamefont {G.}~\bibnamefont {Lamporesi}}, \bibinfo
  {author} {\bibfnamefont {F.}~\bibnamefont {Rabatti}}, \bibinfo {author}
  {\bibfnamefont {G.}~\bibnamefont {Thalhammer}}, \bibinfo {author}
  {\bibfnamefont {F.}~\bibnamefont {Minardi}}, \bibinfo {author} {\bibfnamefont
  {S.}~\bibnamefont {Stringari}},\ and\ \bibinfo {author} {\bibfnamefont
  {M.}~\bibnamefont {Inguscio}},\ }\bibfield  {title} {\bibinfo {title}
  {Entropy {Exchange} in a {Mixture} of {Ultracold} {Atoms}},\ }\href
  {https://doi.org/10.1103/PhysRevLett.103.140401} {\bibfield  {journal}
  {\bibinfo  {journal} {Phys. Rev. Lett.}\ }\textbf {\bibinfo {volume} {103}},\
  \bibinfo {pages} {140401} (\bibinfo {year} {2009})}\BibitemShut {NoStop}%
\bibitem [{\citenamefont {Schmidt}\ \emph {et~al.}(2016)\citenamefont
  {Schmidt}, \citenamefont {Mayer}, \citenamefont {Hohmann}, \citenamefont
  {Lausch}, \citenamefont {Kindermann},\ and\ \citenamefont
  {Widera}}]{schmidt_precision_2016}%
  \BibitemOpen
  \bibfield  {author} {\bibinfo {author} {\bibfnamefont {F.}~\bibnamefont
  {Schmidt}}, \bibinfo {author} {\bibfnamefont {D.}~\bibnamefont {Mayer}},
  \bibinfo {author} {\bibfnamefont {M.}~\bibnamefont {Hohmann}}, \bibinfo
  {author} {\bibfnamefont {T.}~\bibnamefont {Lausch}}, \bibinfo {author}
  {\bibfnamefont {F.}~\bibnamefont {Kindermann}},\ and\ \bibinfo {author}
  {\bibfnamefont {A.}~\bibnamefont {Widera}},\ }\bibfield  {title} {\bibinfo
  {title} {Precision measurement of the $^{87}\mathrm{Rb}$ tune-out wavelength
  in the hyperfine ground state ${F}=1$ at 790 nm},\ }\href
  {https://doi.org/10.1103/PhysRevA.93.022507} {\bibfield  {journal} {\bibinfo
  {journal} {Phys. Rev. A}\ }\textbf {\bibinfo {volume} {93}},\ \bibinfo
  {pages} {022507} (\bibinfo {year} {2016})}\BibitemShut {NoStop}%
\bibitem [{\citenamefont {Kao}\ \emph {et~al.}(2017)\citenamefont {Kao},
  \citenamefont {Tang}, \citenamefont {Burdick},\ and\ \citenamefont
  {Lev}}]{kao:2017}%
  \BibitemOpen
  \bibfield  {author} {\bibinfo {author} {\bibfnamefont {W.}~\bibnamefont
  {Kao}}, \bibinfo {author} {\bibfnamefont {Y.}~\bibnamefont {Tang}}, \bibinfo
  {author} {\bibfnamefont {N.~Q.}\ \bibnamefont {Burdick}},\ and\ \bibinfo
  {author} {\bibfnamefont {B.~L.}\ \bibnamefont {Lev}},\ }\bibfield  {title}
  {\bibinfo {title} {Anisotropic dependence of tune-out wavelength near {Dy}
  741-nm transition},\ }\href {https://doi.org/10.1364/OE.25.003411} {\bibfield
   {journal} {\bibinfo  {journal} {Opt. Express}\ }\textbf {\bibinfo {volume}
  {25}},\ \bibinfo {pages} {3411} (\bibinfo {year} {2017})}\BibitemShut
  {NoStop}%
\bibitem [{\citenamefont {Ratkata}\ \emph {et~al.}(2021)\citenamefont
  {Ratkata}, \citenamefont {Gregory}, \citenamefont {Innes}, \citenamefont
  {Matthies}, \citenamefont {McArd}, \citenamefont {Mortlock}, \citenamefont
  {Safronova}, \citenamefont {Bromley},\ and\ \citenamefont
  {Cornish}}]{ratkata:2021}%
  \BibitemOpen
  \bibfield  {author} {\bibinfo {author} {\bibfnamefont {A.}~\bibnamefont
  {Ratkata}}, \bibinfo {author} {\bibfnamefont {P.~D.}\ \bibnamefont
  {Gregory}}, \bibinfo {author} {\bibfnamefont {A.~D.}\ \bibnamefont {Innes}},
  \bibinfo {author} {\bibfnamefont {J.~A.}\ \bibnamefont {Matthies}}, \bibinfo
  {author} {\bibfnamefont {L.~A.}\ \bibnamefont {McArd}}, \bibinfo {author}
  {\bibfnamefont {J.~M.}\ \bibnamefont {Mortlock}}, \bibinfo {author}
  {\bibfnamefont {M.~S.}\ \bibnamefont {Safronova}}, \bibinfo {author}
  {\bibfnamefont {S.~L.}\ \bibnamefont {Bromley}},\ and\ \bibinfo {author}
  {\bibfnamefont {S.~L.}\ \bibnamefont {Cornish}},\ }\bibfield  {title}
  {\bibinfo {title} {Measurement of the tune-out wavelength for
  $^{133}\mathrm{Cs}$ at 880 nm},\ }\href
  {https://doi.org/10.1103/PhysRevA.104.052813} {\bibfield  {journal} {\bibinfo
   {journal} {Phys. Rev. A}\ }\textbf {\bibinfo {volume} {104}},\ \bibinfo
  {pages} {052813} (\bibinfo {year} {2021})}\BibitemShut {NoStop}%
\bibitem [{\citenamefont {Leonard}\ \emph {et~al.}(2015)\citenamefont
  {Leonard}, \citenamefont {Fallon}, \citenamefont {Sackett},\ and\
  \citenamefont {Safronova}}]{leonard_high-precision_2015}%
  \BibitemOpen
  \bibfield  {author} {\bibinfo {author} {\bibfnamefont {R.~H.}\ \bibnamefont
  {Leonard}}, \bibinfo {author} {\bibfnamefont {A.~J.}\ \bibnamefont {Fallon}},
  \bibinfo {author} {\bibfnamefont {C.~A.}\ \bibnamefont {Sackett}},\ and\
  \bibinfo {author} {\bibfnamefont {M.~S.}\ \bibnamefont {Safronova}},\
  }\bibfield  {title} {\bibinfo {title} {High-precision measurements of the
  $^{87}\mathrm{Rb}$ ${D}$-line tune-out wavelength},\ }\href
  {https://doi.org/10.1103/PhysRevA.92.052501} {\bibfield  {journal} {\bibinfo
  {journal} {Phys. Rev. A}\ }\textbf {\bibinfo {volume} {92}},\ \bibinfo
  {pages} {052501} (\bibinfo {year} {2015})}\BibitemShut {NoStop}%
\bibitem [{\citenamefont {Henson}\ \emph {et~al.}(2015)\citenamefont {Henson},
  \citenamefont {Khakimov}, \citenamefont {Dall}, \citenamefont {Baldwin},
  \citenamefont {Tang},\ and\ \citenamefont {Truscott}}]{henson:2015}%
  \BibitemOpen
  \bibfield  {author} {\bibinfo {author} {\bibfnamefont {B.~M.}\ \bibnamefont
  {Henson}}, \bibinfo {author} {\bibfnamefont {R.~I.}\ \bibnamefont
  {Khakimov}}, \bibinfo {author} {\bibfnamefont {R.~G.}\ \bibnamefont {Dall}},
  \bibinfo {author} {\bibfnamefont {K.~G.~H.}\ \bibnamefont {Baldwin}},
  \bibinfo {author} {\bibfnamefont {L.-Y.}\ \bibnamefont {Tang}},\ and\
  \bibinfo {author} {\bibfnamefont {A.~G.}\ \bibnamefont {Truscott}},\
  }\bibfield  {title} {\bibinfo {title} {Precision measurement for metastable
  helium atoms of the 413 nm tune-out wavelength at which the atomic
  polarizability vanishes},\ }\href
  {https://doi.org/10.1103/PhysRevLett.115.043004} {\bibfield  {journal}
  {\bibinfo  {journal} {Phys. Rev. Lett.}\ }\textbf {\bibinfo {volume} {115}},\
  \bibinfo {pages} {043004} (\bibinfo {year} {2015})}\BibitemShut {NoStop}%
\bibitem [{\citenamefont {Bause}\ \emph {et~al.}(2020)\citenamefont {Bause},
  \citenamefont {Li}, \citenamefont {Schindewolf}, \citenamefont {Chen},
  \citenamefont {Duda}, \citenamefont {Kotochigova}, \citenamefont {Bloch},\
  and\ \citenamefont {Luo}}]{bause:2020}%
  \BibitemOpen
  \bibfield  {author} {\bibinfo {author} {\bibfnamefont {R.}~\bibnamefont
  {Bause}}, \bibinfo {author} {\bibfnamefont {M.}~\bibnamefont {Li}}, \bibinfo
  {author} {\bibfnamefont {A.}~\bibnamefont {Schindewolf}}, \bibinfo {author}
  {\bibfnamefont {X.-Y.}\ \bibnamefont {Chen}}, \bibinfo {author}
  {\bibfnamefont {M.}~\bibnamefont {Duda}}, \bibinfo {author} {\bibfnamefont
  {S.}~\bibnamefont {Kotochigova}}, \bibinfo {author} {\bibfnamefont
  {I.}~\bibnamefont {Bloch}},\ and\ \bibinfo {author} {\bibfnamefont {X.-Y.}\
  \bibnamefont {Luo}},\ }\bibfield  {title} {\bibinfo {title} {Tune-{Out} and
  {Magic} {Wavelengths} for {Ground}-{State} $^{23}\mathrm{Na}^{40}\mathrm{K}$
  {Molecules}},\ }\href {https://doi.org/10.1103/PhysRevLett.125.023201}
  {\bibfield  {journal} {\bibinfo  {journal} {Phys. Rev. Lett.}\ }\textbf
  {\bibinfo {volume} {125}},\ \bibinfo {pages} {023201} (\bibinfo {year}
  {2020})}\BibitemShut {NoStop}%
\bibitem [{\citenamefont {Savard}\ \emph {et~al.}(1997)\citenamefont {Savard},
  \citenamefont {O'Hara},\ and\ \citenamefont {Thomas}}]{savard:1997}%
  \BibitemOpen
  \bibfield  {author} {\bibinfo {author} {\bibfnamefont {T.~A.}\ \bibnamefont
  {Savard}}, \bibinfo {author} {\bibfnamefont {K.~M.}\ \bibnamefont {O'Hara}},\
  and\ \bibinfo {author} {\bibfnamefont {J.~E.}\ \bibnamefont {Thomas}},\
  }\bibfield  {title} {\bibinfo {title} {Laser-noise-induced heating in far-off
  resonance optical traps},\ }\href {https://doi.org/10.1103/PhysRevA.56.R1095}
  {\bibfield  {journal} {\bibinfo  {journal} {Phys. Rev. A}\ }\textbf {\bibinfo
  {volume} {56}},\ \bibinfo {pages} {R1095} (\bibinfo {year}
  {1997})}\BibitemShut {NoStop}%
\bibitem [{\citenamefont {Guo}\ \emph {et~al.}(2010)\citenamefont {Guo},
  \citenamefont {Wang},\ and\ \citenamefont {Ye}}]{guo:2010}%
  \BibitemOpen
  \bibfield  {author} {\bibinfo {author} {\bibfnamefont {K.}~\bibnamefont
  {Guo}}, \bibinfo {author} {\bibfnamefont {G.}~\bibnamefont {Wang}},\ and\
  \bibinfo {author} {\bibfnamefont {A.}~\bibnamefont {Ye}},\ }\bibfield
  {title} {\bibinfo {title} {Dipole polarizabilities and magic wavelengths for
  a {Sr} and {Yb} atomic optical lattice clock},\ }\href
  {https://doi.org/10.1088/0953-4075/43/13/135004} {\bibfield  {journal}
  {\bibinfo  {journal} {J. Phys. B: At. Mol. Opt. Phys.}\ }\textbf {\bibinfo
  {volume} {43}},\ \bibinfo {pages} {135004} (\bibinfo {year}
  {2010})}\BibitemShut {NoStop}%
\bibitem [{\citenamefont {Le~Kien}\ \emph {et~al.}(2013)\citenamefont
  {Le~Kien}, \citenamefont {Schneeweiss},\ and\ \citenamefont
  {Rauschenbeutel}}]{kien:2013}%
  \BibitemOpen
  \bibfield  {author} {\bibinfo {author} {\bibfnamefont {F.}~\bibnamefont
  {Le~Kien}}, \bibinfo {author} {\bibfnamefont {P.}~\bibnamefont
  {Schneeweiss}},\ and\ \bibinfo {author} {\bibfnamefont {A.}~\bibnamefont
  {Rauschenbeutel}},\ }\bibfield  {title} {\bibinfo {title} {Dynamical
  polarizability of atoms in arbitrary light fields: general theory and
  application to cesium},\ }\href {https://doi.org/10.1140/epjd/e2013-30729-x}
  {\bibfield  {journal} {\bibinfo  {journal} {Eur. Phys. J. D}\ }\textbf
  {\bibinfo {volume} {67}} (\bibinfo {year} {2013})}\BibitemShut {NoStop}%
\bibitem [{\citenamefont {Manakov}\ \emph {et~al.}(1986)\citenamefont
  {Manakov}, \citenamefont {Ovsiannikov},\ and\ \citenamefont
  {Rapoport}}]{manakov:1986}%
  \BibitemOpen
  \bibfield  {author} {\bibinfo {author} {\bibfnamefont {N.}~\bibnamefont
  {Manakov}}, \bibinfo {author} {\bibfnamefont {V.}~\bibnamefont
  {Ovsiannikov}},\ and\ \bibinfo {author} {\bibfnamefont {L.}~\bibnamefont
  {Rapoport}},\ }\bibfield  {title} {\bibinfo {title} {Atoms in a laser
  field},\ }\href {https://doi.org/10.1016/S0370-1573(86)80001-1} {\bibfield
  {journal} {\bibinfo  {journal} {Physics Reports}\ }\textbf {\bibinfo {volume}
  {141}},\ \bibinfo {pages} {320} (\bibinfo {year} {1986})}\BibitemShut
  {NoStop}%
\bibitem [{\citenamefont {Steck}(2007)}]{steck}%
  \BibitemOpen
  \bibfield  {author} {\bibinfo {author} {\bibfnamefont {D.~A.}\ \bibnamefont
  {Steck}},\ }\bibfield  {title} {\bibinfo {title} {Quantum and {Atom}
  {Optics}},\ }\href {http://steck.us/teaching} {\  (\bibinfo {year}
  {2007})}\BibitemShut {NoStop}%
\bibitem [{\citenamefont {Blagoev}\ and\ \citenamefont
  {Komarovskii}(1994)}]{blagoev:1994}%
  \BibitemOpen
  \bibfield  {author} {\bibinfo {author} {\bibfnamefont {K.}~\bibnamefont
  {Blagoev}}\ and\ \bibinfo {author} {\bibfnamefont {V.}~\bibnamefont
  {Komarovskii}},\ }\bibfield  {title} {\bibinfo {title} {Lifetimes of {Levels}
  of {Neutral} and {Singly} {Ionized} {Lanthanide} {Atoms}},\ }\href
  {https://doi.org/https://doi.org/10.1006/adnd.1994.1001} {\bibfield
  {journal} {\bibinfo  {journal} {Atomic Data and Nuclear Data Tables}\
  }\textbf {\bibinfo {volume} {56}},\ \bibinfo {pages} {1} (\bibinfo {year}
  {1994})}\BibitemShut {NoStop}%
\bibitem [{\citenamefont {Takasu}\ \emph {et~al.}(2004)\citenamefont {Takasu},
  \citenamefont {Komori}, \citenamefont {Honda}, \citenamefont {Kumakura},
  \citenamefont {Yabuzaki},\ and\ \citenamefont {Takahashi}}]{takasu:2004}%
  \BibitemOpen
  \bibfield  {author} {\bibinfo {author} {\bibfnamefont {Y.}~\bibnamefont
  {Takasu}}, \bibinfo {author} {\bibfnamefont {K.}~\bibnamefont {Komori}},
  \bibinfo {author} {\bibfnamefont {K.}~\bibnamefont {Honda}}, \bibinfo
  {author} {\bibfnamefont {M.}~\bibnamefont {Kumakura}}, \bibinfo {author}
  {\bibfnamefont {T.}~\bibnamefont {Yabuzaki}},\ and\ \bibinfo {author}
  {\bibfnamefont {Y.}~\bibnamefont {Takahashi}},\ }\bibfield  {title} {\bibinfo
  {title} {Photoassociation {Spectroscopy} of {Laser}-{Cooled} {Ytterbium}
  {Atoms}},\ }\href {https://doi.org/10.1103/PhysRevLett.93.123202} {\bibfield
  {journal} {\bibinfo  {journal} {Phys. Rev. Lett.}\ }\textbf {\bibinfo
  {volume} {93}},\ \bibinfo {pages} {123202} (\bibinfo {year}
  {2004})}\BibitemShut {NoStop}%
\bibitem [{\citenamefont {Beloy}\ \emph {et~al.}(2012)\citenamefont {Beloy},
  \citenamefont {Sherman}, \citenamefont {Lemke}, \citenamefont {Hinkley},
  \citenamefont {Oates},\ and\ \citenamefont {Ludlow}}]{beloy:2012}%
  \BibitemOpen
  \bibfield  {author} {\bibinfo {author} {\bibfnamefont {K.}~\bibnamefont
  {Beloy}}, \bibinfo {author} {\bibfnamefont {J.~A.}\ \bibnamefont {Sherman}},
  \bibinfo {author} {\bibfnamefont {N.~D.}\ \bibnamefont {Lemke}}, \bibinfo
  {author} {\bibfnamefont {N.}~\bibnamefont {Hinkley}}, \bibinfo {author}
  {\bibfnamefont {C.~W.}\ \bibnamefont {Oates}},\ and\ \bibinfo {author}
  {\bibfnamefont {A.~D.}\ \bibnamefont {Ludlow}},\ }\bibfield  {title}
  {\bibinfo {title} {Determination of the $5d6s$ ${}^{3}\mathrm{D}_{1}$ state
  lifetime and blackbody-radiation clock shift in {Yb}},\ }\href
  {https://doi.org/10.1103/PhysRevA.86.051404} {\bibfield  {journal} {\bibinfo
  {journal} {Phys. Rev. A}\ }\textbf {\bibinfo {volume} {86}},\ \bibinfo
  {pages} {051404} (\bibinfo {year} {2012})}\BibitemShut {NoStop}%
\bibitem [{\citenamefont {Baumann}\ \emph {et~al.}(1985)\citenamefont
  {Baumann}, \citenamefont {Braun}, \citenamefont {Gaiser},\ and\ \citenamefont
  {Liening}}]{baumann:1985}%
  \BibitemOpen
  \bibfield  {author} {\bibinfo {author} {\bibfnamefont {M.}~\bibnamefont
  {Baumann}}, \bibinfo {author} {\bibfnamefont {M.}~\bibnamefont {Braun}},
  \bibinfo {author} {\bibfnamefont {A.}~\bibnamefont {Gaiser}},\ and\ \bibinfo
  {author} {\bibfnamefont {H.}~\bibnamefont {Liening}},\ }\bibfield  {title}
  {\bibinfo {title} {{Radiative lifetimes and $g_J$ factors of low-lying
  even-parity levels in the {Yb} I spectrum}},\ }\href
  {https://doi.org/10.1088/0022-3700/18/17/001} {\bibfield  {journal} {\bibinfo
   {journal} {J. Phys. B: Atom. Mol. Phys.}\ }\textbf {\bibinfo {volume}
  {18}},\ \bibinfo {pages} {L601} (\bibinfo {year} {1985})}\BibitemShut
  {NoStop}%
\bibitem [{\citenamefont {Jaksch}\ and\ \citenamefont
  {Zoller}(2003)}]{jaksch_creation_2003}%
  \BibitemOpen
  \bibfield  {author} {\bibinfo {author} {\bibfnamefont {D.}~\bibnamefont
  {Jaksch}}\ and\ \bibinfo {author} {\bibfnamefont {P.}~\bibnamefont
  {Zoller}},\ }\bibfield  {title} {\bibinfo {title} {Creation of effective
  magnetic fields in optical lattices: the {Hofstadter} butterfly for cold
  neutral atoms},\ }\href {https://doi.org/10.1088/1367-2630/5/1/356}
  {\bibfield  {journal} {\bibinfo  {journal} {New J. Phys.}\ }\textbf {\bibinfo
  {volume} {5}},\ \bibinfo {pages} {56} (\bibinfo {year} {2003})}\BibitemShut
  {NoStop}%
\bibitem [{\citenamefont {Surace}\ \emph {et~al.}(2023)\citenamefont {Surace},
  \citenamefont {Fromholz}, \citenamefont {Darkwah~Oppong}, \citenamefont
  {Dalmonte},\ and\ \citenamefont {Aidelsburger}}]{surace_abinitio_2023}%
  \BibitemOpen
  \bibfield  {author} {\bibinfo {author} {\bibfnamefont {F.~M.}\ \bibnamefont
  {Surace}}, \bibinfo {author} {\bibfnamefont {P.}~\bibnamefont {Fromholz}},
  \bibinfo {author} {\bibfnamefont {N.}~\bibnamefont {Darkwah~Oppong}},
  \bibinfo {author} {\bibfnamefont {M.}~\bibnamefont {Dalmonte}},\ and\
  \bibinfo {author} {\bibfnamefont {M.}~\bibnamefont {Aidelsburger}},\
  }\bibfield  {title} {\bibinfo {title} {${Ab}\,initio$ derivation of lattice
  gauge theory dynamics for cold gases in optical lattices},\ }\href
  {https://doi.org/10.1103/PRXQuantum.4.020330} {\bibfield  {journal} {\bibinfo
   {journal} {PRX Quantum}\ }\textbf {\bibinfo {volume} {4}},\ \bibinfo {pages}
  {020330} (\bibinfo {year} {2023})}\BibitemShut {NoStop}%
\bibitem [{\citenamefont {González-Tudela}\ and\ \citenamefont
  {Cirac}(2019)}]{gonzalez-tudela_cold_2019}%
  \BibitemOpen
  \bibfield  {author} {\bibinfo {author} {\bibfnamefont {A.}~\bibnamefont
  {González-Tudela}}\ and\ \bibinfo {author} {\bibfnamefont {J.~I.}\
  \bibnamefont {Cirac}},\ }\bibfield  {title} {\bibinfo {title} {Cold atoms in
  twisted-bilayer optical potentials},\ }\href
  {https://doi.org/10.1103/PhysRevA.100.053604} {\bibfield  {journal} {\bibinfo
   {journal} {Phys. Rev. A}\ }\textbf {\bibinfo {volume} {100}},\ \bibinfo
  {pages} {053604} (\bibinfo {year} {2019})}\BibitemShut {NoStop}%
\bibitem [{\citenamefont {Luo}\ and\ \citenamefont
  {Zhang}(2021)}]{luo_spin-twisted_2021}%
  \BibitemOpen
  \bibfield  {author} {\bibinfo {author} {\bibfnamefont {X.-W.}\ \bibnamefont
  {Luo}}\ and\ \bibinfo {author} {\bibfnamefont {C.}~\bibnamefont {Zhang}},\
  }\bibfield  {title} {\bibinfo {title} {Spin-{Twisted} {Optical} {Lattices}:
  {Tunable} {Flat} {Bands} and {Larkin}-{Ovchinnikov} {Superfluids}},\ }\href
  {https://doi.org/10.1103/PhysRevLett.126.103201} {\bibfield  {journal}
  {\bibinfo  {journal} {Phys. Rev. Lett.}\ }\textbf {\bibinfo {volume} {126}},\
  \bibinfo {pages} {103201} (\bibinfo {year} {2021})}\BibitemShut {NoStop}%
\bibitem [{\citenamefont {Meng}\ \emph {et~al.}(2023)\citenamefont {Meng},
  \citenamefont {Wang}, \citenamefont {Han}, \citenamefont {Liu}, \citenamefont
  {Wen}, \citenamefont {Gao}, \citenamefont {Wang}, \citenamefont {Chin},\ and\
  \citenamefont {Zhang}}]{meng_atomic_2021}%
  \BibitemOpen
  \bibfield  {author} {\bibinfo {author} {\bibfnamefont {Z.}~\bibnamefont
  {Meng}}, \bibinfo {author} {\bibfnamefont {L.}~\bibnamefont {Wang}}, \bibinfo
  {author} {\bibfnamefont {W.}~\bibnamefont {Han}}, \bibinfo {author}
  {\bibfnamefont {F.}~\bibnamefont {Liu}}, \bibinfo {author} {\bibfnamefont
  {K.}~\bibnamefont {Wen}}, \bibinfo {author} {\bibfnamefont {C.}~\bibnamefont
  {Gao}}, \bibinfo {author} {\bibfnamefont {P.}~\bibnamefont {Wang}}, \bibinfo
  {author} {\bibfnamefont {C.}~\bibnamefont {Chin}},\ and\ \bibinfo {author}
  {\bibfnamefont {J.}~\bibnamefont {Zhang}},\ }\bibfield  {title} {\bibinfo
  {title} {Atomic {Bose}-{Einstein} condensate in twisted-bilayer optical
  lattices},\ }\href {https://doi.org/10.48550/arXiv.2110.00149} {\bibfield
  {journal} {\bibinfo  {journal} {Nature}\ }\textbf {\bibinfo {volume} {615}},\
  \bibinfo {pages} {231–236} (\bibinfo {year} {2023})}\BibitemShut {NoStop}%
\bibitem [{\citenamefont {Du}\ \emph {et~al.}(2023)\citenamefont {Du},
  \citenamefont {Barral}, \citenamefont {Cantara}, \citenamefont {de~Hond},
  \citenamefont {Lu},\ and\ \citenamefont {Ketterle}}]{du_atomic_2023}%
  \BibitemOpen
  \bibfield  {author} {\bibinfo {author} {\bibfnamefont {L.}~\bibnamefont
  {Du}}, \bibinfo {author} {\bibfnamefont {P.}~\bibnamefont {Barral}}, \bibinfo
  {author} {\bibfnamefont {M.}~\bibnamefont {Cantara}}, \bibinfo {author}
  {\bibfnamefont {J.}~\bibnamefont {de~Hond}}, \bibinfo {author} {\bibfnamefont
  {Y.-K.}\ \bibnamefont {Lu}},\ and\ \bibinfo {author} {\bibfnamefont
  {W.}~\bibnamefont {Ketterle}},\ }\bibfield  {title} {\bibinfo {title} {Atomic
  physics on a 50 nm scale: {Realization} of a bilayer system of dipolar
  atoms},\ }\href {http://arxiv.org/abs/2302.07209} {\bibfield  {journal}
  {\bibinfo  {journal} {arXiv:2302.07209}\ } (\bibinfo {year}
  {2023})}\BibitemShut {NoStop}%
\bibitem [{\citenamefont {Arg\"uello-Luengo}\ \emph {et~al.}(2019)\citenamefont
  {Arg\"uello-Luengo}, \citenamefont {González-Tudela}, \citenamefont {Shi},
  \citenamefont {Zoller},\ and\ \citenamefont {Cirac}}]{arguello-luengo:2019}%
  \BibitemOpen
  \bibfield  {author} {\bibinfo {author} {\bibfnamefont {J.}~\bibnamefont
  {Arg\"uello-Luengo}}, \bibinfo {author} {\bibfnamefont {A.}~\bibnamefont
  {González-Tudela}}, \bibinfo {author} {\bibfnamefont {T.}~\bibnamefont
  {Shi}}, \bibinfo {author} {\bibfnamefont {P.}~\bibnamefont {Zoller}},\ and\
  \bibinfo {author} {\bibfnamefont {J.~I.}\ \bibnamefont {Cirac}},\ }\bibfield
  {title} {\bibinfo {title} {Analogue quantum chemistry simulation},\ }\href
  {https://doi.org/10.1038/s41586-019-1614-4} {\bibfield  {journal} {\bibinfo
  {journal} {Nature}\ }\textbf {\bibinfo {volume} {574}},\ \bibinfo {pages}
  {215} (\bibinfo {year} {2019})}\BibitemShut {NoStop}%
\bibitem [{\citenamefont {Safronova}\ \emph
  {et~al.}(2015{\natexlab{b}})\citenamefont {Safronova}, \citenamefont
  {Zuhrianda}, \citenamefont {Safronova},\ and\ \citenamefont
  {Clark}}]{safronova:2015}%
  \BibitemOpen
  \bibfield  {author} {\bibinfo {author} {\bibfnamefont {M.~S.}\ \bibnamefont
  {Safronova}}, \bibinfo {author} {\bibfnamefont {Z.}~\bibnamefont
  {Zuhrianda}}, \bibinfo {author} {\bibfnamefont {U.~I.}\ \bibnamefont
  {Safronova}},\ and\ \bibinfo {author} {\bibfnamefont {C.~W.}\ \bibnamefont
  {Clark}},\ }\bibfield  {title} {\bibinfo {title} {Extracting transition rates
  from zero-polarizability spectroscopy},\ }\href
  {https://doi.org/10.1103/PhysRevA.92.040501} {\bibfield  {journal} {\bibinfo
  {journal} {Phys. Rev. A}\ }\textbf {\bibinfo {volume} {92}},\ \bibinfo
  {pages} {040501} (\bibinfo {year} {2015}{\natexlab{b}})}\BibitemShut
  {NoStop}%
\bibitem [{\citenamefont {Sperling}\ \emph {et~al.}(2021)\citenamefont
  {Sperling}, \citenamefont {Schubert}, \citenamefont {Wenderoth},\ and\
  \citenamefont {Hens}}]{sperling_breakthrough_2021}%
  \BibitemOpen
  \bibfield  {author} {\bibinfo {author} {\bibfnamefont {J.}~\bibnamefont
  {Sperling}}, \bibinfo {author} {\bibfnamefont {M.-H.}\ \bibnamefont
  {Schubert}}, \bibinfo {author} {\bibfnamefont {M.}~\bibnamefont
  {Wenderoth}},\ and\ \bibinfo {author} {\bibfnamefont {K.}~\bibnamefont
  {Hens}},\ }\bibfield  {title} {\bibinfo {title} {Breakthrough instruments and
  products: {Laser} light tunable across the visible up to mid-infrared:
  {Novel} turnkey cw {OPO} with efficiency-optimized design},\ }\href
  {https://doi.org/10.1063/5.0080023} {\bibfield  {journal} {\bibinfo
  {journal} {Rev. Sci. Instrum.}\ }\textbf {\bibinfo {volume} {92}},\ \bibinfo
  {pages} {129502} (\bibinfo {year} {2021})}\BibitemShut {NoStop}%
\bibitem [{\citenamefont {Blatt}\ \emph {et~al.}(2009)\citenamefont {Blatt},
  \citenamefont {Thomsen}, \citenamefont {Campbell}, \citenamefont {Ludlow},
  \citenamefont {Swallows}, \citenamefont {Martin}, \citenamefont {Boyd},\ and\
  \citenamefont {Ye}}]{blatt:2009}%
  \BibitemOpen
  \bibfield  {author} {\bibinfo {author} {\bibfnamefont {S.}~\bibnamefont
  {Blatt}}, \bibinfo {author} {\bibfnamefont {J.~W.}\ \bibnamefont {Thomsen}},
  \bibinfo {author} {\bibfnamefont {G.~K.}\ \bibnamefont {Campbell}}, \bibinfo
  {author} {\bibfnamefont {A.~D.}\ \bibnamefont {Ludlow}}, \bibinfo {author}
  {\bibfnamefont {M.~D.}\ \bibnamefont {Swallows}}, \bibinfo {author}
  {\bibfnamefont {M.~J.}\ \bibnamefont {Martin}}, \bibinfo {author}
  {\bibfnamefont {M.~M.}\ \bibnamefont {Boyd}},\ and\ \bibinfo {author}
  {\bibfnamefont {J.}~\bibnamefont {Ye}},\ }\bibfield  {title} {\bibinfo
  {title} {Rabi spectroscopy and excitation inhomogeneity in a one-dimensional
  optical lattice clock},\ }\href {https://doi.org/10.1103/PhysRevA.80.052703}
  {\bibfield  {journal} {\bibinfo  {journal} {Phys. Rev. A}\ }\textbf {\bibinfo
  {volume} {80}},\ \bibinfo {pages} {052703} (\bibinfo {year}
  {2009})}\BibitemShut {NoStop}%
\bibitem [{\citenamefont {Reinaudi}\ \emph {et~al.}(2007)\citenamefont
  {Reinaudi}, \citenamefont {Lashazye}, \citenamefont {Wang},\ and\
  \citenamefont {Gu\'{e}ry-Odelin}}]{reinaudi:2007}%
  \BibitemOpen
  \bibfield  {author} {\bibinfo {author} {\bibfnamefont {G.}~\bibnamefont
  {Reinaudi}}, \bibinfo {author} {\bibfnamefont {T.}~\bibnamefont {Lashazye}},
  \bibinfo {author} {\bibfnamefont {Z.}~\bibnamefont {Wang}},\ and\ \bibinfo
  {author} {\bibfnamefont {D.}~\bibnamefont {Gu\'{e}ry-Odelin}},\ }\bibfield
  {title} {\bibinfo {title} {Strong saturation absorption imaging of dense
  clouds of ultracold atoms},\ }\href {https://doi.org/10.1364/OL.32.003143}
  {\bibfield  {journal} {\bibinfo  {journal} {Optics Letters}\ }\textbf
  {\bibinfo {volume} {32}},\ \bibinfo {pages} {3143} (\bibinfo {year}
  {2007})}\BibitemShut {NoStop}%
\bibitem [{\citenamefont {Drever}\ \emph {et~al.}(1983)\citenamefont {Drever},
  \citenamefont {Hall}, \citenamefont {Kowalski}, \citenamefont {Hough},
  \citenamefont {Ford}, \citenamefont {Munley},\ and\ \citenamefont
  {Ward}}]{drever:1983}%
  \BibitemOpen
  \bibfield  {author} {\bibinfo {author} {\bibfnamefont {R.~W.~P.}\
  \bibnamefont {Drever}}, \bibinfo {author} {\bibfnamefont {J.~L.}\
  \bibnamefont {Hall}}, \bibinfo {author} {\bibfnamefont {F.~V.}\ \bibnamefont
  {Kowalski}}, \bibinfo {author} {\bibfnamefont {J.}~\bibnamefont {Hough}},
  \bibinfo {author} {\bibfnamefont {G.~M.}\ \bibnamefont {Ford}}, \bibinfo
  {author} {\bibfnamefont {A.~J.}\ \bibnamefont {Munley}},\ and\ \bibinfo
  {author} {\bibfnamefont {H.}~\bibnamefont {Ward}},\ }\bibfield  {title}
  {\bibinfo {title} {Laser phase and frequency stabilization using an optical
  resonator},\ }\href {https://doi.org/10.1007/BF00702605} {\bibfield
  {journal} {\bibinfo  {journal} {Appl. Phys. B}\ }\textbf {\bibinfo {volume}
  {31}},\ \bibinfo {pages} {97} (\bibinfo {year} {1983})}\BibitemShut {NoStop}%
\bibitem [{\citenamefont {Kitagawa}\ \emph {et~al.}(2008)\citenamefont
  {Kitagawa}, \citenamefont {Enomoto}, \citenamefont {Kasa}, \citenamefont
  {Takahashi}, \citenamefont {Ciury\l{}o}, \citenamefont {Naidon},\ and\
  \citenamefont {Julienne}}]{kitagawa:2008}%
  \BibitemOpen
  \bibfield  {author} {\bibinfo {author} {\bibfnamefont {M.}~\bibnamefont
  {Kitagawa}}, \bibinfo {author} {\bibfnamefont {K.}~\bibnamefont {Enomoto}},
  \bibinfo {author} {\bibfnamefont {K.}~\bibnamefont {Kasa}}, \bibinfo {author}
  {\bibfnamefont {Y.}~\bibnamefont {Takahashi}}, \bibinfo {author}
  {\bibfnamefont {R.}~\bibnamefont {Ciury\l{}o}}, \bibinfo {author}
  {\bibfnamefont {P.}~\bibnamefont {Naidon}},\ and\ \bibinfo {author}
  {\bibfnamefont {P.~S.}\ \bibnamefont {Julienne}},\ }\bibfield  {title}
  {\bibinfo {title} {Two-color photoassociation spectroscopy of ytterbium atoms
  and the precise determinations of $s$-wave scattering lengths},\ }\href
  {https://doi.org/10.1103/PhysRevA.77.012719} {\bibfield  {journal} {\bibinfo
  {journal} {Phys. Rev. A}\ }\textbf {\bibinfo {volume} {77}},\ \bibinfo
  {pages} {012719} (\bibinfo {year} {2008})}\BibitemShut {NoStop}%
\bibitem [{\citenamefont {Fukuhara}\ \emph {et~al.}(2009)\citenamefont
  {Fukuhara}, \citenamefont {Sugawa}, \citenamefont {Takasu},\ and\
  \citenamefont {Takahashi}}]{fukuhara:2009}%
  \BibitemOpen
  \bibfield  {author} {\bibinfo {author} {\bibfnamefont {T.}~\bibnamefont
  {Fukuhara}}, \bibinfo {author} {\bibfnamefont {S.}~\bibnamefont {Sugawa}},
  \bibinfo {author} {\bibfnamefont {Y.}~\bibnamefont {Takasu}},\ and\ \bibinfo
  {author} {\bibfnamefont {Y.}~\bibnamefont {Takahashi}},\ }\bibfield  {title}
  {\bibinfo {title} {All-optical formation of quantum degenerate mixtures},\
  }\href {https://doi.org/10.1103/PhysRevA.79.021601} {\bibfield  {journal}
  {\bibinfo  {journal} {Phys. Rev. A}\ }\textbf {\bibinfo {volume} {79}},\
  \bibinfo {pages} {021601} (\bibinfo {year} {2009})}\BibitemShut {NoStop}%
\bibitem [{\citenamefont {Meggers}\ and\ \citenamefont
  {Tech}(1978)}]{meggers:1978}%
  \BibitemOpen
  \bibfield  {author} {\bibinfo {author} {\bibfnamefont {W.}~\bibnamefont
  {Meggers}}\ and\ \bibinfo {author} {\bibfnamefont {J.}~\bibnamefont {Tech}},\
  }\bibfield  {title} {\bibinfo {title} {The {First} {Spectrum} of {Ytterbium}
  ({Yb} i)},\ }\href {https://doi.org/10.6028/jres.083.003} {\bibfield
  {journal} {\bibinfo  {journal} {J. Res. Natl. Bur. Stand. (1977)}\ }\textbf
  {\bibinfo {volume} {83}},\ \bibinfo {pages} {13} (\bibinfo {year}
  {1978})}\BibitemShut {NoStop}%
\bibitem [{\citenamefont {Lange}\ \emph {et~al.}(1970)\citenamefont {Lange},
  \citenamefont {Luther},\ and\ \citenamefont {Stendel}}]{lange:1970}%
  \BibitemOpen
  \bibfield  {author} {\bibinfo {author} {\bibfnamefont {W.}~\bibnamefont
  {Lange}}, \bibinfo {author} {\bibfnamefont {J.}~\bibnamefont {Luther}},\ and\
  \bibinfo {author} {\bibfnamefont {A.}~\bibnamefont {Stendel}},\ }\href@noop
  {} {\bibfield  {journal} {\bibinfo  {journal} {Proceedings, 2nd Conference of
  European Group for Atomic Spectroscopy}\ } (\bibinfo {year}
  {1970})}\BibitemShut {NoStop}%
\bibitem [{\citenamefont {Cho}\ \emph {et~al.}(2012)\citenamefont {Cho},
  \citenamefont {Lee}, \citenamefont {Lee}, \citenamefont {Ahn}, \citenamefont
  {Lee}, \citenamefont {Yu}, \citenamefont {Lee},\ and\ \citenamefont
  {Park}}]{cho:2012}%
  \BibitemOpen
  \bibfield  {author} {\bibinfo {author} {\bibfnamefont {J.~W.}\ \bibnamefont
  {Cho}}, \bibinfo {author} {\bibfnamefont {H.-g.}\ \bibnamefont {Lee}},
  \bibinfo {author} {\bibfnamefont {S.}~\bibnamefont {Lee}}, \bibinfo {author}
  {\bibfnamefont {J.}~\bibnamefont {Ahn}}, \bibinfo {author} {\bibfnamefont
  {W.-K.}\ \bibnamefont {Lee}}, \bibinfo {author} {\bibfnamefont {D.-H.}\
  \bibnamefont {Yu}}, \bibinfo {author} {\bibfnamefont {S.~K.}\ \bibnamefont
  {Lee}},\ and\ \bibinfo {author} {\bibfnamefont {C.~Y.}\ \bibnamefont
  {Park}},\ }\bibfield  {title} {\bibinfo {title} {Optical repumping of
  triplet-$p$ states enhances magneto-optical trapping of ytterbium atoms},\
  }\href {https://doi.org/10.1103/PhysRevA.85.035401} {\bibfield  {journal}
  {\bibinfo  {journal} {Phys. Rev. A}\ }\textbf {\bibinfo {volume} {85}},\
  \bibinfo {pages} {035401} (\bibinfo {year} {2012})}\BibitemShut {NoStop}%
\end{thebibliography}
\end{document}